\documentclass{article}

\usepackage{epsf}
\usepackage{longtable}
\usepackage{amssymb,amsmath,amsfonts}

\usepackage{epubtk}
\usepackage{graphicx}

\newcommand{\be}{\begin{equation}}
\newcommand{\ee}{\end{equation}}
\newcommand{\beq}{\begin{equation}}
\newcommand{\eeq}{\end{equation}}
\newcommand{\beqa}{\begin{eqnarray}}
\newcommand{\eeqa}{\end{eqnarray}}
\newcommand{\bea}{\begin{eqnarray}}
\newcommand{\eea}{\end{eqnarray}}

\newcommand{\gsim}{\mathrel{\raisebox{-.6ex}{$\stackrel{\textstyle>}{\sim}$}}}

\begin{document}

\title{\bf Black Holes in Higher Dimensions}

\author{%
\epubtkAuthorData{\bf Roberto Emparan}{%
Instituci\'o Catalana de Recerca i Estudis
Avan\c cats (ICREA)\\
and\\
Departament de F{\'\i}sica Fonamental, Universitat de Barcelona\\
Marti i Franqu{\`e}s 1, E-08028 Barcelona, Spain}{%
emparan@ub.edu}{%
}%
\\
~\\
\epubtkAuthorData{\bf Harvey S. Reall}{%
Department of Applied Mathematics and Theoretical Physics\\
University of Cambridge, Centre for Mathematical Sciences\\
Wilberforce Road, Cambridge CB3 0WA \\
United Kingdom}{%
hsr1000@cam.ac.uk}{%
}%
}

\date{}
\maketitle

\begin{abstract}

We review black hole solutions of higher-dimensional vacuum gravity, and
of higher-dimensional supergravity theories. The discussion of vacuum
gravity is pedagogical, with detailed reviews of Myers-Perry solutions, black
rings, and solution-generating techniques. We discuss black
hole solutions of maximal supergravity theories, including black holes
in anti-de Sitter space. General results and open problems are discussed
throughout.

\end{abstract}

\epubtkKeywords{black holes, string theory, supergravity}

\newpage

\small
\tableofcontents
\normalsize

\newpage


\section{Introduction}
\label{section:introduction}

Classical General Relativity in more than four spacetime dimensions has
been the subject of increasing attention in recent years. Among the
reasons why it should be interesting to study this extension of
Einstein's theory, and in particular its black hole solutions, we may
mention that

\begin{itemize}

\item String theory contains gravity and requires more than four
dimensions. In fact, the first successful statistical counting of black
hole entropy in string theory was performed for a five-dimensional black
hole \cite{stromingervafa}. This example provides the best laboratory for the microscopic
string theory of black holes.

\item The AdS/CFT correspondence 
relates the properties of a
$d$-dimensional black hole with those of a quantum field theory in $d-1$
dimensions \cite{adscftreview}. 

\item The production of higher-dimensional black holes in future
colliders becomes a conceivable possibility in scenarios involving large
extra dimensions and TeV-scale gravity \cite{Cavaglia:2002si,Kanti:2004nr}.

\item As mathematical objects, black hole spacetimes are among the most
important Lorentzian Ricci-flat manifolds in any dimension.

\end{itemize}

These, however, refer to {\it applications} of the subject--important
though they are-- but we believe that higher-dimensional gravity is also
of intrinsic interest. Just as the study of quantum field theories
with a field content very different than any conceivable extension of
the Standard Model has been a very useful endeavour throwing light on
general features of quantum fields, we believe that endowing General
Relativity with a tunable parameter--namely the spacetime dimensionality
$d$--should also lead to valuable insights into the nature of the
theory, in particular of its most basic objects: the black holes. For
instance, four-dimensional black holes are known to possess a number of
remarkable features, such as uniqueness, spherical topology, dynamical
stability, and the laws of black hole mechanics. One would like to know
which of these are peculiar to four-dimensions, and which hold more
generally. At the very least, this study will lead to a deeper
understanding of classical black holes and of what spacetime can do at
its most extreme.

There is a growing awareness that the physics of higher-dimensional
black holes can be markedly different, and much richer, than in four
dimensions. Arguably, two advances are largely responsible for this
perception: the discovery of dynamical instabilities of extended black
hole horizons \cite{Gregory:1993vy}, and the discovery of black hole
solutions with horizons of non-spherical topology and not fully
characterized by their conserved charges \cite{Emparan:2001wn}.

At the risk of anticipating results and concepts that will be developed
only later in this review, in the following we try to give
simple answers to two frequently asked questions: 1) why should one expect
any interesting new dynamics in higher-dimensional General Relativity,
and 2) what are the main obstacles to a direct generalization of the
four-dimensional techniques and results. A straightforward answer to
both questions is to simply say that as the number of dimensions grows,
the number of degrees of freedom of the gravitational field also
increases, but more specific yet intuitive answers are possible. 

\subsection{Why gravity is richer in $d>4$}

The novel features of higher-dimensional black holes that have been
identified so far can be understood in physical terms as due to the
combination of two main ingredients: different rotation dynamics, and
the appearance of extended black objects. 

There are two aspects of rotation that change significantly when
spacetime has more than four dimensions. First, there is the possibility
of rotation in several independent rotation planes \cite{myersperry}.
The rotation group
$SO(d-1)$ has Cartan subgroup $U(1)^{N}$, 
with
\beq
N \equiv \left\lfloor
\frac{d-1}{2} \right\rfloor\,, 
\eeq
hence there is the possibility of $N$
independent angular momenta. In simpler and
more explicit terms, group the $d-1$ spatial dimensions (say, at
asymptotically flat infinity) in pairs $(x_1,x_2)$, $(x_3,x_4)$,\dots, 
each pair defining a plane, and choose polar coordinates in each plane,
$(r_1,\phi_1)$, $(r_2,\phi_2)$,\dots. Here we see the possibility of having
$N$ independent (commuting) rotations associated to the vectors
$\partial_{\phi_1}$, $\partial_{\phi_2}$ \dots. To each of these
rotations we associate an angular momentum
component $J_i$.

The other aspect of rotation that changes qualitatively as the number of
dimensions increases is the relative competition between the
gravitational and centrifugal potentials. The radial fall-off of the
Newtonian potential
\beq
-\frac{GM}{r^{d-3}}
\label{newtpot}
\eeq
depends on the number of dimensions, whereas the
centrifugal barrier
\beq
\frac{J^2}{M^2 r^2}
\label{rotpot}
\eeq
does not, since rotation is confined to a plane. We see that the
competition between (\ref{newtpot}) and (\ref{rotpot}) is different in
$d=4$, $d=5$, and $d\geq 6$. In Newtonian physics this is well-known to
result in a different stability of Keplerian orbits, but this precise
effect is not directly relevant to the black hole dynamics we are
interested in. Still, the same kind of dimension-dependence will have
rather dramatic consequences for the behavior of black holes.

The other novel ingredient that appears in $d>4$ but is absent in lower
dimensions (at least in vacuum gravity) is the presence of black objects
with extended horizons, i.e., black strings and in general black
$p$-branes. Although these are not asymptotically flat solutions, they
provide the basic intuition for understanding novel kinds of
asymptotically flat black holes. 

Let us begin from the simple observation that, given a black hole
solution of the vacuum Einstein equations in $d$ dimensions, with
horizon geometry $\Sigma_H$, then we can immediately construct a vacuum
solution in $d+1$ dimensions by simply adding a flat spatial
direction\epubtkFootnote{This is no longer true if the field equations involve
not only the Ricci tensor but also the Weyl tensor, such as in Lovelock
theories.}. The new horizon geometry is then a \emph{black string} with
horizon $\Sigma_H\times {\mathbb R}$. Since the Schwarzschild solution
is easily generalized to any $d\geq 4$, it follows that black strings
exist in any $d\geq 5$. In general, adding $p$ flat directions we find
that black $p$-branes with horizon $S^q\times {\mathbb R}^p$ (with
$q\geq 2$) exist in any $d\geq 6+p-q$.

How are these related to new kinds of asymptotically flat black holes?
Heuristically, take a piece of black string, with $S^q\times {\mathbb
R}$ horizon, and curve it to form a {\it black ring} with horizon
topology $S^q\times S^1$. Since the black string has a tension, then the
$S^1$, being contractible, will tend to collapse. But we may try to set the
ring into rotation and in this way provide a centrifugal repulsion that
balances the tension. This turns out to be possible in any $d\geq 5$, so
we expect that non-spherical horizon topologies are a generic feature of
higher-dimensional General Relativity. 

It is also natural to try to apply this heuristic construction to black
$p$-branes with $p>1$, namely, to bend the worldvolume spatial
directions into a compact manifold, and balance the tension by
introducing suitable rotations. The possibilities are still under
investigation, but it is clear that an increasing variety of black holes
should be expected as $d$ grows. Observe again that the underlying
reason is a combination of extended horizons with rotation.

Horizon topologies other than spherical are forbidden in $d=4$ by
well-known theorems \cite{hawkingellis}. 
These are rigorous, but also rather technical and
formal results. Can we find a simple, intuitive explanation for the {\it
absence} of vacuum black rings in $d=4$? The previous argument would
trace this fact back to the absence of asymptotically flat vacuum black
holes in $d=3$. This is often attributed to the absence of propagating
degrees of freedom for the three-dimensional graviton (or one of its
paraphrases: $2+1$-gravity is topological, the Weyl tensor vanishes
identically, etc), but here we shall use the simple observation that the
quantity $GM$ is dimensionless in $d=3$. Hence, given any amount of
mass, there is no length scale to tell us where the black hole horizon
could be\epubtkFootnote{It follows that the introduction of a length scale,
for instance in the form of a (negative) cosmological constant, is a
necessary condition for the existence of a black hole in $2+1$
dimensions. But gravity may still remain topological.}. So we would
attribute the absence of black strings in $d=4$ to the lack of such a
scale. This observation goes some way towards understanding the absence
of vacuum black rings with horizon topology $S^1\times S^1$ in four
dimensions: it implies that there cannot exist black ring solutions
with different scales for each of the two circles, and in particular one
could not make one radius arbitrarily larger than the
other. This argument, though, could still allow for black rings where
the radii of the two $S^1$ are set by the same scale, i.e., the black
rings should be plump. The horizon topology theorems then tell us that
plump black rings do not exist: they would actually be within a
spherical horizon.

Extended horizons also introduce a feature absent in $d=4$: dynamical
horizon instabilities \cite{Gregory:1993vy}. Again, this is to some
extent an issue of scales. Black brane horizons can be much larger in
some of their directions than in others, and so perturbations with
wavelength of the order of the `short' horizon length can fit several
times along the `long' extended directions. Since the horizon area tends
to increase by dividing up the extended horizon into black holes of
roughly the same size in all its dimensions, this provides grounds to
expect an instability of the extended horizon (however, when other
scales are present, as in charged solutions, the situation can become
quite more complicated). It turns out that higher-dimensional rotation
can make the horizon much more extended in some directions than in
others, which is expected to trigger this kind of instability
\cite{Emparan:2003sy}. At the threshold of the instability, a zero-mode
deformation of the horizon has been conjectured to lead to new `pinched'
black holes that do not have four-dimensional counterparts.

Finally, an important question raised in higher dimensions refers to the
rigidity of the horizon. In four dimensions, stationarity implies the
existence of a $U(1)$ rotational isometry \cite{hawkingellis}. In higher
dimensions stationarity has been proven to imply one rigid rotation
symmetry too \cite{hiw}, but not (yet?) more than one. However, all {\it
known} higher-dimensional black holes have multiple rotational
symmetries. Are there stationary black holes with less symmetry, for
example just the single $U(1)$ isometry guaranteed in general? Or are
black holes always as rigid as can be? This is, in our opinion, the main
unsolved problem on the way to a complete classification of
five-dimensional black holes, and an important issue in understanding
the possibilities for black holes in higher dimensions.

\subsection{Why gravity is more difficult in $d>4$}

Again, the simple answer to this question is the larger number of
degrees of freedom. However, this can not be an entirely satisfactory
reply, since one often restricts to solutions with a large degree of
symmetry for which the number of actual degrees of freedom may not depend on
the dimensionality of spacetime. A more satisfying answer should explain
why the methods that are so successful in $d=4$ become harder, less
useful, or even inapplicable, in higher dimensions. 

Still, the larger number of metric components and of equations
determining them is the main reason for the failure so far to find a useful
extension of the Newman-Penrose (NP) formalism to $d>4$. This formalism,
in which all the Einstein equations and Bianchi identities are written
out explicitly, was instrumental in deriving the Kerr solution and
analyzing its perturbations. The formalism is tailored to deal with
algebraically special solutions, but even if algebraic classifications
have been developed for higher dimensions \cite{coley}, and applied to known black
hole solutions, no practical extension of the NP formalism has appeared
yet that can be used to derive the solutions nor to study their
perturbations.

Then, it seems natural to restrict to solutions with a high degree of
symmetry. Spherical symmetry yields easily by force of Birkhoff's
theorem. The next simplest possibility is to impose stationarity and
axial symmetry. In four dimensions this implies the existence of two
commuting abelian isometries--time translation and axial rotation--,
which is extremely powerful: by integrating out the two isometries from
the theory we obtain an integrable two-dimensional $GL(2,\mathbb{R})$
sigma-model. The literature on these theories is enormous and many
solution-generating techniques are available, which provide a variety of
derivations of the Kerr solution. 

There are two natural ways of extending axial symmetry to higher
dimensions. We may look for solutions invariant under the group $O(d-2)$
of spatial rotations around a given line axis, where the orbits of $O(d-2)$ are $(d-3)$-spheres. However, in more than
four dimensions these orbits have non-zero curvature. As a
consequence, after dimensional reduction on these orbits, the sigma-model acquires terms (of exponential type) that prevent a straightforward integration of the equations (see \cite{Charmousis:2003wm,Charmousis:2006fx} for an investigation of these equations). 

This suggests looking for a different higher-dimensional extension of
the four-dimensional axial symmetry. Instead of rotations around a line,
consider rotations around (spatial) codimension-2 hypersurfaces. 
These are $U(1)$ symmetries. If we assume $d-3$ commuting $U(1)$ symmetries, so that we have a spatial $U(1)^{d-3}$ symmetry in addition to the 
timelike symmetry ${\mathbb
R}$, then the vacuum Einstein
equations again reduce to an integrable two-dimensional
$GL(d-2,\mathbb{R})$ sigma-model with powerful
solution-generating techniques. 

However, there is an important limitation: only in $d=4,5$ can these
geometries be globally asymptotically flat. Global asymptotic flatness
implies an asymptotic factor $S^{d- 2}$ in the spatial geometry, whose
isometry group $O(d-1)$ has a Cartan subgroup $U(1)^{N}$. If, as above,
we demand $d-3$ axial isometries, then, asymptotically, these symmetries
must approach elements of $O(d-1)$, so we need $U(1)^{d-3} \subset
U(1)^N$, i.e.,
\beq 
d-3 \le  N = \left\lfloor \frac{d-1}{2} \right\rfloor,
\eeq 
which is only possible in $d=4,5$. This is the main reason for the
recent great progress in the construction of exact five-dimensional
black holes, and the failure to extend it to $d>5$.

Finally, the classification of possible horizon topologies becomes
increasingly complicated in higher dimensions \cite{gallowayschoen}. In four spacetime
dimensions the (spatial section of the) horizon is a two-dimensional
surface, so the possible topologies can be easily characterized and
restricted. Much less restriction is possible as $d$ is increased.

All these aspects will be discussed in more detail below.

\newpage


\section{Scope and organization of this article}
\label{section:organisation}

\subsection{Scope}

The emphasis of this article is on classical properties of exact
higher-dimensional black hole solutions. We devote most space to a
rather pedagogical discussion of vacuum solutions. Since this includes
black rings, there is some overlap with our earlier review
\cite{Emparan:2006mm}. The present review discusses material that has
appeared since \cite{Emparan:2006mm}, in particular the ``doubly
spinning" black ring solution of \cite{Pomeransky:2006bd}. However, we
shall not discuss several aspects of black ring physics that were dealt
with at length in \cite{Emparan:2006mm}, for example, black ring
microphysics. On the other hand, we present some new material: figures
\ref{figure:7D8Dphasespace}, \ref{figure:doublering},
\ref{figure:5Dphasespace}, \ref{fig:ads4bh}, \ref{fig:ads5bh} describing
the physical parameter ranges (phase space) of higher-dimensional black
holes, and figure \ref{figure:mp5d} for the area of 5D Myers-Perry
solutions, have not been presented earlier. Some of our discussion of
the properties of the solutions is also new.

Our discussion of non-vacuum black holes is less pedagogical than that
of the vacuum solutions. It is essentially a survey of the literature.
In going beyond vacuum solutions, we had to decide what kinds of matter
field to consider. Since much of the motivation for the study of extra
dimensions comes from string theory, we have restricted ourselves to
considering black hole solutions of supergravity theories known to arise
as consistent truncations of $d=10,11$ supergravity. We consider both
asymptotically flat, and asymptotically anti-de Sitter black holes. 

In the asymptotically flat case, we consider only solutions of maximal
supergravity theories arising from toroidal reduction of $d=10,11$
supergravity to five or more dimensions. In particular, this implies
that in five dimensions we demand the presence of a Chern-Simons term
for the gauge field, with a precise coefficient. A review of charged
rotating black holes with other values for the Chern-Simons coupling can
be found in \cite{Kleihaus:2007kc}.

In the asymptotically AdS case, we consider solutions of gauged
supergravity theories arising from dimensional reduction of $d=10,11$
supergravity on spheres, in particular the maximal gauged supergravity
theories in $d=4,5,7$. Obviously $d=4$ does not fall within our
``higher-dimensional" remit but asymptotically AdS $d=4$ black holes are
not as familiar as their asymptotically flat cousins so it seems
worthwhile reviewing them here. In AdS, several different asymptotic
boundary conditions are of physical interest. We consider only black
holes obeying standard ``normalizable" boundary conditions
\cite{adscftreview}. Note that all known black hole solutions satisfying
these restrictions involve only abelian gauge fields.

Important related subjects that we do not discuss include: black holes
in brane-world scenarios \cite{Maartens:2003tw}; black holes in
spacetimes with Kaluza-Klein asymptotics \cite{Harmark:2005pp}, and in
general black holes with different asymptotics than flat or AdS; black
holes in higher-derivative theories \cite{myersreview}; black hole
formation at the LHC or in cosmic rays, and the spectrum of their
radiation \cite{Cavaglia:2002si,Kanti:2004nr}.

\subsection{Organization}

Sections~\ref{section:equations} to \ref{section:vachigherd} are devoted
to asymptotically flat vacuum solutions: section~\ref{section:equations}
introduces basic notions and solutions, in particular the
Schwarzschild-Tangherlini black hole. Section~\ref{section:myersperry}
presents the Myers-Perry solutions, first with a single angular
momentum, then with arbitrary rotation. Section~\ref{section:vac5d}
reviews the great recent progress in five-dimensional vacuum black
holes: first we discuss black rings, with one and two angular momenta;
then we introduce the general analysis of solutions with two rotational
isometries (or $d-3$, in general). In section~\ref{section:vachigherd}
we briefly describe a first attempt at understanding $d\geq 6$ vacuum
black holes beyond the MP solutions.

Section~\ref{section:charged} reviews asymptotically flat black holes
with gauge fields (within the restricted class mentioned above).
Section~\ref{section:general} concludes our overview of asymptotically
flat solutions (vacuum and charged) with a discussion of general results
and some open problems.
Finally, section~\ref{section:cosconst} reviews asymptotically AdS black hole solutions of gauged supergravity theories.

\newpage


\section{Basic concepts and solutions}
\label{section:equations}

In this section we present the basic framework for General Relativity in
higher dimensions, beginning with the definition of conserved charges in
vacuum --i.e., mass and angular momentum--, and the introduction of a
set of dimensionless variables that are convenient to describe the phase
space and phase diagram of higher-dimensional rotating black holes. Then
we introduce the Tangherlini solutions that generalize the
four-dimensional Schwarzschild solution. The analysis that proves their
classical stability is then reviewed. Black strings and black
$p$-branes, and their Gregory-Laflamme instability, are briefly
discussed for their relevance to novel kinds of rotating black holes.


\subsection{Conserved charges}

The Einstein-Hilbert action is generalized to higher dimensions in the form
\beq
I=\frac{1}{16\pi G}\int d^d x \sqrt{-g}R +I_{matter}.
\eeq
This is a straightforward generalization, and the only aspect that deserves some
attention is the implicit definition of Newton's constant $G$ in $d$ dimensions.
It enters the Einstein equations in the conventional form
\beq
R_{\mu\nu}-\frac{1}{2}g_{\mu\nu}R=8\pi G T_{\mu\nu}\,.
\eeq
where $T_{\mu\nu}=2(-g)^{-1/2} (\delta I_{matter}/\delta g^{\mu\nu})$. This
definition of the gravitational coupling constant, without any
additional dimension-dependent factors, has the notable advantage that
the Bekenstein-Hawking entropy formula takes the same form
\beq\label{bhentropy}
S=\frac{{\mathcal A}_H}{4 G}
\eeq
in every dimension. This follows, e.g., from the standard Euclidean
quantum gravity calculation of the entropy.

Mass, angular momenta, and other conserved charges of isolated systems
are defined by comparing to the field created near asymptotic infinity
by a weakly gravitating system (ref.~\cite{Jamsin:2007qh} gives a careful
Hamiltonian analysis of conserved charges in higher-dimensional
asymptotically flat spacetimes). The Einstein equations for a small
perturbation around flat Minkowski space
\beq
g_{\mu\nu}=\eta_{\mu\nu}+h_{\mu\nu}
\eeq
in linearized approximation take the conventional form
\beq
\Box \bar h_{\mu\nu}=-16\pi G T_{\mu\nu}
\eeq
where $\bar h_{\mu\nu}=h_{\mu\nu}-\frac{1}{2}h \eta_{\mu\nu}$ and we have
imposed the transverse gauge
condition $\nabla_{\mu}\bar h^{\mu\nu}=0$.

Since the sources are localized and we work at linearized perturbation
order, the fields in the asymptotic region are the same as created by
pointlike sources of mass $M$ and
angular momentum with antisymmetric matrix $J_{ij}$, at the origin
$x^k=0$ of flat
space in Cartesian coordinates,
\beqa
T_{tt}&=& M \delta^{(d-1)}(x^k),\\
T_{ti}&=& -\frac{1}{2}J_{ij}\nabla_{j}\delta^{(d-1)}(x^k).
\eeqa
The equations are easily integrated, assuming stationarity, to find
\beqa
\bar h_{tt}&=& \frac{16\pi G}{(d-3)\Omega_{d-2}}\frac{M}{r^{d-3}},\\
\bar h_{ti}&=& -\frac{8\pi G}{\Omega_{d-2}}\frac{x^k J_{ki}}{r^{d-1}},
\eeqa
where $r=\sqrt{x^i x^i}$, and
$\Omega_{d-2}=2\pi^{(d-1)/2}/\Gamma\left(\frac{d-1}{2}\right)$ is the
area of a unit $(d-2)$-sphere. From
here we recover the metric perturbation
$h_{\mu\nu}=\bar h_{\mu\nu}-\frac{1}{d-2}\bar h
\eta_{\mu\nu}$ as
\beqa\label{hmunu}
h_{tt}&=& \frac{16\pi G}{(d-2)\Omega_{d-2}}\frac{M}{r^{d-3}},\\
h_{ij}&=& \frac{16\pi G}{(d-2)(d-3)\Omega_{d-2}}\frac{M}{r^{d-3}}\delta_{ij},\\
h_{ti}&=& -\frac{8\pi G}{\Omega_{d-2}}\frac{x^k J_{ki}}{r^{d-1}}.
\eeqa
It is often convenient to have the off-diagonal rotation components of
the metric in a different form. By making a suitable coordinate rotation
the angular momentum matrix $J_{ij}$ can be put in block-diagonal form,
each block being a $2\times 2$ antisymmetric matrix with parameter 
\beq
J_a\equiv J_{2a-1,2a}.
\eeq
Here $a=1,\dots, N$ labels the different
independent rotation planes. If we introduce polar coordinates on each
of the planes 
\beq
(x_{2a-
1},x_{2a})=(r_a\cos\phi_a,r_a\sin\phi_a)
\eeq
then (no sum over $a$)
\beq\label{hti}
h_{t\phi_a}=-\frac{8\pi G J_a}{\Omega_{d-2}}\frac{r_a^2}{r^{d-1}}
=-\frac{8\pi G J_a}{\Omega_{d-2}}\frac{\mu_a^2}{r^{d-3}}.
\eeq 
In the last expression we have introduced the `direction cosines'
\beq
\mu_a=\frac{r_a}{r}\,.
\eeq

Given the abundance of black hole solutions in higher dimensions, one is
interested in comparing properties, such as the horizon area
$\mathcal{A}_H$, of different solutions characterized by the same set of
parameters $(M, J_a)$. A meaningful comparison between dimensionful magnitudes
requires the introduction of a common scale, so the comparison is made
between dimensionless magnitudes obtained by factoring out this scale.
Since classical General Relativity in vacuum is scale-invariant, the
common scale must be one of the physical parameters of the solutions,
and a natural choice is the mass. Thus we introduce dimensionless quantities
for the spins $j_a$ and the area $a_H$, 
\beq
\label{jaHdef}
j^{d-3}_a=c_J\,
\frac{J^{d-3}_a}{GM^{d-2}} \,,\qquad
a_H^{d-3}=c_\mathcal{A}\,\frac{\mathcal{A}_H^{d-3}}{(GM)^{d-2}}\,,
\eeq
where the numerical constants are
\beq
c_J =\frac{\Omega_{d-3}}{2^{d+1}}\frac{(d-2)^{d-2}}{(d-3)^{\frac{d-3}{2}}}\,,\qquad
c_\mathcal{A}=\frac{\Omega_{d-3}}{2(16\pi)^{d-3}}(d-2)^{d-2}
\left(\frac{d-4}{d-3}\right)^{\frac{d-3}{2}}\,
\eeq
(these definitions follow the choices in \cite{Emparan:2007wm}).
Studying the entropy, or the area $\mathcal{A}_H$, as a function
of $J_a$ for fixed mass is equivalent to finding the function $a_H (j_a)$.

Note that with our definition of the gravitational constant $G$ 
both the Newtonian gravitational potential energy,
\beq
\Phi=-\frac{1}{2}h_{tt},
\eeq
and the force law (per unit mass) 
\beq
{\mathbf F}=-\nabla\Phi=\frac{(d-3)8\pi G}{(d-2)\Omega_{d-
2}}\frac{M}{r^{d-2}}\hat{\mathbf r}
\eeq
acquire $d$-dependent numerical prefactors. Had we chosen to
define Newton's constant so as to absorb these factors in the expressions for
$\Phi$ or ${\mathbf F}$, eq.~(\ref{bhentropy}) would have been more
complicated. 

To warm up before dealing with black holes, we follow John Michell and
Simon de Laplace and compute, using Newtonian mechanics, the radius at
which the escape velocity of a test particle in this field reaches the
speed of light. The kinetic energy of a particle of unit mass with
velocity $v=c=1$ is $K=1/2$, so the equation $K+\Phi=0$ that determines
the Michell-Laplace `horizon' radius is
\beq\label{michell}
h_{tt}(r=r_{ML})=1\quad \Rightarrow \quad r_{ML}=\left(\frac{16\pi G M}{(d-
2)\Omega_{d-2}}\right)^{\frac{1}{d-3}}.
\eeq
We will see in the next section that, just like in four dimensions, this is
precisely equal to the horizon radius for a static black hole in higher dimensions.


\subsection{The Schwarzschild-Tangherlini solution and black $p$-branes}

Consider the linearized solution above for a static source (\ref{hmunu})
in spherical coordinates, and pass to a gauge where $r$
is the
area radius,
\beq
r\to r-\frac{8\pi G}{(d-2)(d-3)\Omega_{d-2}}\frac{M}{r^{d-3}}.
\eeq
The linearized approximation to the field of a static source is then
\beq
ds^2_{(lin)}=-\left(1-\frac{\mu}{r^{d-
3}}\right)dt^2+\left(1+\frac{\mu}{r^{d-3}}\right)dr^2+r^2 d\Omega_{d-
2}^2,
\eeq
where, to lighten the notation, we have introduced the `mass parameter'
\beq
\mu=\frac{16\pi G M}{(d-2)\Omega_{d-2}}.
\eeq
This suggests that the
Schwarzschild solution generalizes to higher dimensions in the form
\beq
ds^2=-\left(1-\frac{\mu}{r^{d-3}}\right)dt^2+\frac{dr^2}{1-
\frac{\mu}{r^{d-3}}}+r^2 d\Omega_{d-2}^2\,.
\eeq
In essence, all we have done is change the radial fall-off $1/r$ of the
Newtonian potential to the $d$-dimensional one, $1/r^{d-3}$. As
Tangherlini found in 1963 \cite{Tangherlini:1963bw}, this turns out to
give the correct solution: it is straightforward to check that this
metric is indeed Ricci-flat. It is apparent that there is an event
horizon at $r_0=\mu^{1/(d-3)}$, which coincides with the
Michell-Laplace result \eqref{michell}.

Having this elementary class of black hole solutions, it is easy to construct other
vacuum solutions with event horizons in $d\geq 5$. The direct product of
two Ricci-flat manifolds is itself a Ricci-flat manifold. So, given any
vacuum black hole solution $\mathcal{B}$ of the Einstein equations in
$d$ dimensions, the metric
\beq
ds^2_{d+p}=ds^2_{d}(\mathcal{B})+\sum_{i=1}^{p}dx^i dx^i
\eeq
describes a black $p$-brane, in which the black hole horizon $\mathcal
H\subset\mathcal{B}$ is extended to a horizon ${\mathcal H}\times
{\mathbb R}^p$, or ${\mathcal H}\times {\mathbb T}^p$ if we identify
periodically $x^i\sim x^i + L_i$. A simple way of obtaining another kind
of vacuum solutions is the following: unwrap one of the directions
$x^i$; perform a boost $t\to \cosh\alpha t +\sinh\alpha x^i$, $x^i\to
\sinh\alpha t +\cosh\alpha x^i$, and re-identify points periodically
along the new coordinate $x^i$. Although locally equivalent to the
static black brane, the new boosted black brane solution is globally
different from it.

These black brane spacetimes are not (globally) asymptotically flat, so
we only introduce them insofar as they are relevant for understanding
the physics of asymptotically flat black holes.

\subsection{Stability of the static black hole}

The stability of the $d>4$ Schwarzschild solution against linearized 
gravitational perturbations can be analyzed by decomposing such 
perturbations into scalar, vector and tensor types according to how they 
transform under the rotational symmetry group $SO(d-1)$ 
\cite{HG,kodama:03a,kodama:03}. Assuming a time dependence $e^{-i\omega 
t}$ and expanding in spherical harmonics on $S^{d-2}$, the equations 
governing each type of perturbation reduce to a single ODE governing the 
radial dependence. This equation can be written in the form of a 
time-independent Schr\"odinger equation with ``energy" eigenvalue 
$\omega^2$. 

In investigating stability, we consider perturbations that are regular
on the future horizon and outgoing at infinity. An instability would
correspond to a mode with $\mathrm{Im}\; \omega > 0$. For such modes,
the boundary conditions at the
horizon and infinity imply that the LHS of the Schr\"odinger equation is
self-adjoint, and hence $\omega^2$ is real. Therefore an unstable mode
must have negative imaginary $\omega$. For tensor modes, the potential
in the Schr\"odinger equation is manifestly positive, hence $\omega^2>0$
and there is no instability \cite{HG}. For vectors and scalars, the
potential is not everywhere positive. Nevertheless, it can be shown that
the operator appearing on the LHS of the Schr\"odinger equation is
positive, hence $\omega^2>0$ and there is no instability
\cite{kodama:03}. In conclusion, the $d>4$ Schwarzschild solution is
stable against linearized gravitational perturbations.


\subsection{Gregory-Laflamme instability}
\label{sec:GLinstab}

The instabilities of black strings and black branes
\cite{Gregory:1993vy,Gregory:1994bj} have been reviewed in
\cite{Kol:2004ww,Harmark:2007md}, so we shall be brief in this section and only
mention the features that are most relevant to our subject. We shall
only discuss neutral black holes and black branes: when charges are
present, the problem becomes quite more complex.

This instability is the prototype for situations where the size of the
horizon is much larger in some directions than in others. Consider, as a
simple, extreme case of this, the black string obtained by adding a flat
direction $z$ to the Schwarzschild solution. One can decompose linearized gravitational perturbations into scalar, vector and tensor types according to how they transform with respect to transformations of the Schwarzschild coordinates.
Scalar and vector perturbations of this solution are stable \cite{Gregory:1987nb}. Tensor perturbations that are homogeneous along the $z$-direction are also
stable, since they are the same as tensor perturbations of the
Schwarzschild black hole. However, there appears an instability for
long-wavelength tensor perturbations with non-trivial dependence on $z$:
the frequency $\omega$ of perturbations $\sim e^{-i(\omega t -kz)}$
acquires a positive imaginary part when $k<k_{GL}\sim 1/r_0$, where
$r_0$ is the Schwarzschild horizon radius. Thus, if the string is
compactified on a circle of length $L >2\pi/k_{GL} \sim r_0$, it becomes
unstable. Of the unstable modes, the fastest one (with the largest
imaginary frequency) occurs for $k$ roughly one half of $k_{GL}$. The
instability creates inhomogeneities along the direction of the string.
Their evolution beyond the linear approximation has been followed
numerically in \cite{Choptuik:2003qd}. It is unclear yet what the
endpoint is: the inhomogeneities may well grow until a sphere pinches
down to a singularity.\epubtkFootnote{It has been shown that this requires infinite affine parameter distance along the null geodesics generators  of the horizon \cite{Horowitz:2001cz}. However, it may still take finite time as measured by an external observer \cite{Marolf:2005vn}.} In
this case, the Planck scale will be reached along the evolution, and
fragmentation of the black string into black holes, consistently with an
increase in the total horizon entropy, may occur. 

Another important feature of this phenomenon is the appearance of a
zero-mode (i.e., static) perturbation with $k=k_{GL}$. Perturbing the
black string with this mode yields a new static solution with
inhomogeneities along the string direction
\cite{Gregory:1987nb,Gubser:2001ac}. Following numerically these static
perturbations beyond the linear approximation has given a new class of
inhomogeneous black strings \cite{Wiseman:2002zc}.

These results easily generalize to black $p$-branes: for a wavevector
$\mathbf{k}$ along the $p$ directions
tangent to the brane, the perturbations  $\sim
\exp\left(-i\omega t +i
{\mathbf k}\cdot{\mathbf z})\right)$ with $|\mathbf{k}|\leq
k_{GL}$ are unstable. The value of
$k_{GL}$ depends on the codimension of the black brane but not on $p$.

\newpage


\section{Myers-Perry solutions}
\label{section:myersperry}

The generalization of the Schwarzschild solution to $d>4$ is, as we have
seen, a rather straightforward problem. However, in General Relativity
it is often very difficult to extend a solution from the static case to
the stationary one (as exemplified by the Kerr solution). Impressively, in
1986 Myers and Perry (MP) managed to find exact solutions for black holes in
any dimension $d>4$, rotating in all possible independent rotation
planes \cite{myersperry}. The feat was possible since the solutions
belong in the
Kerr-Schild class
\beq
g_{\mu\nu}=\eta_{\mu\nu}+2H(x^\rho) k_\mu k_\nu
\eeq
where $k_\mu$ is a null vector with respect to both $g_{\mu\nu}$ and the
Minkowski metric $\eta_{\mu\nu}$. This entails a sort of linearization
of the problem, which facilitates greatly the resolution of the
equations. Of all known vacuum black holes in $d>4$, only the
Myers-Perry solutions seem to have this property.

In this section we review these solutions and their properties,
beginning from the black holes with a single rotation, and then extend
them to arbitrary rotation. The existence of ultra-spinning regimes in
$d\geq 6$ is emphasized. The symmetries and stability of the MP solutions are
also discussed.


\subsection{Rotation in a single plane}
\label{subsec:singlespin}

Let us begin with the solutions that rotate in a single plane. These are
not only simpler, but they also exhibit more clearly the
qualitatively new physics afforded by the additional dimensions.

The metric takes the form
\begin{eqnarray}
ds^2&=& -dt^2 + \frac{\mu}{r^{d-5}\Sigma}\left( dt-a\sin^2\theta
\,d\phi\right)^2 +{\Sigma\over\Delta}dr^2+\Sigma d\theta^2
+(r^2+a^2)\sin^2\theta\, d\phi^2 \nonumber\\
&&
+ r^2\cos^2\theta\, d\Omega^2_{(d-4)}\,,
\label{mphole}
\end{eqnarray}
where
\beq \Sigma=r^2+a^2\cos^2\theta\,,\qquad
\Delta=r^2+a^2-\frac{\mu}{r^{d-5}}\,. 
\eeq
The physical mass and angular momentum are easily obtained by comparing
the asymptotic field to (\ref{hmunu}) and (\ref{hti}), and are given in
terms of the
parameters $\mu$ and $a$ by
\beq\label{mandj} 
M= \frac{(d-2)\Omega_{d-2}}{16\pi G}\mu\,,\qquad
J=\frac{2}{d-2}M a\,. 
\eeq
Hence
one can think of $a$ as essentially the angular momentum per unit
mass. We can choose $a\geq 0$ without loss of generality.\epubtkFootnote{This
choice corresponds to rotation in the positive sense (i.e. increasing
$\phi$). The solution presented in \cite{myersperry} is obtained by
$\phi \rightarrow -\phi$, which gives rotation in a negative sense.}

As in Tangherlini's solution, this metric seems to follow from a rather
straightforward extension of the Kerr solution, which is recovered when
$d=4$. The first line in eq.~(\ref{mphole}) looks indeed like the Kerr
solution, with the $1/r$ fall-off replaced, in appropriate places, by
$1/r^{d-3}$. The second line contains the line element on a
($d-4$)-sphere which accounts for the additional spatial dimensions. It
might therefore seem that, again, the properties of these black holes
should not differ much from their four-dimensional
counterparts.

However, this is not the case. Heuristically, we can see the competition
between gravitational attraction and centrifugal repulsion
in the expression
\beq
\frac{\Delta}{r^2}-1=-\frac{\mu}{r^{d-3}}+\frac{a^2}{r^2}\,. 
\eeq
Roughly, the first term on the right-hand side corresponds to the
attractive gravitational potential and falls off in a
dimension-dependent fashion. In contrast, the repulsive
centrifugal barrier described by the second term does not depend
on the total number of dimensions, since rotations always refer to
motions in a plane. 

Given the similarities between
\eqref{mphole} and the Kerr solution it is clear that the
outer event horizon lies at the largest (real)
root $r_0$ of $g_{rr}^{-1}=0$, i.e., $\Delta(r)=0$. 
Thus, we expect that the features of the event
horizons will be strongly dimension-dependent, and this is indeed the
case. If there is an event horizon at $r=r_0$,
\beq\label{horz} 
r_0^2+a^2 -\frac{\mu}{r_0^{d-5}}=0\,,
\eeq
its area will be
\beq\label{mparea}
\mathcal{A}_H=r_0^{d-4}(r_0^2+a^2)\Omega_{d-2}\,.
\eeq
For $d=4$, a regular horizon is present for values of the spin parameter
$a$ up to the Kerr bound: $a=\mu/2$ (or $a=GM$), which corresponds to an
extremal black hole with a single degenerate horizon (with vanishing
surface gravity). Solutions with $a>GM$ correspond to naked
singularities. In $d=5$, the situation is apparently quite similar since
the real root at $r_0=\sqrt{\mu-a^2}$ exists only up to the extremal
limit $\mu=a^2$. However, this extremal solution has zero area, and in
fact, has a naked ring singularity. 

For $d\geq 6$, $\Delta(r)$ is always positive at large values of $r$,
but the term $-\mu/r^{d-5}$ makes it negative at small $r$ (we are
assuming positive mass). Therefore
$\Delta$ always has a (single) positive real root independently of the
value of $a$. Hence regular black hole solutions exist with arbitrarily
large $a$.
Solutions with large angular momentum per unit mass are referred to as
``ultra-spinning''. 

An analysis of the shape of the horizon in the ultra-spinning regime
$a\gg r_0$ shows that the black holes flatten along the plane of
rotation \cite{Emparan:2003sy}: the extent of the horizon along this
plane is $\sim a$, while in directions transverse to this plane its size
is $\sim r_0$. In fact, a limit can be taken in which the ultraspinning
black hole becomes a black membrane with horizon geometry
$\mathbb{R}^2\times S^{d-4}$. This turns out to have important
consequences for black holes in $d\geq 6$, as we will discuss later. The
transition between the regime in which the black hole behaves like a
fairly compact, Kerr-like object, and the regime in which it is better
characterized as a membrane, is most clearly seen by analyzing the black
hole temperature
\begin{equation}
\label{TMP}
T_H = \frac{1}{4 \pi} \left( \frac{2 r_0}{r_0^2+a^2} + \frac{d-5}{r_0}
\right)\,.
\end{equation}
At
\beq\label{onset}
\left(\frac{a}{r_0}\right)_{\rm mem}=\sqrt{\frac{d-3}{d-5}}\,,
\eeq
this temperature reaches a minimum. For $a/r_0$ smaller than this value, quantities
like $T_H$ and $\mathcal{A}_H$ decrease, in a manner similar to the Kerr
solution.
However, past this point they rapidly approach the black membrane results in
which $T_H\sim 1/r_0$ and $\mathcal{A}_H\sim a^2 r_0^{d-4}$, with $a^2$
characterizing the area of the membrane worldvolume.

The properties of the solutions are conveniently encoded using the
dimensionless variables $a_H$, $j$ introduced in \eqref{jaHdef}.
For the solutions \eqref{mphole} the curve $a_H(j)$ can be found in parametric
form, in terms of the dimensionless `shape' parameter $\nu=\frac{r_0}{a}$, as
\beqa
j ^{d-3} &=& \frac{\pi}{(d-3)^\frac{d-3}{2}}
\frac{\Omega_{d-3}}{\Omega_{d-2}}\;
\frac{\nu^{5-d}}{1+\nu^2}\,,\label{jnu}\\
a_H^{d-3}&=&8\pi\left(\frac{d-4}{d-3}\right)^{\frac{d-3}{2}}
\frac{\Omega_{d-3}}{\Omega_{d-
2}}\;\frac{\nu^2}{1+\nu^2}\,.\label{ahnu}
\eeqa
The static and ultra-spinning limits correspond to $\nu\to\infty$ and
$\nu\to 0$, respectively. These curves are represented for $d=5,6,10$ in
figure~\ref{figure:aHj}. The
inflection point where $d^2 a_H/d j^2$ changes sign when
$d\geq 6$, occurs at the value \eqref{onset}.
\epubtkImage{}{%
\begin{figure}[h]
  \def\epsfsize#1#2{1#1}
  \centerline{\epsfbox{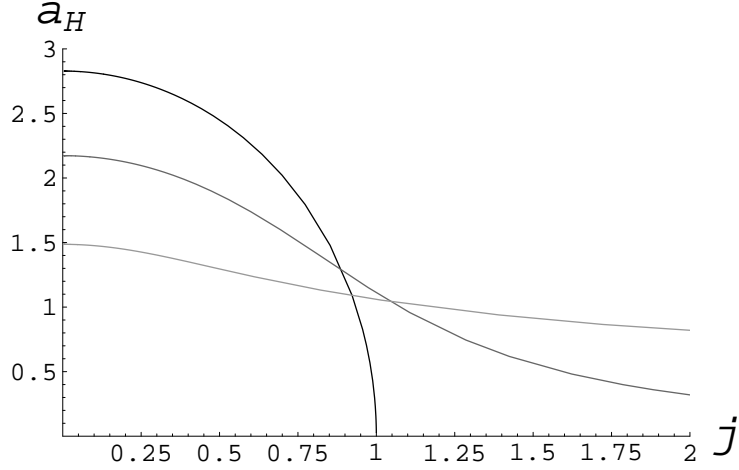}}
\caption{\it Horizon area vs.~angular momentum for Myers-Perry black
holes with a single spin in $d=5$ (black), $d=6$ (dark gray), $d=10$ (light
gray).}
  \label{figure:aHj}
\end{figure}}

 
\subsection{General solution}
\label{sec:generalMP}

Ref.~\cite{myersperry} also gave black hole solutions with arbitrary rotation in
each of the $N \equiv \left\lfloor \frac{d-1}{2} \right\rfloor$ independent
rotation planes. The cases of odd and even 
$d$ are slightly different.
When $d$ is odd the solution is 
\beq\label{odd}
ds^2=-dt^2+(r^2+a_i^2)(d\mu_i^2+\mu_i^2 d\phi_i^2) +\frac{\mu
r^2}{\Pi F}(dt-a_i\mu_i^2 d\phi_i)^2+\frac{\Pi F}{\Pi-\mu
r^2}dr^2\,. 
\eeq
Here and below $i=1,\dots, N$ and
we assume summation over $i$. The mass parameter is $\mu$, not to be
confused with the direction cosines $\mu_i$, which satisfy $\mu_i^2=1$.
For even $d$ the general solution is
\beq \label{even} 
ds^2=-dt^2+r^2
d\alpha^2+(r^2+a_i^2)(d\mu_i^2+\mu_i^2 d\phi_i^2) +\frac{\mu
r}{\Pi F}(dt-a_i\mu_i^2 d\phi_i)^2+\frac{\Pi F}{\Pi-\mu
r}dr^2\,,
\eeq
where now  $\mu_i^2+\alpha^2=1$. 

For both cases we can write the functions $F(r, \mu_i)$ and $\Pi(r)$ as
\beq\label{fpi}
F(r, \mu_i)=1-\frac{a_i^2\mu_i^2}{r^2+a_i^2}\,,\qquad
\Pi(r)=\prod_{i=1}^{N}(r^2+a_i^2)\,. 
\eeq
The relation between $\mu$ and $a_i$ and the mass and angular momenta is
the same as \eqref{mandj}. The event horizon is again at the largest
real root of $g^{rr}$, this is,
\beq\label{MPhorizon}
\Pi(r_0)-\mu r_0^2=0\quad
(\mathrm{odd}\; d)\,,\qquad 
\Pi(r_0)-\mu r_0=0\quad
(\mathrm{even}\; d)\,.
\eeq
The horizon area is
\beq\label{MParea}
\mathcal{A}_H=\frac{\Omega_{d-2}}{2\kappa}\mu\left(d-3
-\frac{2a_i^2}{r_0^2+a_i^2}\right)\,,
\eeq
and the surface gravity $\kappa$ is
\beq\label{MPkappa}
\kappa=\lim_{r\to r_0}\frac{\Pi'-2\mu r}{2\mu r^2}\quad
(\mathrm{odd}\; d)\,,\qquad 
\kappa=\lim_{r\to r_0}\frac{\Pi'-\mu}{2\mu r}\quad
(\mathrm{even}\; d)\,.
\eeq
Extremal solutions are obtained when $\kappa=0$ at the event horizon.

\subsubsection{Phase space}

The determination of $r_0$ involves an equation of degree $2N$ which in
general is difficult, if not impossible, to solve algebraically. So the
presence of horizons for generic parameters in \eqref{odd} and
\eqref{even} is difficult to ascertain. Nevertheless, a number of
features, in particular the ultra-spinning regimes that are important in
the determination of the allowed parameter range, can be analyzed.

Following eqs.~\eqref{jaHdef}, we can fix the mass and define
dimensionless quantities $j_i$ for each of the angular momenta. Up to a
normalization constant, the rotation parameters $a_i$ at fixed mass are
equivalent to the $j_i$. We take $(j_1,\dots,j_{N})$ as the coordinates
in the phase space of solutions. We aim to determine the region in this
space that corresponds to actual black hole solutions.

Consider first the case in which all spin parameters are non-zero. Then
an upper extremality bound on
a combination of the spins arises. If it is exceeded, naked
singularities appear, as in the $d=4$ Kerr black hole \cite{myersperry}.
So we can expect that, as long as all spin parameters take values not
too dissimilar, $j_1\sim j_2 \sim \dots \sim j_N$, all spins must remain
parametrically $O(1)$, i.e., there is no ultraspinning regime in which {\em
all} $j_i\gg 1$.

Next, observe that for odd $d$, a sufficient (but not necessary) condition for the
existence of a horizon is that any two of the spin parameters vanish,
i.e., if two $a_i$ vanish, a horizon will always exist irrespective of
how large the remaining spin parameters are. For even $d$, the existence
of a horizon is guaranteed if any one of the spins vanishes. Thus,
arbitrarily large (i.e., ultraspinning) values can be achieved for all but two (one) of the
$j_i$ in odd (even) dimensions. 

Assume then an ultraspinning regime in which $n$ rotation parameters are
comparable among themselves, and much larger than the remaining $N-n$
ones. A limit then exists to a black $2n$-brane of limiting horizon topology
$S^{d-2}\to \mathbb{R}^{2n}\times S^{d-2(n+1)}$. The limiting geometry
is in fact the direct product of $\mathbb{R}^{2n}$ and a
$(d-2n)$-dimensional Myers-Perry black hole \cite{Emparan:2003sy}.
Thus, in an ultraspinning
regime the allowed phase space of $d$-dimensional black holes can be
inferred from that of $(d-2n)$-dimensional black holes. Let us then begin
from $d=5,6$ and then proceed to higher $d$.

The phase space is fairly easy to determine in $d=5,6$, see
figure~\ref{figure:phasespace}. In $d=5$ eq.~\eqref{MPhorizon} admits a real
root for
\beq\label{5dphase}
|j_1|+|j_2|\leq 1\,,
\eeq
which is a square, with extremal solutions at the boundaries where the
inequality is saturated. These extremal solutions have regular horizons
if, and only if, both angular momenta are non-vanishing. There are no
ultra-spinning regimes: our arguments above relate this fact
to the non-existence of three-dimensional vacuum black holes.

In $d=6$ the phase space of regular black hole solutions is again
bounded by a curve of extremal black holes. In terms of the
dimensionless parameter $\nu=r_0/\mu^{1/3}$, the extremal curve is
\beqa\label{6DextMP}
|j_1|=\left(\frac{\pi}{2\sqrt{3}}\right)^{1/3}
\sqrt{\frac{1-4\nu^3\pm\sqrt{1-16\nu^3}}{4\nu}}\,,\qquad
|j_2|=\left(\frac{\pi}{2\sqrt{3}}\right)^{1/3}
\sqrt{\frac{1-4\nu^3\mp\sqrt{1-16\nu^3}}{4\nu}}\,,
\eeqa
with $0\leq\nu\leq
2^{-4/3}$. As $\nu\to 0$ we get into the ultra-spinning regimes, in which one of
the spins diverges while the other vanishes, according to the general
behavior discussed above. In this regime, at, say, constant
large $j_1$, the solutions approach a Kerr black membrane and thus the
available phase space is of the form $j_2\leq f(j_1)$ i.e., a rescaled version of
the Kerr bound $j\leq 1/4$. Functions such as $a_H$ can be
recovered from the four-dimensional solutions.

\epubtkImage{}{%
\begin{figure}[h]
  \def\epsfsize#1#2{1.3#1}
  \centerline{\epsfbox{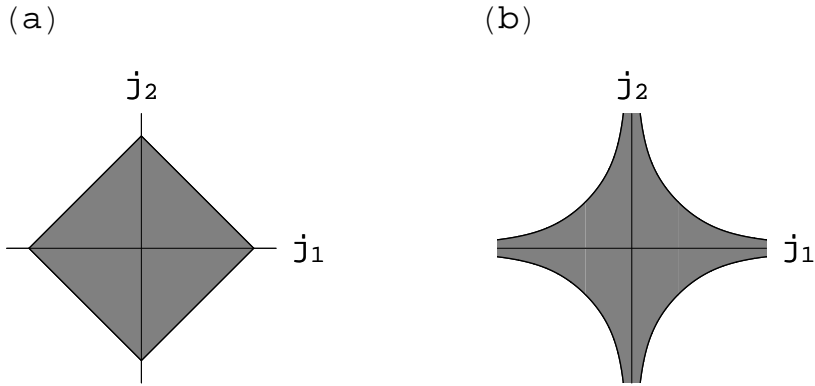}}
\caption{\it Phase space of (a) five-dimensional and (b) six-dimensional
MP rotating black holes: black holes exist for parameters within the
shaded regions. The boundaries of the phase space correspond to extremal
black holes with regular horizons, except at the corners of the square
in five dimensions where they become
naked singularities. The six-dimensional phase space extends along the axes
to arbitrarily large
values of each of the two angular momenta (ultra-spinning regimes).}
  \label{figure:phasespace}
\end{figure}}
In $d=7$, with three angular momenta $j_1,j_2,j_3$ it is more
complicated to obtain the explicit form of the surface of extremal
solutions that bound the phase space of MP black holes, but it is still
possible to sketch it, see figure~\ref{figure:7D8Dphasespace}(a). There
are ultra-spinning regimes in which one of the angular momenta becomes
much larger than the other two. In this limit the phase space of
solutions at, say, large $j_3$, becomes asymptotically of the form
$|j_1|+|j_2|\leq f(j_3)$, i.e., of the same form as the five-dimensional
phase space \eqref{5dphase}, only rescaled by a factor $f(j_3)$ (which
vanishes as $j_3\to \infty$). 

A similar `reduction' to a phase space in
two fewer dimensions along ultraspinning directions
appears in the phase space of $d=8$ MP black holes,
figure~\ref{figure:7D8Dphasespace}(b): a section at constant large $j_3$
becomes asymptotically of the same shape as the six-dimensional diagram
\eqref{6DextMP}, rescaled by a $j_3$-dependent factor. 
\epubtkImage{}{%
\begin{figure}[h]
  \def\epsfsize#1#2{.36#1}
  \centerline{\epsfbox{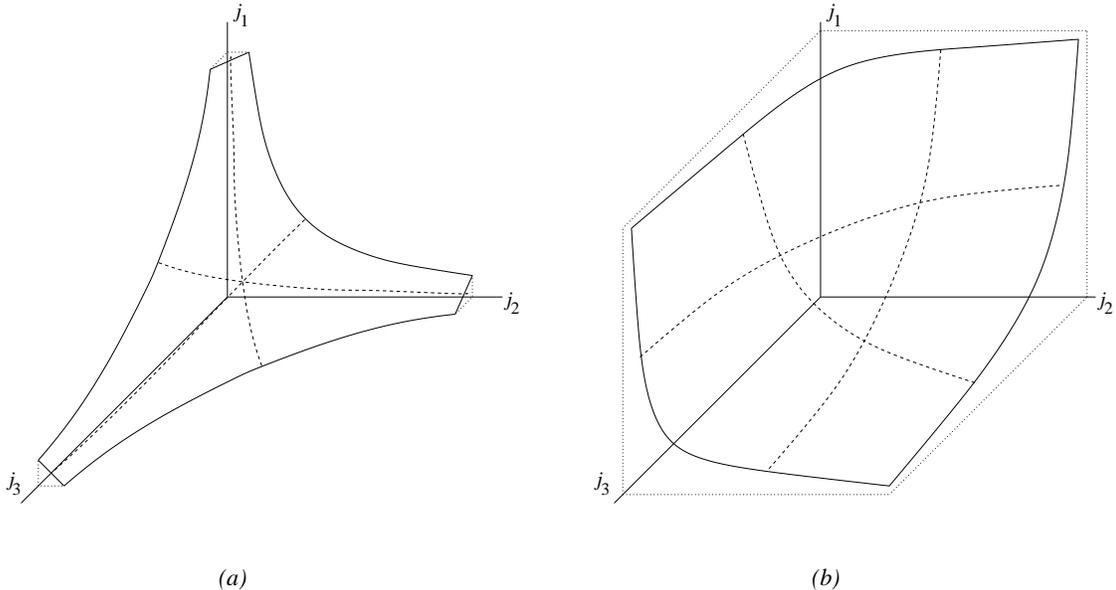}}
\caption{\it Phase space of (a) seven-dimensional, and (b) eight-dimensional
MP rotating black holes (in a representative quadrant $j_i\geq 0$). The
surfaces for extremal black holes are
represented: black holes exist in the region bounded by these surfaces.
(a) $d=7$: the hyperbolas where the surface
intersects the planes $j_i=0$ (which are $j_k j_l=1/\sqrt{6}$, i.e., $a_k
a_l=\sqrt{\mu}$, and $r_0=0$) correspond to naked singularities with
zero area, otherwise the extremal solutions are non-singular.
The three prongs extend to infinity: these are the ultra-spinning
regimes in which one spin is much larger than the other two. The prong
along $j_i$ becomes asymptotically of the form
$|j_k|+|j_l|\leq f(j_i)$, i.e., the same shape as the five-dimensional
diagram fig.~\ref{figure:phasespace}(a). 
(b) $d=8$: ultra-spinning regimes exist in which two spins are much larger
than the third one. The sections at large constant $j_i$ approach
asymptotically the same shape as the six-dimensional phase space
fig.~\ref{figure:phasespace}(b).}
  \label{figure:7D8Dphasespace}
\end{figure}}

These examples illustrate how we can infer the qualitative form of the
phase space in dimension $d$ if we know it in $d-2$. E.g., in $d=9,10$,
with four angular momenta, the sections of the phase space at large
$j_4$ approach the shapes in figure~\ref{figure:7D8Dphasespace} (a) and
(b), respectively.

If we manage to determine the regime of parameters where regular black
holes exist, we can express other (dimensionless) physical magnitudes as
functions of the phase space variables $j_a$. Figure~\ref{figure:mp5d}
is a plot of the area function $a_H(j_1,j_2)$ in $d=5$, showing only the
quadrant $j_1,j_2\geq 0$: the complete surface allowing $j_1,j_2<0$ is a
tent-like dome. In $d=6$ the shape of the area surface is a little more
complicated to draw, but it can be visualized by combining the
information from the plots we have presented in this section. In
general, the `ultra-spinning reduction' to $d-2n$ dimensions also yields
information about the area and other properties of the black holes.

\epubtkImage{}{%
\begin{figure}[h]
  \def\epsfsize#1#2{.7#1}
  \centerline{\epsfbox{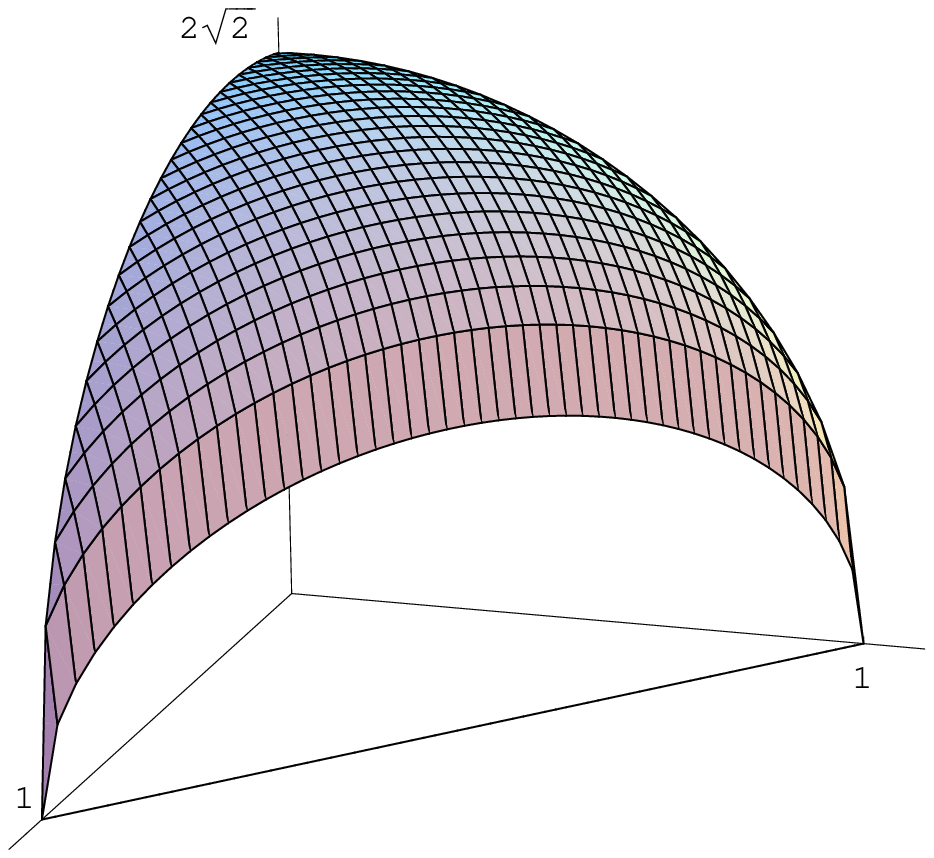}}
\caption{\it Horizon area $a_H(j_1,j_2)$ of five-dimensional MP black holes. We
only display a representative quadrant $j_1,j_2\geq 0$ of the full phase space of
figure~\ref{figure:phasespace}(a), the rest of the surface being obtained by
reflection along the planes $j_1=0$ and $j_2=0$.}
  \label{figure:mp5d}
\end{figure}}

\subsubsection{Global structure}

Let us now discuss briefly the global structure of these solutions,
following \cite{myersperry}.
The global topology of the solutions outside the event horizon is
essentially the same as for the Kerr solution. However, there are cases
where there can be only one non-degenerate horizon: even $d$ with at least
one spin vanishing; odd $d$ with at least two spins vanishing; odd $d$
with one $a_i=0$ and $\mu>\sum_i \Pi_{j\neq i}a_j^2$
There is also the possibility, for odd $d$ and all spin parameters
non-vanishing, of solutions with event horizons with negative $\mu$.
However, they contain naked closed causal curves.

The MP solutions have singularities where $\mu r/\Pi F\to\infty$ for
even $d$, $\mu r^2/\Pi F\to\infty$ for odd $d$. For even $d$ and all
spin parameters non-vanishing, the solution has a curvature singularity
where $F=0$, which is the boundary of a $(d-2)$-ball at $r=0$, thus
generalizing the ring singularity of the Kerr solution; as in the
latter, the solution can be extended to negative $r$. If one of the
$a_i=0$, then $r=0$ itself is singular. For odd $d$ and all $a_i\neq 0$,
there is no curvature singularity at any $r^2\geq 0$. The extension to
$r^2<0$ contains singularities, though. If one spin parameter vanishes,
say $a_1=0$, then there is a curvature singularity at the edge of a
$(d-3)$-ball at $r=0$, $\mu_1=0$; however, in this case this ball itself
is the locus of a conical singularity. If more than one spin parameter
vanishes then $r=0$ is singular. The causal nature of these
singularities varies according to the number of horizons that the
solution possesses, see \cite{myersperry} for further details.


\subsection{Symmetries}

The Myers-Perry solutions are manifestly invariant under time
translations, and also under the rotations generated by the $N$ Killing
vector fields $\partial/\partial \phi_i$. These symmetries form a $\mathbb{R}
\times U(1)^N$ isometry group. In general, this is the full isometry
group (up to discrete factors). However, the solutions exhibit symmetry
enhancement for special values of the angular momentum. For example, the
solution rotating in a single plane (\ref{mphole}) has a manifest
$\mathbb{R}
\times U(1) \times SO(d-3)$ symmetry. If $n$ angular momenta are equal
and non-vanishing then the $U(1)^n$ associated with the corresponding
2-planes is enhanced to a non-abelian $U(n)$ symmetry. This reflects the
freedom to rotate these 2-planes into each other. If $n$ angular momenta
vanish then the symmetry enhancement is from $U(1)^n$ to an orthogonal
group $SO(2n)$ or $SO(2n+1)$ for $d$ odd or even respectively
\cite{vasudevan}. Enhancement of symmetry is reflected in the metric
depending on fewer coordinates. For example, in the most extreme case of
$N$ equal angular momenta in $2N+1$ dimensions, the solution has
isometry group $\mathbb{R} \times U(N)$ and is cohomogeneity-1, i.e., it depends
on a single (radial) coordinate \cite{glpp}. 

In addition to isometries, the Kerr solution possesses a ``hidden"
symmetry associated with the existence of a second rank Killing tensor,
i.e., a symmetric tensor $K_{\mu\nu}$ obeying $K_{(\mu\nu;\rho)}=0$
\cite{walker}. This gives rise to an extra constant of the motion along
geodesics, rendering the geodesic equation integrable. It turns out that
the general Myers-Perry solution also possesses hidden symmetries
\cite{kubiznakfrolov} (this was first realized for the special case of
$d=5$ \cite{frolovstojk}). In fact, it has sufficiently many hidden
symmetries to render the geodesic equation integrable
\cite{kubiznakpage}. In addition, the Klein-Gordon equation governing a
free massive scalar field is separable in the Myers-Perry background
\cite{kubiznakfrolovsep}. These developments have been reviewed in
\cite{frolovreview}.


\subsection{Stability}
\label{subsec:MPstab}

The classical linearized stability of these black holes remains largely
an open problem. As just mentioned, it is possible to separate variables
in the equation governing scalar field perturbations
\cite{Ida:2002zk,Cardoso:2004cj,Morisawa:2004fs}. However, little
progress has been made with the study of linearized gravitational
perturbations. For Kerr, the study of gravitational perturbations is
analytically tractable because of a seemingly-miraculous decoupling of
the components of the equation governing such perturbations, allowing it
to be reduced to a single scalar equation \cite{teukolsky}. An analogous
decoupling has not been achieved for Myers-Perry black holes, except in
a particular case that we discuss below.

Nevertheless, it has been possible to infer the appearance of an
instability in the ultra-spinning regime of black holes in $d\geq 6$
\cite{Emparan:2003sy}. We have seen that in this regime, when $n$
rotation parameters $a_i$ become much larger than the mass parameter
$\mu$ and the rest of the $a_i$, the geometry of the black hole horizon
flattens out along the fast-rotation planes and approaches a black
$2n$-brane. As discussed in section~\ref{sec:GLinstab} black $p$-branes
are unstable to developing ripples along their spatial worldvolume
directions. Therefore, in the limit of infinite rotation the MP black
holes go over to unstable configurations. Then it is natural to
conjecture that the instability sets in already at finite values of the
rotation parameters. In fact, the rotation may not need to be too large
in order for the instability to appear. The GL instability of a neutral
black brane horizon $\mathbb{R}^p\times S^q$ appears when the size $L$
of the horizon along the brane directions is larger than the size $r_0$
of the $S^q$. We have seen that the sizes of the horizon along
directions parallel and transverse to the rotation plane are $\sim a_i$
and $\sim r_0$, respectively. This brane-like behavior of MP black holes
sets in when $a_i \gsim r_0$, which suggests that the instability will
appear shortly after crossing thresholds like \eqref{onset}. This idea
is supported by the study of the possible fragmentation of the
rotating MP black hole: the total horizon area can increase by splitting
into smaller black holes whenever $a_i \gsim r_0$ \cite{Emparan:2003sy}.
The analysis of \cite{Emparan:2003sy} indicates that the instability
should be triggered by gravitational perturbations. It is therefore not
surprising that scalar field perturbations appear to remain stable even in the
ultraspinning regime \cite{Cardoso:2004cj,Morisawa:2004fs}.

This instability has also played a central role in proposals for
connecting MP black holes to new black hole phases in $d\geq 6$. We
discuss this in section~\ref{section:vachigherd}.

The one case in which progress has been made with the analytical
study of linearized gravitational perturbations is the case of odd
dimensionality, $d=2N+1$, with equal angular momenta \cite{klrperturb,murata}.
As discussed above, this Myers-Perry solution is cohomogeneity-1, which
implies that the equations governing perturbations of this background
are just ODEs. There are two different approaches to this problem, one for $N>1$ \cite{klrperturb} and one for $N=1$ (i.e. $d=5$) \cite{murata}.

For $d=5$, the spatial geometry of the horizon is described by a homogeneous metric on $S^3$, with $SU(2) \times U(1)$ isometry group. Since $S^3 \sim SU(2)$, one can define a basis of $SU(2)$-invariant 1-forms and expand the components of the metric perturbation using this basis \cite{murata}. The equations governing gravitational perturbations will then reduce to a set of coupled scalar ODEs. These equations have not yet been derived for the Myers-Perry solution (however, this method has been applied to study perturbations of a static Kaluza-Klein black hole with $SU(2) \times U(1)$ symmetry \cite{murata2}). 

For $N>1$, gravitational perturbations can be classified into
scalar, vector and tensor types according to how they transform with
respect to the $U(N)$ isometry group. The different types of
perturbation decouple from each other. Tensor perturbations are governed
by a single ODE that is almost identical to that governing a massless
scalar field. Numerical studies of this ODE give no sign of any
instability \cite{klrperturb}. Vector and scalar type perturbations
appear to give coupled ODEs; the analysis of these has not yet been
completed.

It seems likely that other MP solutions with enhanced symmetry will also
lead to more tractable equations for gravitational perturbations. For
example, it would be interesting to consider the cases of equal angular
momenta in even dimensions (which resemble the Kerr solution in many
physical properties), and MP solutions with a single non-zero angular
momentum (whose geometry \eqref{mphole} contains a four-dimensional
factor, at a constant angle in the $S^{d-4}$, mathematically similar to
the Kerr metric; in fact this four-dimensional geometry is type D). The
latter case would allow one to test whether the ultraspinning
instability is present.

\newpage


\section{Vacuum solutions in five dimensions}
\label{section:vac5d}

In section \ref{section:myersperry} we have discussed the MP solutions,
which can be regarded as the higher-dimensional versions of the Kerr
solution. However, in recent years it has been realized that higher
dimensions allow for a much richer landscape of black hole solutions
that do not have four-dimensional counterparts. In particular, there has
been great progress in our understanding of five-dimensional vacuum
black holes, insofar as we consider stationary solutions with two
rotational Killing vectors. The reason is that this sector of the theory
is completely integrable, and solution-generating techniques are
available. We begin by analyzing in section~\ref{subsec:blackrings} a
qualitatively new class of solutions with connected horizons: black
rings, with one and two angular momenta. Then, in
section~\ref{subsec:weyl} we present the general study of stationary
solutions with two rotational symmetries; actually we can discuss the
general case of $d-3$ commuting $U(1)$ spatial isometries. The simplest
of these are the generalized Weyl solutions
(section~\ref{subsubsec:weyl}). The general case is addressed in
section~\ref{subsubsec:genaxi}, and we discuss the characterization of
solutions by their rod structures. The powerful solution-generating
technique of Belinsky and Zakharov, based on inverse-scattering methods,
is then introduced: the emphasis is on its practical application to generate
old and new black hole solutions. Section~\ref{subsec:multibhs}
discusses multi-black hole solutions (black Saturns, di-rings and
bicycling black rings) obtained in this way. Work towards determining the
stability properties of black rings and multi-black holes is reviewed
in section~\ref{subsec:5Dstability}.


\subsection{Black rings}
\label{subsec:blackrings}

\subsubsection{One angular momentum}

Five-dimensional black rings are black holes with horizon topology
$S^1\times S^2$ in asymptotically flat spacetime. The $S^1$ describes a
contractible circle, not stabilized by topology but by the centrifugal
force provided by rotation. An exact solution for a black ring with
rotation along this $S^1$ was presented in \cite{Emparan:2001wn}. Its
most convenient form was given in \cite{Emparan:2004wy}
as\epubtkFootnote{An alternative form was found in \cite{Hong:2003gx}. The relation
between the two is given in \cite{Emparan:2006mm}.}
\beqa\label{neutral}
ds^2&=&-\frac{F(y)}{F(x)}\left(dt-C\: R\:\frac{1+y}{F(y)}\:
d\psi\right)^2\nonumber\\[2mm]
&&+\frac{R^2}{(x-y)^2}\:F(x)\left[
-\frac{G(y)}{F(y)}d\psi^2-\frac{dy^2}{G(y)}
+\frac{dx^2}{G(x)}+\frac{G(x)}{F(x)}d\phi^2\right]\,,
\eeqa
where
\beq\label{fandg}
F(\xi)=1+\lambda\xi,\qquad G(\xi)=(1-\xi^2)(1+\nu\xi)\,,
\eeq
and
\beq\label{coeff}
C=\sqrt{\lambda(\lambda-
\nu)\frac{1+\lambda}{1-\lambda}}\,.
\eeq
The dimensionless parameters $\lambda$ and $\nu$ must lie in the range 
\beq\label{lanurange}
0< \nu\leq\lambda<1\,.
\eeq 
The coordinates vary in the ranges
$-\infty \leq y \leq -1$ and $-1 \leq x \leq 1$,
with asymptotic infinity
recovered as $x \to y\to -1$. The axis of rotation around the $\psi$
direction is at $y = -1$, and the axis of rotation around $\phi$ is
divided
into two pieces: $x = 1$ is the disk bounded by the ring, and $x = -1$ is
its complement from the ring to infinity. The horizon lies at $y=-1/\nu$. Outside
it, at $y=-1/\lambda$, lies an
ergosurface.
A detailed analysis of this solution and its properties can be found in
\cite{Emparan:2006mm} and \cite{Elvang:2006dd}, so we shall only
discuss it briefly. 

In the form given above the solution possesses three independent
parameters: $\lambda$, $\nu$, and $R$. Physically, this sounds like one
too many: given a ring with mass $M$ and angular momentum $J$, we
expect its radius to be dynamically fixed by the balance between
the centrifugal and tensional forces. This is also the case for the
black ring (\ref{neutral}): in general it has a conical defect on the
plane of the ring, $x=\pm 1$. In order to avoid it,
 the angular variables must be identified with periodicity 
\beq\label{period0}
\Delta\psi=\Delta\phi=4\pi\frac{\sqrt{F(-1)}}{|G'(-1)|}=
2\pi\frac{\sqrt{1-\lambda}}{1-\nu}\,
\eeq
and the two parameters $\lambda$, $\nu$ must satisfy
\beq\label{equil0}
\lambda=\frac{2\nu}{1+\nu^2}\,.
\eeq
This eliminates one parameter, and leaves the expected
two-parameter $(\nu,R)$ family of solutions. The mechanical
interpretation of this balance of forces for thin rings is discussed in
\cite{Emparan:2007wm}. The Myers-Perry solution with a single rotation
is obtained as a limit of the general solution (\ref{neutral})
\cite{Emparan:2004wy}, but cannot be recovered if $\lambda$ is
eliminated through (\ref{equil0}).

The physical parameters of the solution (mass, angular momentum, area,
angular velocity, surface gravity) in terms of $\nu$ and $R$ can be found
in \cite{Emparan:2006mm}. It can be seen that while $R$ provides a
measure of the radius of the ring's $S^1$, the parameter $\nu$ can be
interpreted as a `thickness' parameter characterizing its shape,
corresponding roughly to the ratio
between the $S^2$ radius and the $S^1$ radius. 

More precisely, one finds two branches of solutions, whose physical
differences are seen most
clearly in terms of the dimensionless variables $j$ and $a_H$ introduced
above. For a black ring in equilibrium, the phase curve $a_H(j)$ 
can be expressed in parametric form as
\beq\label{zetaeta}
{a_H}= 2\sqrt{\nu(1-\nu)}\,,\qquad
j= \sqrt{\frac{(1+\nu)^3}{8\nu}}\,,
\eeq
and is depicted in figure~\ref{figure:ringah}.
\epubtkImage{}{%
\begin{figure}[h]
  \def\epsfsize#1#2{1#1}
  \centerline{\epsfbox{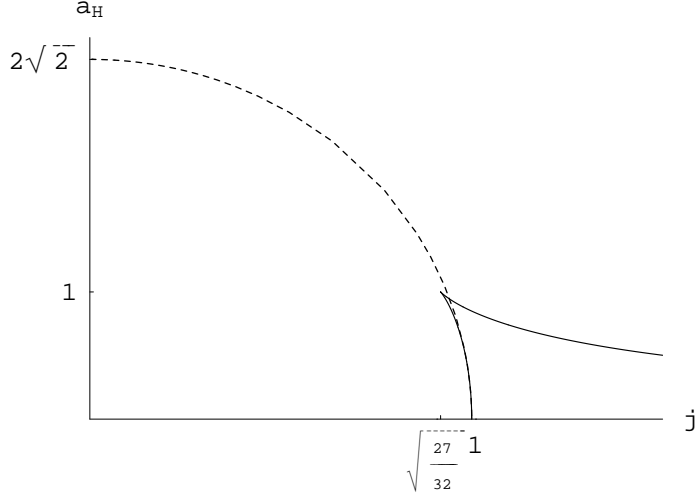}}
\caption{\it Curve $a_H(j)$ of horizon area vs.~spin for
five-dimensional black rings rotating along their $S^1$ (solid). The
dashed curve corresponds to five-dimensional MP black holes (see
figure~\ref{figure:aHj}). The solid curve for black rings has two
branches that meet at a regular, non-extremal minimally rotating black ring at
$j=\sqrt{27/32}$: an upper branch of thin black rings, and a lower
branch of fat black rings. Fat black rings always have smaller
area than MP black
holes. Their curves meet at the same zero-area naked singularity at
$j=1$.}
  \label{figure:ringah}
\end{figure}}
This curve is easily seen to have a cusp at $\nu=1/2$, which corresponds
to a minimum value of $j=\sqrt{27/32}$ and a maximum $a_H=1$. Branching off
from this cusp, the {\em thin black ring} solutions ($0<\nu<1/2$) extend
to $j\to\infty$ as $\nu \to 0$, with asymptotic $a_H\to 0$. The {\em fat
black ring} branch ($1/2\leq \nu <1$) has lower area and extends only to
$j\to 1$, ending at $\nu\to 1$ at the same zero-area singularity as the
MP solution. This implies that in the range $\sqrt{27/32}\leq j<1$ there exist
three different solutions (thin and fat black rings, and MP black hole)
with the same value of $j$. The notion of black hole uniqueness that was
proven to hold in four dimensions does not extend to five dimensions.

Refs.~\cite{Elvang:2006dd} and \cite{Frolov:2006pu} contain detailed
analyses of the geometrical features of black ring horizons. Some
geodesics of the black ring metric have been studied with a view towards
different applications: \cite{Nozawa:2005eu} study them in the context
of the Penrose process, and \cite{Elvang:2006dd} consider them for tests
of stability. Ref.~\cite{Hoskisson:2007zk} is a more complete analysis
of geodesics.

\subsubsection{Two angular momenta}
\label{subsubsec:twospinring}

Rotation in the second independent plane corresponds to rotation of the
$S^2$ of the ring. In the limit of infinite $S^1$ radius, a section
along the length of the ring gives an $S^2$ that is essentially like
that of a four-dimensional black hole: setting it into rotation is thus
similar to having a Kerr-like black hole. Thus, an upper, extremal bound
on the rotation of the $S^2$ is expected (actually, the motion of the
ring along its $S^1$ yields a momentum that can be viewed as electric
Kaluza-Klein charge, so instead of a Kerr solution, the $R\to\infty$
limit yields a rotating electric KK black hole). 

Solutions with rotation only along the $S^2$, but not on the $S^1$, are
fairly straightforward to construct and have been given in
\cite{Mishima:2005id,Figueras:2005zp}. However, these black rings cannot
support themselves against the centripetal tension and thus possess
conical singularities on the plane of the ring. Constructing the exact
solution for a black ring with both rotations is a much more complicated
task, which has been achieved by Pomeransky and Sen'kov in
\cite{Pomeransky:2006bd} (the techniques employed are reviewed below in
section~\ref{subsec:weyl}). They have furthermore managed to present it
in a fairly compact form:
\bea\label{twospinmetric}
 ds^2&=&-\frac{H(y,x)}{H(x,y)}(dt+\Omega)^2-\frac{F(x,y)}{H(y,x)}d\psi^2
	-2\frac{J(x,y)}{H(y,x)} d\psi d\phi
  	+\frac{F(y,x)}{H(y,x)}d\phi^2\nonumber\\
&&-\frac{2 k^2 H(x,y)}{(x-y)^2(1-\nu)^2}\left(\frac{dx^2}{G(x)}-
\frac{dy^2}{G(y)}\right)\,.
 \eea
Here we follow the notation introduced in \cite{Pomeransky:2006bd},
except that we have chosen mostly plus signature, and exchanged
$\phi\leftrightarrow\psi$ to conform to the notation in (\ref{neutral}).
The reader should be warned that although the meaning of $x$ and $y$ is
essentially the same in both solutions, the same letters are used in
(\ref{twospinmetric}) as in (\ref{neutral}) for different parameters
and functions. In particular, the angles $\phi$ and $\psi$ have been
rescaled here to have canonical periodicity $2\pi$.

The metric functions take a very complicated form in the
general case in which the black ring is not in equilibrium (their
explicit forms can be found in \cite{Morisawa:2007di}), but they
simplify significantly when balance of forces (i.e., cancellation of
conical singularities) is imposed. In this case the one-form $\Omega$
characterizing the rotation is \cite{Pomeransky:2006bd}
\bea
 \Omega&=&-\frac{2 k \lambda \sqrt{(1+\nu )^2-\lambda ^2}}{H(y,x)}\bigl[ (1-x^2) y \sqrt{\nu}d\phi
 \nonumber\\
 &&+\frac{1+y}{1-\lambda +\nu}
 \left(1+\lambda -\nu +x^2 y \nu (1-\lambda-\nu)+2\nu x(1-y)\right)\,d\psi\bigr],
 \eea
 and the functions $G$, $H$, $J$, $F$ become
\bea\label{GHJF}
 G(x)&=&(1-x^2)\left(1+\lambda x+\nu x^2\right)\,,\nonumber\\
 H(x,y)&=& 1+\lambda ^2-\nu^2+2\lambda\nu (1-x^2)y
      +2x\lambda(1-y^2\nu^2)+ x^2 y^2 \nu(1-\lambda^2-\nu^2)\,,\nonumber\\
 J(x,y)&=&\frac{2 k^2 (1-x^2) (1-y^2) \lambda  \sqrt{\nu}}{(x-y) (1-\nu)^2}
 \,\left(1+\lambda ^2 -\nu ^2
  + 2 (x+y) \lambda  \nu-x y \nu (1-\lambda ^2-\nu^2)\right),\\
 F(x,y) &=& \frac{2 k^2}{(x-y)^2 (1-\nu)^2} \Bigl[G(x) (1-y^2)\left[\left((1-\nu)^2-\lambda ^2\right)
  (1+\nu )+y \lambda (1-\lambda ^2+2 \nu -3 \nu ^2)\right]\nonumber\\
	&&+ G(y) \bigl[2 \lambda ^2
    + x \lambda ((1-\nu )^2+\lambda ^2)
   + x^2\left((1-\nu )^2-\lambda ^2\right) (1+\nu)
    +x^3\lambda(1-
	\lambda^2-3\nu^2+2\nu^3)\nonumber\\
   &&- x^4 (1-\nu ) \nu (-1+\lambda ^2+\nu ^2)\bigr]\Bigr].\nonumber
\eea
When $\lambda=0$ we find flat spacetime.
In order to recover the metric (\ref{neutral}) 
one must take $\nu\to 0$, identify $R^2= 2k^2 (1+\lambda)^2$ and rename
$\lambda\to \nu$. 

The parameters $\lambda$ and $\nu$ are restricted to
\beq\label{twospinrange}
0\leq\nu<1\,,\qquad 2\sqrt{\nu}\leq\lambda<1+\nu
\eeq
for the existence of
regular black hole horizons. The bound $\lambda\geq
2\sqrt{\nu}$ is actually a Kerr-like bound on the rotation of the $S^2$.
To see this, consider the equation for vanishing $G(y)$,
\beq\label{twospinhorizon}
 1+\lambda y+\nu y^2=0\,,
\eeq
which determines the position of the horizon within the allowed range
$-\infty<y<-1$. If we identify $y\to -k/r$, $\lambda\to 2m/k$ and
$\nu\to a^2/k^2$, this becomes the familiar $r^2-2mr+a^2=0$ (this is not
to say that $m$ and $a$ correspond to the physical mass and angular
momentum parameters, although they are related to them). Imposing that
the roots of (\ref{twospinhorizon}) are real yields the required bound.
When it is saturated, $\lambda= 2\sqrt{\nu}$, the horizon is degenerate,
and when exceeded it becomes a naked singularity. The parameter $k$ sets
a scale in the solution and gives (roughly) a measure of the ring
radius. The extremal Myers-Perry solution is recovered as a limit of the
extremal solutions in which $\nu\to 1$, $\lambda\to 2$. However, in
order to recover the general Myers-Perry solution as a limit, one needs
to relax the equilibrium condition that has been imposed to obtain
(\ref{GHJF}), and use the more general form of these functions given in
\cite{Morisawa:2007di}.

The physical parameters $M$, $J_{\psi}=J_1$, $J_\phi\equiv J_2$,
$\mathcal{A}_H$ of the solution have been computed in
\cite{Pomeransky:2006bd}. An analysis of the physical properties of the
solution, and in particular the phase space has been presented in
\cite{Elvang:2007hs}. To plot the parameter region where black rings
exist, we fix the mass and employ the dimensionless angular momentum
variables $j_1$, $j_2$ introduced in \eqref{jaHdef}.
The phase space of
doubly-spinning black rings is in figure~\ref{figure:doublering} for the region
$j_1>j_2\geq 0$ (the rest of the plane is obtained by iterating and
exchanging $\pm j_{1,2}$). It is bounded by three curves (besides the
axis, which is not a boundary in the full phase plane): 
\epubtkImage{}{%
\begin{figure}[h]
  \def\epsfsize#1#2{1.2#1}
  \centerline{\epsfbox{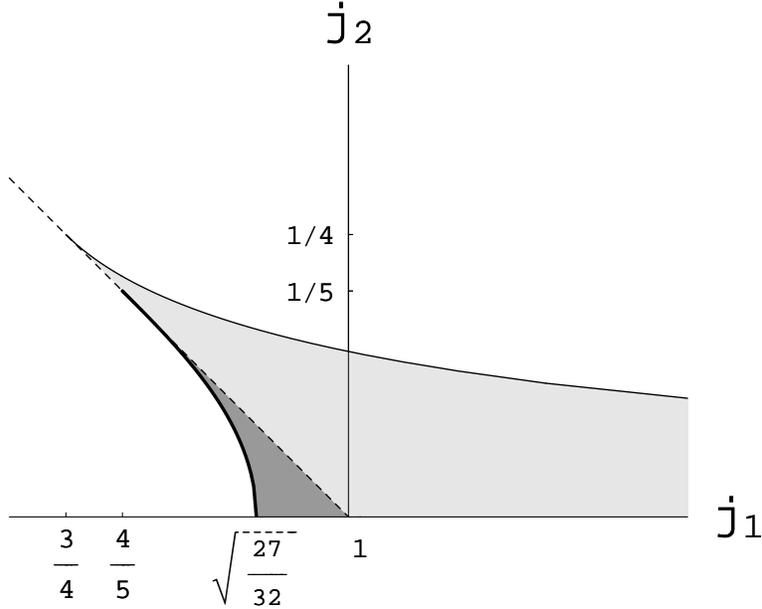}}
\caption{\it Phase space of doubly-spinning black rings ($j_1\equiv
j_\psi$, $j_2\equiv j_\phi$), restricted to the representative region
$j_1>j_2\geq 0$. The dashed line $j_1+j_2=1$ corresponds to extremal MP
black holes (see figure~\ref{figure:phasespace}(a)). The (upper) thin
black curve corresponds to regular extremal black rings with degenerate
horizons at maximal $S^2$ spin $j_2$, for given $S^1$ rotation $j_1$. It
ends on the extremal MP curve at $(3/4,1/4)$. The (lower) thick black
curve corresponds to regular non-extremal black rings with minimal spin
$j_1$ along $S^1$ for given $j_2$ on $S^2$. It ends on the extremal MP
curve at $(4/5,1/5)$. Black rings exist in the gray-shaded parameter
regions bounded by the black curves, the segment $j_1\in [3/4,4/5]$ of
the extremal MP dashed line, and the $j_1$ axis with
$j_1\geq\sqrt{27/32}$. In the light-gray region there exist only thin
black rings. In the dark-gray spandrel between the dashed MP line, the
thick black curve, and the axis $\sqrt{27/32}<j_1<1$, there exist thin
and fat black rings, and MP black holes:
there is discrete three-fold non-uniqueness.
}
  \label{figure:doublering}
\end{figure}}
\begin{enumerate}
\item Extremal black
rings, with maximal $J_2$ for given $J_1$, along the curve
\beqa
j_1=\frac{1+4\sqrt{\nu}+\nu}{4\nu^{1/4}(1+\sqrt{\nu})}\,,\qquad
j_2=\frac{\nu^{1/4}}{2(1+\sqrt{\nu})}\,,\qquad 0\leq \nu\leq 1\,
\eeqa
(thin solid curve
in fig.~\ref{figure:doublering}). This curve extends between $j_1=3/4$,
$j_2=1/4$ (as $\nu\to 1$) to
$j_1\to\infty$, $j_2\to
0$ (as $\nu\to 0$). (See \cite{reallentropy} for more discussion of extremal rings.)

\item Non-extremal
minimally spinning black rings, with minimal $J_1$ for given
$J_2$, along the curve
\beqa
j_1&=&\frac{
   \left(3(1+\nu^2)+(1+\nu)\sqrt{(9+\nu)(1+9\nu)}-26\nu\right)^{3/2}}
	{8 (1-\nu)^2 
\sqrt{\left(\sqrt{(9+\nu)(1+9\nu)}-1-\nu\right)
\left(5(1+\nu)-\sqrt{(9+\nu)(1+9 \nu)}\right)}}
\,,\nonumber\\
j_2&=&\frac{\sqrt{\frac{\nu  
	\left(3+\sqrt{(9+\nu)(1+9\nu)}+
	\nu  
	\left(18 \sqrt{(9+\nu)(1+9\nu)}-103+
	\nu
	\left( 3 \nu +\sqrt{(9+\nu)(1+9\nu)}-103\right)
	\right)
	\right)}
		{\sqrt{(9+\nu)(1+9\nu)}-1-\nu}}}{2 \sqrt{2} (1-\nu)^2}\,
\eeqa
(thick solid curve
in fig.~\ref{figure:doublering}). 
This curve extends between $j_1=4/5$, $j_2=1/5$ (as $\nu\to 1$) and
$j_1=\sqrt{27/32}$, $j_2= 0$ (as $\nu\to 0$).

\item Limiting extremal MP black holes, with $j_1+j_2=1$
within the range $j_1\in [3/4,4/5]$ (dashed line 
in fig.~\ref{figure:doublering}). 
There is a discontinuous
increase in the area when the black
rings reach the extremal MP line.
\end{enumerate}

For doubly-spinning black rings the angular momentum along the $S^2$ is
always bounded above by the one in the $S^1$ as
\beq
|j_2|<\frac{1}{3}|j_{1}|\,.
\eeq
This is saturated at the endpoint of the extremal black ring curve
$j_1=3/4$, $j_2=1/4$. 

Figure~\ref{figure:5Dphasespace} shows the phase space covered by all 5D
black holes with a single horizon. Two kinds of black rings (thin and
fat) and one MP black hole, the three of them with the same values of
$(M,J_1,J_2)$, exist in small spandrels near the corners of the MP phase
space square. It is curious that, once black rings are included, the
available phase space for five-dimensional black holes resembles more
closely that of six-dimensional MP black holes,
figure~\ref{figure:phasespace}(b). 
\epubtkImage{}{%
\begin{figure}[h]
  \def\epsfsize#1#2{1#1}
  \centerline{\epsfbox{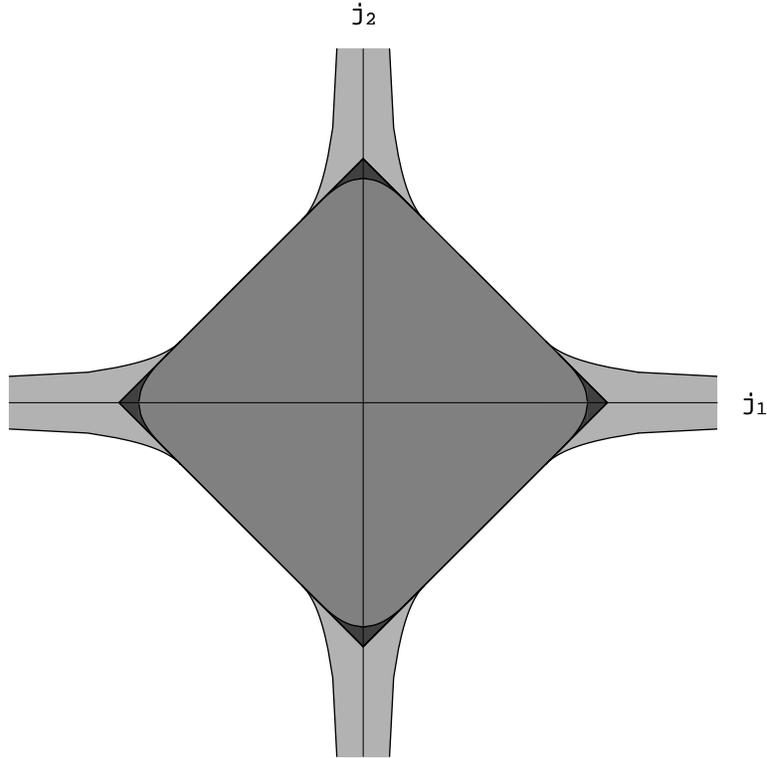}}
\caption{\it The phase space $(j_1,j_2)$ covered by doubly-spinning MP
black holes and black rings, obtained by replicating
figure~\ref{figure:doublering} taking $\pm j_1 \leftrightarrow \pm j_2$. The
square $|j_1|+|j_2|\leq 1$ corresponds to MP black holes (see
figure~\ref{figure:phasespace}(a)). The light-gray zones contain thin black
rings only, and the medium-gray zone contains MP black holes only. At
each point in the
dark-gray spandrels near the corners of the square 
there exist one thin and one fat black ring, and one MP black hole. }
  \label{figure:5Dphasespace}
\end{figure}}

Ref.~\cite{Elvang:2007hs} contains sectional plots of the surface
$a_H(j_1,j_2)$ for black rings at constant $j_2$, for $j_1>j_2\geq 0$,
from which it is possible to obtain an idea of the shape of the 
surface. 
In the complete range of $j_1$ and $j_2$ the phase space of
five-dimensional black holes (with connected event horizons) consists of
the `dome' of MP black holes (fig.~\ref{figure:mp5d} replicated on all
four quadrants), with `romanesque vaults' of black rings protruding from
its corners, and additional substructure in the region of non-uniqueness
--our knowledge of architecture is insufficient to describe it in words.

It is also interesting to study other properties of black rings, such as
temperature and horizon angular velocities, expressions for which can be found in
\cite{Elvang:2007hs}. It is curious to notice that even if the two
angular momenta can never be equal, the two angular velocities
$\Omega_1$ and $\Omega_2$ have equal values, for a given mass, when 
\beq
\lambda=\sqrt{2\nu-\sqrt{\nu}(1+\nu)}
\eeq
which lies in the allowed range \eqref{twospinrange}. We can easily
understand why this is possible: $\Omega_1$ becomes arbitrarily small for
thin rings, even if $j_1$ is large, so it can be made equal to any given
$\Omega_2$.

On the other hand, the temperature of the black ring --which for thin
rings with a single spin is bounded below and diverges as the ring
becomes infinitely long and thin (at fixed mass)-- decreases to zero when
the second spin is taken to the extremal limit, so there exist `cold'
thin black rings.

Some consequences of
these features to properties of multi-ring solutions will be discussed in
section~\ref{subsec:multibhs}.


\subsection{Stationary axisymmetric solutions with $d-3$ rotational symmetries}
\label{subsec:weyl}

A sector of five-dimensional vacuum General Relativity where a complete
classification of black hole solutions may soon be achieved is the class
of stationary solutions with two angular Killing vectors. Integration of
the three Killing directions yields a two-dimensional non-linear sigma
model that is completely integrable. Solutions can be characterized in
terms of their {\em rod structure} along multiple directions, introduced
in \cite{Emparan:2001wk} and extended in \cite{Harmark:2004rm}. It has
been proven that these data (whose relation to physical parameters is
unfortunately not quite direct), in addition to the total mass and
angular momenta, uniquely characterize asymptotically flat solutions
\cite{hollandsyazad}.

Since most of the analysis is applicable to any number of dimensions we
will keep $d$ arbitrary, although only in $d=4,5$ can the solutions be
globally asymptotically flat. So henceforth we assume that the spacetime
admits $d-2$ commuting, non-null, Killing vectors
$\xi_{(a)}=\partial/\partial x^a$, $a=0,\dots, d-3$ (we assume, although
this is not necessary, that the zero-th vector is asymptotically
timelike and all other vectors are asymptotically spacelike). Then it is
possible to prove that, under natural suitable conditions, the
two-dimensional spaces orthogonal to all three Killing vectors are
integrable \cite{Emparan:2001wk}. In this case the metrics admit the
form
\begin{equation}\label{aximetrics}
ds^2=g_{ab}(r,z)\,dx^a dx^b+e^{2\nu(r,z)}\left(dr^2+dz^2\right)\;. 
\end{equation}
Without loss of generality
we can choose coordinates so that
\beq\label{rconstraint}
\det g_{ab}=-r^2\,.
\eeq
For this class of geometries, the Einstein equations divide in two
groups, one for the matrix 
$g$,
\begin{equation}
\partial_r U+\partial_z V=0\;, \label{eqforg}
\end{equation}
with
\beq\label{UVdefs}
U=r(\partial_r g)g^{-1}\,,\qquad V=r(\partial_z g)g^{-1}\,,
\eeq
and a second group of equations for $\nu$,
\begin{equation}\label{eqfornu}
\partial_r\nu=-\frac{1}{2r}
  +\frac{1}{8r}\;\mathrm{Tr}(U^2-V^2)\,,
  \qquad
  \partial_z\nu=\frac{1}{4r}\;\mathrm{Tr}(UV)\, . 
\end{equation}
The equations for $\nu$ satisfy the integrability condition
$\partial_r\partial_z\nu =\partial_z\partial_r
\nu$ as a consequence of (\ref{eqforg}). Therefore, once 
$g_{ab}(r,z)$ is determined, the function $\nu(r,z)$ is determined by a
line integral, up to an integration constant that can be
absorbed by rescaling the coordinates.

Equations (\ref{eqforg}) and (\ref{UVdefs}) are the equations for the
principal chiral field model, a non-linear sigma model with group
$GL(d-2,{\mathbb R})$, which is a completely integrable system. In the
present case, it is also subject to the constraint (\ref{rconstraint}).
This introduces additional features, some of which will be discussed
below.

In order to understand the structure of the solutions of this system, it
is convenient to analyze first a simple particular case \cite{Emparan:2001wk}.

\subsubsection{Weyl solutions}
\label{subsubsec:weyl}

Consider the simplest situation in which the Killing vectors are
mutually orthogonal. In this case the solutions admit the diagonal
form\epubtkFootnote{An equivalent system, but with a
cosmological
interpretation under a Wick rotation of the coordinates $(t,r,z)\to
(x,t,\tilde z)$, is discussed in \cite{Feinstein:1999ij}, along with some simple
solution-generating techniques.}
\beq
\label{weyl}
ds^2 = - e^{2U_0} dt^2 + \sum_{a=1}^{d-3} e^{2U_a} (dx^a)^2
+ e^{2\nu} (dr^2+dz^2)\,.
\eeq
Equations (\ref{eqforg}) require that $U_a(r,z)$, $a=0,\dots,d-3$, be
axisymmetric solutions of the
Laplace equation
\begin{equation}
\label{Ulap}
\left( \partial_r^2 + \frac{1}{r} \partial_r + \partial_z^2 \right) U_a
=0 \ ,
\end{equation}
in the auxiliary three-dimensional flat space
\beq
ds^2=dz^2+dr^2+r^2 d\phi^2\,,
\eeq
while equations (\ref{eqfornu}) for $\nu(r,z)$  become
\beqa
\label{nusol}
\partial_r \nu &=& - \frac{1}{2r }
+ \frac{r}{2} \sum_{a=0}^{d-3} \left[ (\partial_r U_a)
- (\partial_z U_a)^2 \right]\,,
\\
\partial_z \nu &=& r \sum_{i=0}^{d-3} \partial_r U_a \partial_z U_a \,,
\eeqa
The constraint (\ref{rconstraint}) implies that only $d-3$ of the $U_a$
are independent, since they must satisfy
\beq\label{rodsum}
\sum_{a=0}^{d-3} U_a = \log r\,.
\eeq
Thus we see that solutions are fully determined once the boundary
conditions for the $U_i$ are specified at infinity and at the $z$-axis.
Note that $\log r$ is the solution that corresponds to an infinite rod
of zero thickness and linear mass density $1/2$ along the axis $r=0$.
The solutions are in fact characterized by the `rod' sources
of $U_i$ along the axis, which are subject to add up to an infinite rod
(\ref{rodsum}). The potential for a semi-infinite rod along
$[a_k,+\infty)$ with linear density $\varrho$ is
\beq
U=\varrho \log \mu_k\,,
\eeq 
where
\beq\label{mua}
\mu_k=\sqrt{r^2+(z-a_k)^2}-(z-a_k)\,.
\eeq
If the rod instead extends along $(-\infty,a_k]$ then
\beq
U=\varrho \log \bar\mu_k\,,
\eeq 
where
\beq\label{barmua}
\bar\mu_k=-\sqrt{r^2+(z-a_k)^2}-(z-a_k)=-\frac{r^2}{\mu_k}\,.
\eeq
Given the linearity of
equations (\ref{Ulap}), one can immediately construct the potential for a finite
rod of density $\varrho$ along $[a_{k-1},a_k]$ as
\beq
U=\varrho\log\left(\frac{\mu_{k-1}}{\mu_k}\right)\,.
\eeq
The functions $U_i$ for any choice of rods are sums of these.
Integration of (\ref{nusol}) is then a straightforward, if tedious
matter, see appendix G in \cite{Emparan:2001wk}. Ref.~\cite{Koikawa:2005ia}
has applied the inverse scattering method (to be reviewed below) to
provide explicit diagonal solutions with an arbitrary number of rods.

At a rod source for $U_i$, the orbits of the corresponding Killing
vector vanish: if it is an angular Killing vector $\partial_{\phi_i}$,
then the corresponding one-cycles shrink to zero size, and the
periodicity of $\phi_{i}$ must be chosen appropriately in order to avoid
conical singularities; if it is instead the timelike Killing
$\partial_t$, then it becomes null there. In both cases, a necessary
(but not sufficient) condition for regularity at the rod is that the
linear density be 
\beq
\varrho=\frac{1}{2}\,.
\eeq
Otherwise when $r\to 0$ at the rod, the curvature diverges. If all rods
are of density $1/2$, then given the constraint (\ref{rodsum}), at any
given point on the axis there will be one cycle of zero (or null) length
with all others having finite size. This phenomenon of some cycles
smoothly shrinking to zero with other cycles blowing up to finite size
as one moves along the axis (essentially discovered by Weyl in 1917
\cite{Weyl:1917gp}), is
referred to in string-theoretical contexts as `bubbling'. When it is
$\partial_t$ that becomes null, the rod corresponds to a horizon.

The rod structures for the four and five-dimensional
Schwarzschild and Tangherlini solutions are depicted in
figure~\ref{figure:rods_bh}. The
Rindler space of uniformly accelerated observers is recovered as the
horizon rod becomes semi-infinite, $a_2\to \infty$.
Ref.~\cite{Emparan:2001wk} gives a number of `thumb rules' for
interpreting rod diagrams.

\epubtkImage{}{%
\begin{figure}[h]
  \def\epsfsize#1#2{.55#1}
  \centerline{\epsfbox{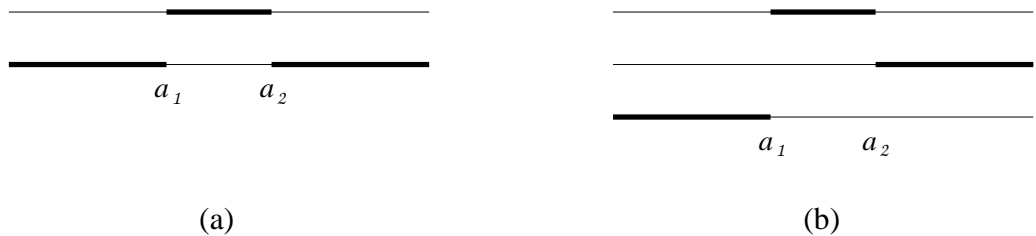}}
\caption{\it Rod structures for the (a) 4D Schwarzschild and (b) 5D
Tangherlini black holes. From top to bottom, the lines represent the
sources for the time, $\phi$ and $\psi$ (in 5D) potentials $U_t$,
$U_\phi$, $U_\psi$.}
  \label{figure:rods_bh}
\end{figure}}

\subsubsection{General axisymmetric class}
\label{subsubsec:genaxi}

In the general case where the Killing vectors are not orthogonal to each
other the simple construction in terms of solutions of the linear
Laplace equation does not apply. Nevertheless, the equations can still
be completely integrated, and the characterization of solutions in terms
of rod structure can be generalized.

\paragraph{Rod structure and regularity.}
\label{para:rodstructure}

Let us begin by extending the characterization of rods
\cite{Harmark:2004rm}. In general a rod is an interval along the
$z$-axis where the action of a Killing vector $v$ has fixed points (for
a spacelike vector) or it becomes null (for a timelike vector). In both
cases $|v|^2\to 0$ at the rod. In general, $v$ will be a linear
combination $v=v^a\xi_{(a)}$ in a given Killing basis $\xi_{(a)}$,
$a=0,\dots,d-3$: usually, this basis is chosen so that it becomes a
coordinate basis of orthogonal vectors at asymptotic infinity. In the
orthogonal case of the previous section
one could assign a basis vector (and only one, for rod density $1/2$)
$\xi_{(a)}$ to each rod, but this is not possible in general if the
vectors are not orthogonal. For instance, at a rod corresponding to a
rotating horizon, the Killing vector that vanishes is typically of the form
$\partial_t+\Omega_a \partial_{\phi_a}$. 

More precisely, the condition (\ref{rconstraint}) implies that the
matrix $g(r=0,z)$ must have at least one zero eigenvalue. It can be
shown that regularity of the solution (analogous to the requirement of
density $1/2$ in the orthogonal case) requires that only one eigenvalue
is zero at any given interval on the axis \cite{Harmark:2004rm}. Each
such interval is called a rod, and for each rod $z\in [a,b]$ we assign a
{\em direction} vector $v$ such that
\begin{equation}
\label{vrod}
g(0,z) v = 0 \quad \mathrm{for} \quad z \in [a,b]\,.
\end{equation}
At
isolated points on the axis the kernel of $g(r=0,z)$ is spanned by two
vectors instead of only one: this will happen at the points where two
intervals with different eigenvectors meet. 
The rod structure of the solution consists of the specification of
intervals $a_1<a_2<\dots<a_n$, plus the $n+1$ directions $v_{(k)} \in
\ker ( g(0,z) )$ for each rod $[a_{k-1},a_k]$, $k=1,...,n+1$
(with $a_0=-\infty$, $a_{n+1}=+\infty$) \cite{Harmark:2004rm}. The
vector $v$ is defined up to an arbitrary normalization constant.

The rod is referred to as timelike or spacelike according to the
character of the rescaled norm $g_{ab}(r,z)v^a v^b/r^2$ at the rod
$r\to0$, $z\in[a_{k-1},a_k]$. For a timelike rod normalized such that the
generator of asymptotic time translations enters with coefficient equal
to one, the rest of the coefficients correspond to the angular
velocities of the horizon (if this satisfies all other regularity
requirements). 

For a spacelike rod, the following two regularity
requirements are important. First, given a  rod-direction vector 
\beq
v=\frac{\partial}{\partial \psi}
\eeq
with norm
\beq
|v|^2=g_{ij}(r,z)v^i v^j
\eeq
the length $L=|v|\Delta\psi$ of its circular orbits at constant $r$, $z$,
vanishes near the rod like $\sim r$. Since the proper radius of these
circles approaches $e^{\nu(0,z)} dr$, a conical singularity will be
present at the rod unless these orbits are periodically identified with
period
\beq\label{deltapsi}
\Delta\psi =2\pi e^{\nu(0,z)}\left(\frac{\partial|v|}{\partial
r}(0,z)\right)^{-1}\,.
\eeq
When several rods are present, it may be impossible to satisfy
simultaneously all the periodicity conditions. The physical
interpretation is that the forces among objects in the configuration
cannot be balanced and as a result, conical singularities appear. If the
geometry admits a Wick rotation to Euclidean
time, then (\ref{deltapsi}) gives the temperature $T_H=2\pi/\Delta\psi$
of the horizon associated to the rod.

Second, the presence of time components on a spacelike rod creates
causal pathologies. Consider a vector $\zeta$ that is timelike
in
some region of spacetime, and whose norm
does
not vanish at a given spacelike rod $[a,b]$. If the direction vector $v$
associated
to this rod is such that 
\beq
v\cdot
\zeta\neq 0 \qquad \mathrm{at}\quad  r=0,\;z\in [a,b]\,,
\eeq 
then making the orbits of $v$ periodic introduces closed timelike
curves: periodicity $\Delta \psi$ along orbits of $v$ requires that the
orbits of $\zeta$ be identified as well with period equal to an integer
fraction of $(v\cdot \zeta)\Delta \psi$. Then, closed timelike curves
will appear wherever $\zeta$ is timelike. These are usually regarded as
pathological if they occur outside the horizon, so such time components
on spacelike rods must be avoided.

Further analysis of the equations \eqref{eqforg}, \eqref{eqfornu} and
their source terms can be found in \cite{Harmark:2005vn}.

\paragraph{The method of Belinsky and Zakharov.} As shown by Belinsky
and Zakharov (BZ), the system of non-linear equations (\ref{eqforg}) and
(\ref{rconstraint}) is completely integrable
\cite{Belinsky:1971nt,Belinski:2001ph}. It admits a Lax pair of linear
equations (spectral equations) whose compatibility conditions coincide
with the original non-linear system. This allows to generate an infinite
number of solutions starting from known ones following a purely
algebraic procedure. Since, as we have seen, we can very easily generate
diagonal solutions, this method allows to construct a vast class of
axisymmetric solutions. It seems likely, but to our knowledge has not
been proven, that all axisymmetric solutions can be generated in this
way. 

The spectral equations for (\ref{rconstraint}) and (\ref{eqforg}) are
\beq\label{eqsPsi}
  D_1\Psi=\frac{r V-\lambda U}{\lambda^2+r^2}~\Psi\,, 
  \qquad D_2\Psi=\frac{r U+\lambda V}{\lambda^2+r^2}~\Psi\,, 
\eeq
where $\lambda$ is the (in general complex) spectral parameter,
independent of $r$ and $z$, and $D_1$, $D_2$ are two commuting
differential operators, 
\beq
  D_1=\partial_z-\frac{2\lambda^2}{\lambda^2+r^2}~\partial_\lambda\,,
\qquad 
  D_2=\partial_r+\frac{2\lambda r}{\lambda^2+r^2}~\partial_\lambda\,.
\eeq
The function $\Psi(\lambda,r,z)$ is a $(d-2)\times(d-2)$ matrix such
that $\Psi(0,r,z)=g(r,z)$, where $g$ is a solution of (\ref{eqforg}).
Compatibility of equations \eqref{eqsPsi} then implies equations
\eqref{eqforg}, \eqref{UVdefs}.

Since (\ref{eqsPsi}) are linear equations, we can construct new
solutions by `dressing' a `seed' solution $g_0$. The seed defines
matrices $U_0$ and $V_0$ through \eqref{UVdefs}. Eqs.~(\ref{eqsPsi}) can
then be solved to determine
$\Psi_0$. Then, we `dress' this solution using a matrix
$\chi=\chi(\lambda,r,z)$ to find a new solution of the form
\begin{equation}\label{newsol}
\Psi=\chi\Psi_0 \,.
\end{equation}
Introducing (\ref{newsol}) in (\ref{eqsPsi}) we find a system of
equations for $\chi$. 
The simplest and most interesting solutions are the {\em solitonic}
ones, for which the matrix $\chi(\lambda,r,z)$ can be written in terms
of simple poles
\begin{equation}\label{chipoles}
\chi=1+\sum_{k=1}^n\frac{R_k}{\lambda-\tilde\mu_k}\;. 
\end{equation}
The residue matrices $R_k$ and the `pole-position' functions
$\tilde\mu_k$ depend only on $r$ and $z$. For a dressing function of
this form, it is fairly straightforward to determine the functions $\tilde\mu_k$
and the matrices $R_k$. The pole-positions are
\begin{equation}\label{muk}
\tilde\mu_k= \left\{
  \begin{array}{cl}
  \mu_k&  \mathrm{for~a~}\mathit{soliton},   \\ 
  \bar\mu_k&  \mathrm{for~an~}\mathit{antisoliton},
  \end{array} 
   \right. 
\end{equation}
where $\mu_k$, $\bar\mu_k$ were introduced in \eqref{mua},
\eqref{barmua}. In principle the
parameters $a_k$ may be complex, and
must appear in conjugate pairs if the metric is to be real.
However, in all the examples that we consider the $a_k$ are real:
complex poles appear to lead to naked singularities, for instance, in
the Kerr solution they occur when the extremality bound on the angular
momentum is violated.

The solution for the matrices $R_k$, 
where $k$ labels the solitons, can
be constructed by first introducing
a set of $d-2$-dimensional vectors $m^{(k)}$  
using the seed as
\begin{equation}\label{mvec}
m_a^{(k)}=m_{0b}^{(k)}\left[\Psi_0^{-1}(\tilde\mu_k,r,z)\right]_{ba}\,.
\end{equation}
The constant vectors $m_{0}^{(k)}$ introduced here are the
crucial new data determining the rod orientations in the new solution. 
Defining now the
symmetric matrix
\begin{equation}\label{gammamat}
\Gamma_{kl}=\frac{m_a^{(k)}\, (g_0)_{ab}\, m_b^{(l)}}{r^2+\tilde\mu_k \tilde\mu_l}\; , 
\end{equation}
the $R_k$ are
\begin{equation}\label{Rk}
(R_k)_{ab}=m_a^{(k)}\sum_{l=1}^n\frac{(\Gamma^{-1})_{lk}m_c^{(l)}(g_0)_{cb}}{\tilde\mu_l}\;. 
\end{equation}

All the information about the solitons that are added to the solution is
contained in the soliton positions $a_k$ and the soliton-orientation
vectors $m_{0}^{(k)}$. This is all we need to determine the dressing
matrix in \eqref{chipoles}, and
then the new metric
$g(r,z)=\Psi(0,r,z)=\chi(0,r,z)\Psi_0(0,r,z)$,
\begin{equation}\label{finsol}
  g_{ab}=(g_0)_{ab}-
  \sum_{k,l=1}^{n}\frac{ 
  (g_0)_{ac}\, m_c^{(k)}\,  (\Gamma^{-1})_{kl}\;  m_d^{(l)}\, (g_0)_{db}}
                       {\tilde\mu_k\tilde\mu_l}\; .  
\end{equation}

There is, however, one problem that turns out to be particularly vexing
in $d>4$: the new metric $g$ in \eqref{finsol} does not satisfy in
general the constraint \eqref{rconstraint}: the introduction of the $n$
solitons gives a determinant for the new metric
\begin{equation}\label{newdet}
 \det g=(-1)^nr^{2n}\left(\prod_{k=1}^n{\tilde\mu_k}^{\;-2}\right)\det g_0\;. 
\end{equation}
A determinant of a new physical solution $g^\textrm{(phys)}$ must
satisfy the constraint \eqref{rconstraint}, but $g$ in \eqref{newdet}
does not. This problem can be expediently resolved by simply multiplying
the metric obtained in \eqref{finsol} by an overall factor 
\beq\label{uniren}
g^\textrm{(phys)}=\pm \left(\frac{r^2}{\pm
\mathrm{det}g}\right)^{\frac{1}{d-2}}\; g
\eeq
(we may ignore here the choice of
$\pm$ signs). Now, observe that the problem that
\eqref{uniren} solves is that of making the rod densities at each point
along the axis to add up to a total density $1/2$. However, recall that
regularity required that individual rod densities, and not just their
sums, be exactly equal to $1/2$. This is a problem for \eqref{uniren}
whenever $d>4$: since \eqref{newdet} contains only solitons and
antisolitons with regular densities $\pm 1/2$ (with $-1/2$ allowed only at intermediate
steps), the fractional power in \eqref{uniren} introduces rods with
fractional densities $1/(d-2)$,
which in $d>4$ will always result
in curvature singularities at the rod. 

A possible way out of this problem is to restrict to transformations
that act only on a $2\times 2$ block of the seed $g_0$
\cite{Koikawa:2005ia}. In this case it is possible to apply the above
renormalization to only this part of the metric--effectively, the same
as in four dimensions-- and thus obtain a solution with the correct,
physical rod densities. However, it is clear that, if we start from
diagonal seeds, this method cannot deal with solutions with off-diagonal
terms in more than one $2\times 2$ block, e.g., with a single rotation.
It cannot be applied to obtain solutions with rotation in several
planes. Still, ref.~\cite{Azuma:2005az} has applied this method to
obtain solutions with arbitrary number of rods, with rotation in a
single plane.

Fortunately, a clever and very practical way out of this problem has
been proposed by Pomeransky \cite{Pomeransky:2005sj} that can deal with
the general case in any number of dimensions. The key idea is the
observation that \eqref{newdet} is independent of the `realignment'
vectors $m_{0b}^{(k)}$. One may then start from a solution with physical
rod densities, `remove' a number of solitons from it (i.e., add solitons
or antisolitons with negative densities $-1/2$), and then readd these
same solitons, but now with different vectors $m_{0b}^{(k)}$, so the
rods affected by these solitons acquire in general new directions. If
the original seed solution satisfied the determinant constraint
\eqref{rconstraint}, then {\em by construction} so will the metric
obtained after re-adding the solitons (including in particular the
sign). And more importantly, if the densities of the initial rods are
all $\pm 1/2$ and negative densities do not appear in the end result,
the final metric will only contain regular $1/2$ densities.

In the simplest form of this method, one starts from a diagonal, hence
static, solution $(g_0,e^{2\nu_0})$ and then removes some solitons or
antisolitons with `trivial' vectors aligned with one of the Killing
basis vectors $m_{0}^{(k)}=\xi_{(a)}$, i.e., $m_{0b}^{(k)} =\delta_{ab}$
(recall that in the diagonal case this alignment of rods is indeed
possible).
Removing a soliton or antisoliton $\tilde\mu_k$ at $z=a_k$ aligned with
the direction $a$ amounts to changing
\beq\label{remsoliton}
(g_0')_{aa}=-\frac{\tilde\mu_k^2}{r^2}(g_0)_{aa}\,,\qquad 
\eeq
while leaving unchanged all other metric components
$(g_0')_{bb}=(g_0)_{bb}$ with $b\neq a$. We now take this new metric
$g_0'$ as the seed to which, following the BZ method, we readd the same
solitons and antisolitons, but with more general vectors
$m_{0}^{(k)}=\sum_b C_b \xi_{(b)}$. Note that there is always the
freedom to rescale each of these vectors by a constant. 
Finally, the two-dimensional conformal factor $e^{2\nu}$ for the new
solution is obtained from the seed as
\begin{equation}\label{e2nu}
 e^{2\nu}=e^{2\nu_0}\;\frac{\det\Gamma}{\det\Gamma_{0}}\,, 
\end{equation}
where the matrices $\Gamma_{0}$ and $\Gamma$ are obtained from
\eqref{gammamat} using $g_0$ and $g$ respectively.

If the vectors $m_{0a}^{(k)}$ for the readded solitons mix the time and
spatial Killing directions, then this procedure may yield a stationary
(rotating) version of the initial static solution. The method requires
the determination of the function $\Psi_0(\lambda,r,z)$ that solves
\eqref{eqsPsi} for the seed. This is straightforward to determine for
diagonal seeds (see the examples below) so for these the method is
completely algebraic. Although even a two-soliton transformation of a
multi-rod metric can easily result in long expressions for the metric
coefficients, the method can be readily implemented in a computer
program for symbolic manipulation. The procedure can also be applied,
although it becomes quite more complicated, to non-diagonal seeds. In
this case, the function $\Psi_0(\lambda,r,z)$ for the non-diagonal seed
is most simply determined if this solution itself is constructed
starting from a diagonal seed. The doubly spinning black ring of
\cite{Pomeransky:2006bd} was obtained in this manner. 

\paragraph{BZ construction of the Kerr, Myers-Perry and black ring solutions.}

We sketch here the method to obtain all known black hole solutions with
connected horizons in four and five dimensions. They illustrate the basic
ideas for how to add rotation to $S^2$, $S^3$ and $S^1\times S^2$
components of multi-black hole horizons. 

The simplest case, which demonstrates one of the basic tools for adding
rotation in more complicated cases, is the Kerr solution--in fact, one
generates the Kerr-Taub-NUT solution, and then sets the nut charge to
zero. Begin from the Schwarzschild solution, generated e.g., using the
techniques available for Weyl solutions \cite{Emparan:2001wk}, and whose
rod structure is depicted in figure~\ref{figure:rods_bh}. The seed metric is
\beq\label{schwg0}
g_0=\mathrm{diag}\left\{ -\frac{\mu_1}{\mu_2},r^2\frac{\mu_2}{\mu_1}\right\}\,,
\eeq
with $a_1<a_2$. Then, remove an anti-soliton at $z=a_1$, with vector
$m_{0}^{(1)}=(1,0)$, and a soliton at $z=a_2$
with the same vector. Following the rule \eqref{remsoliton} we obtain the matrix
\beq
g_0'=\mathrm{diag}\left\{ -\frac{\mu_2}{\mu_1},r^2\frac{\mu_2}{\mu_1}\right\}\,.
\eeq
It is now convenient (but not necessary) to rescale the metric by a
factor $\mu_1/\mu_2$. This yields the new metric
\beq
\tilde g_0'=\mathrm{diag}\left\{ -1,r^2\right\}\,.
\eeq
This is nothing but flat space, so what we have done here is undo the
generation of the Schwarzschild metric out of Minkowski space. But now
the idea is to retrace our previous steps back, after re-adding the
solitons with new vectors $m_{0}^{(1,2)}$. 

So,
following the method above, add an anti-soliton at $z=a_1$ and a soliton
at $z=a_2$, with respective constant vectors
$m_{0}^{(1)}=(1,A_1)$ and
$m_{0}^{(2)}=(1,A_2)$. For this step, we need to construct the matrix
$\Gamma$ in \eqref{gammamat} which in turn requires the matrix 
\beq
\tilde\Psi_0(\lambda,r,z)=\mathrm{diag}\left\{ -1,r^2-2z\lambda-\lambda^2\right\}\,
\eeq
that solves the spectral equations \eqref{eqsPsi} for $\tilde g_0'$.
Equations~\eqref{mvec}, \eqref{gammamat}, and \eqref{finsol}
give then a new metric $\tilde g$. But we
still have to
undo the rescaling we did to get $\tilde g_0'$ from $g_0'$. This is,
\beq
g=\frac{\mu_2}{\mu_1}\tilde g\,,
\eeq
By construction, $g$ is correctly normalized, i.e., satisfies
\eqref{rconstraint}. Finally, the function $e^{2\nu}$ is obtained using
\eqref{e2nu}, which is straightforward since we obtained $\Gamma$ when
the solitons were re-added, and $\Gamma_0=\Gamma(A_1=A_2=0)$. 
The new
solution contains two more parameters than the Schwarzschild seed: these
correspond to the rotation and nut parameters. The latter can be set to
zero once the parameters are correctly identified. For details
about how the Boyer-Lindquist form of the Kerr solution is recovered,
see \cite{Belinski:2001ph}.

The Myers-Perry black hole with two angular momenta is obtained in a
very similar way \cite{Pomeransky:2005sj} starting from the five-dimensional
Schwarzschild-Tangherlini solution, whose rod structure is shown in
figure~\ref{figure:rods_bh}. We immediately see that
\beq
g_0=\mathrm{diag}\left\{ -\frac{\mu_1}{\mu_2},\mu_2,\frac{r^2}{\mu_1}\right\}\,.
\eeq
Now remove an
anti-soliton at $z=a_1$ and a soliton at $z=a_2$, both with vectors $(1,0,0)$, to find
\beq
g_0'=\mathrm{diag}\left\{ -\frac{\mu_2}{\mu_1},\mu_2,\frac{r^2}{\mu_1}\right\}\,.
\eeq
An overall rescaling by $\mu_1/\mu_2$ simplifies the metric to
\beq
\tilde g_0'=\mathrm{diag}\left\{ -1,\mu_1,\frac{r^2}{\mu_2}\right\}=
\mathrm{diag}\left\{ -1,\mu_1,-\bar\mu_2\right\}\,.
\eeq
This is the metric that we dress with solitons applying the BZ method.
Observe that it does not satisfy \eqref{rconstraint} and so it is not a
physical metric, but this is not a problem.
The associated $\tilde\Psi_0$ is
\beq
\tilde\Psi_0(\lambda,r,z)=
\mathrm{diag}\left\{ -1,\mu_1-\lambda,-\bar\mu_2+\lambda\right\}\,
\eeq
Now add the anti-soliton at $z=a_1$ and the soliton at $z=a_2$, with
vectors $m_{0}^{(1)}=(1,0,B_1)$ and $m_{0}^{(2)}=(1,A_2,0)$ (more
general vectors give singular solutions). A final rescaling of the
metric by $\mu_2/\mu_1$ yields a physically normalized solution. The two
new parameters that we have added are associated with the two angular
momenta. The MP solution with a single spin can be obtained through a
one-soliton transformation, which is not possible for Kerr. See
\cite{Pomeransky:2005sj} for the transformation to the coordinates used
in section~\ref{sec:generalMP}.

The black ring with rotation along the $S^1$ requires a more complicated
seed, but on the other hand it requires only a one-soliton
transformation \cite{Elvang:2007rd} (the first systematic derivations of
this solution used a two-soliton transformation
\cite{Iguchi:2006rd,Tomizawa:2006vp}). The seed is described in
figure~\ref{figure:rods_ring}. The static black ring of
\cite{Emparan:2001wk} (which necessarily contains a conical singularity)
is recovered for $a_1=a_2$. However, one needs to introduce a `phantom'
soliton point at $a_1$, and a negative density rod, to eventually obtain
the rotating black ring. Thus we see that the initial solution need not
satisfy any regularity requirements. To obtain the rotating black ring
we remove an anti-soliton at $z=a_1$ with direction $(1,0,0)$, and readd
it with vector $m_{0}^{(1)}=(1,0,B_1)$.
At the end of the process one must adjust the parameters, including
$B_1$ and the rod positions, to remove a possible singularity at the
phantom point $a_1$.

\epubtkImage{}{%
\begin{figure}[h]
  \def\epsfsize#1#2{.55#1}
  \centerline{\epsfbox{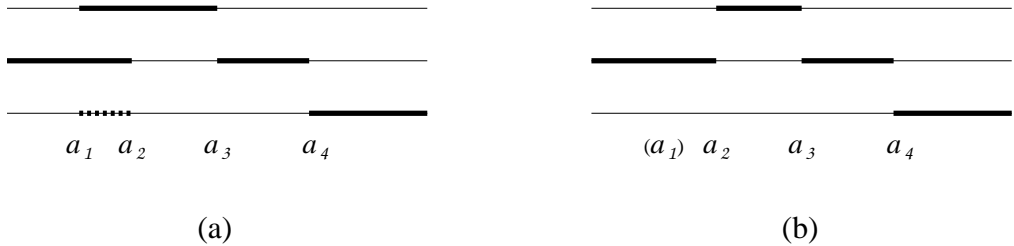}}
\caption{\it Rod structures for (a) the seed used to generate (b) the
rotating black ring. The seed metric is diagonal, and the dotted rod has
negative density $-1/2$. In the
final solution the parameters can be adjusted so the metric at $z=a_1$
on the axis is completely smooth. The (upper) horizon rod in the
final
solution has mixed direction $(1,0,\Omega_\psi)$, while the
other rods are aligned purely along the $\phi$ or $\psi$ directions.}
  \label{figure:rods_ring}
\end{figure}}

The doubly-spinning ring has resisted all attempts at deriving it
directly from a diagonal, static seed. Instead, \cite{Pomeransky:2006bd}
obtained it in a two-step process. Rotation of the $S^2$ of the ring
is similar to the rotation of the Kerr solution. In fact the black ring
solutions with rotation only along the $S^2$ can be obtained by applying
to the static black ring the same kind of two-soliton transformations
that yielded Kerr from a Schwarzschild seed \eqref{schwg0}
\cite{Tomizawa:2005wv}. Hence, if we
begin from the black ring rotating along the $S^1$ and perform similar
soliton and antisoliton transformations at the endpoints of the horizon
rod, we can expect to find a doubly-spinning ring. The main technical
difficulty is in constructing the function $\Psi_0(\lambda,r,z)$ for the
single-spin black ring solution \eqref{neutral}.
However, if we construct this solution via a
one-soliton transformation as we have just explained, this function is
directly obtained from \eqref{newsol}. In this manner, the solution
\eqref{twospinmetric} was derived.

This method has also been applied to construct solutions with
disconnected components of the horizon, which we shall discuss next. The
previous examples provide several `thumb rules' for constructing such
solutions. However, there is no precise recipe for the most efficient
way of generating the sought solution. Quite often, unexpected
pathologies show up, of both local and global type, so a careful
analysis of the solutions generated through this method is always
necessary.

Finally, observe that there are certain arbitrary choices in this method:
it is possible to choose different solitons and anti-solitons, with
different orientations, and still get essentially the same final physical solution.
Also, the intermediate rescaling, and the form for
$\tilde\Psi_0$, admit different choices. All this may lead to
different-looking forms of the final solution, some of them possibly
simpler than others. Occasionally, spurious singularities may be
introduced through bad choices. 

\paragraph{Other methods.}

In four dimensions there exists another technique, akin to the
B\"acklund transformation that adds solitons to a seed solution, to
integrate the equations for the stationary axisymmetric class of vacuum
solutions \cite{backlund}. Refs.~\cite{Mishima:2005id,Iguchi:2006tu}
have extended this to higher dimensions. Unfortunately, even if this
method may produce simpler expressions than the BZ technique, it cannot
deal with more than two off-diagonal terms, and hence no more than a
single angular momentum. Ref.~\cite{Mishima:2005id} applied this method
to derive a black ring with rotation in the $S^2$ (but not along the
$S^1$). The same authors used this technique to provide the first
systematic derivation (i.e. through explicit integration of Einstein's
equations, instead of guesswork) of the black ring with rotation along
$S^1$ \cite{Iguchi:2006rd}. The connection between the B\"acklund
transformation method and the BZ technique has been studied in
\cite{Tomizawa:2006jz}, where it is argued that all the solutions
obtained by a two-soliton B\"acklund transformation on an arbitrary
diagonal seed are contained among those that the BZ method generates
from the same seed. This may not be too surprising in view of similar,
and more general, results in four dimensions \cite{cosgrove}. It may be
interesting to investigate the application of related but more efficient
axisymmetric solution-generating methods \cite{sibgatullin} to higher
dimensions.

Ref.~\cite{Giusto:2007fx} develops a different algebraic method to
obtain stationary axisymmetric solutions in five dimensions from a given
seed. An $SO(2,1)$ subgroup of the ``hidden'' $SL(3,\mathbb{R})$
symmetry of solutions with at least one spatial Killing vector (the
presence of a second one is assumed later) is identified that preserves
the asymptotic boundary conditions, and whose action on a static
solution generates a one-parameter family of stationary solutions with
angular momentum---e.g., one can obtain the Myers-Perry solution from a
Schwarzschild-Tangherlini seed. It is conjectured that all vacuum
stationary axisymmetric solutions can be obtained by repeated
application of these transformations on static seeds.


\subsection{Multi-black hole solutions}
\label{subsec:multibhs}

In $d=4$ it is believed that there are no stationary multi-black hole
solutions of vacuum gravity. However, such solutions do exist in $d=5$.
`Black Saturn' solutions, in which a central MP-type of black hole is
surrounded by a concentric rotating black ring, have been constructed in
\cite{Elvang:2007rd}. They exhibit a number of interesting features,
such as rotational dragging of one black object by the other, as well as
both co- and counter-rotation. For instance, we may start from a static
seed and act with the kind of one-soliton BZ transformation that turns
on the rotation of the black ring. This gives angular momentum (measured
by a Komar integral on the horizon) to the black ring but not to the
central black hole. However, the horizon rod of this black hole is
reoriented and acquires a non-zero angular velocity: the black hole is
dragged along by the black ring rotation. It is also possible (this
needs an additional one-soliton transformation that turns on the
rotation of the MP black hole) to have a central black hole with a
static horizon that nevertheless has non-vanishing angular momentum: the
`proper', inner rotation of the black hole is cancelled at its horizon
by the black ring drag-force.

The explicit solutions are rather complicated, but an intuitive
discussion of their properties is presented in \cite{Elvang:2007hg}. The
existence of black Saturns is hardly surprising: since black rings can
have arbitrarily large radius, it is clear that we can put a small black
hole at the center of a very long black ring, and the interaction
between the two objects will be negligible. In fact, since a black ring
can be made arbitrarily thin and light for any fixed value of its angular
momentum, then for any non-zero values of the total mass and angular
momentum, we can obtain a configuration with larger total area than any
MP black hole or black ring: put almost all the mass in a central,
almost static black hole, and the angular momentum in a very thin and
long black ring. Such configurations can be argued to attain the maximal
area (i.e., entropy) for given values of $M$ and $J$. Observe also that
for fixed total $(M,J)$ we can vary, say, the mass and spin of the black
ring, while adjusting the mass and spin of the central black hole to add
up to the total $M$ and $J$. These configurations, then, exhibit
doubly-continuous non-uniqueness.

We can similarly consider multi-ring solutions. Di-rings, with two
concentric black rings rotating on the same plane, were first
constructed in \cite{Iguchi:2007is}; ref.~\cite{Evslin:2007fv} rederived
them using the BZ approach. Each new ring adds two parameters to the
continuous degeneracy of solutions with given total $M$ and $J$. 

Note that the surface gravities (i.e., temperatures) and angular
velocities of disconnected components of the horizon are in general
different. Equality of these `intensive parameters' is a necessary
condition for thermal equilibrium --and presumably also for mergers in
phase space to solutions with connected horizon components, see
section~\ref{subsec:hidphasediag}. So these multi-black hole
configurations cannot in general exist in thermal equilibrium (this is
besides the problems in constructing a Hartle-Hawking state
when ergoregions are present \cite{Kay:1988mu}). The curves for
solutions where all disconnected components of the horizon have the same
surface gravities and angular velocities are presented in
figure~\ref{figure:phases5D} (see \cite{Elvang:2007hg}). All continuous
degeneracies are removed, and black Saturns are always subdominant in
total horizon area. It is expected that no multi-ring solutions exist in
this class.

\epubtkImage{}{%
\begin{figure}[h]
  \def\epsfsize#1#2{1.2#1}
  \centerline{\epsfbox{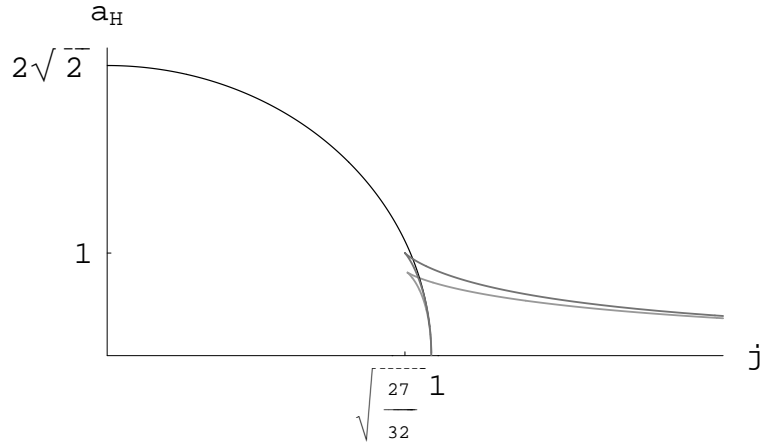}}
\caption{\it Curves $a_H(j)$ for phases of five-dimensional black holes
with a single angular momentum: MP black hole (black), black ring (dark
gray), black Saturn (light gray). We only include those black Saturns
where the central black hole and the black ring have equal surface
gravities and angular velocities. The three curves meet tangentially at
a naked singularity at $j=1$, $a_H=0$. The cusp of the black ring curve
occurs at $j=\sqrt{27/32}\approx 0.9186$, $a_H=1$. The cusp of the black
Saturn curve is at $j\approx 0.9245$, with area $a_H\approx
0.81$.}
  \label{figure:phases5D}
\end{figure}}

It is also possible to have two black rings lying and rotating on
orthogonal, independent planes. Such {\em bicycling black rings} have
been constructed using the BZ method \cite{Izumi:2007qx,Elvang:2007hs},
and provide a way of obtaining configurations with arbitrarily large
values of both angular momenta for fixed mass -- which cannot be
achieved, simultaneously for both spins, either by the MP black holes
or by doubly-spinning black rings. The solutions in
\cite{Izumi:2007qx,Elvang:2007hs} are obtained by applying to each of
the two rings the kind of transformations that generate the
singly-spinning black ring. Thus each black ring possesses angular
momentum only on its plane, along the $S^1$, but not in the orthogonal
plane, on the $S^2$--nevertheless, they drag each other so the two
horizon angular velocities are both non-zero on each of the two
horizons. The solutions contain four free parameters, corresponding to
e.g., the masses of each of the rings and their two angular momenta. It is clear
that a more general, six-parameter solution must exist in which
each black ring has both angular momenta turned on. 

It can be argued, extending the arguments in \cite{Elvang:2007hg}, that
multi-black hole solutions allow to cover the entire $(j_1,j_2)$ phase
plane of five-dimensional solutions. It would be interesting to
determine for which parameter values these multi-black holes have the
same surface gravity and angular velocities on all disconnected components
of the horizon. With this constraint, multi-black hole solutions still
allow to cover a larger region of the $(j_1,j_2)$ plane than already
covered by solutions with connected horizon,
fig.~\ref{figure:5Dphasespace}. For instance, it can be argued that some
bicycling black rings (within the six-parameter family mentioned above for
which the four angular velocities of the entire system can be varied
independently) should satisfy these `thermal equilibrium' conditions: as
we have seen, a doubly-spinning black ring can have $\Omega_1=\Omega_2$.
Thus, if we consider two identical doubly-spinning thin black rings, one
on each of the two planes, then we can make them have $S^1$-angular
velocity equal to the $S^2$-angular velocity of the other ring in the
orthogonal plane. These solutions then lie along the lines $|j_1|=
|j_2|$, reaching arbitrarily large $|j_{1,2}|$, which is not covered by
the single-black hole phases in figure~\ref{figure:5Dphasespace}.
Clearly, there will also exist configurations extending continuously
away from this line.

Black Saturns with a single black ring should also exist that satisfy
`thermal equilibrium' conditions. In fact, the possibility of varying
the temperature of the ring by tuning the rotation in the $S^2$ might
 allow to cover portions of the $(j_1,j_2)$ plane beyond
fig.~\ref{figure:5Dphasespace}. If so, this would be unlike the
situation with a single rotation, where thermal-equilibrium Saturns lie
within the range of $j$ of black rings, fig.~\ref{figure:phases5D}.

The Weyl ansatz of sec.~\ref{subsubsec:weyl} enables one to generate easily
solutions in $d=5$ with multiple black holes of horizon topology $S^3$ which are
asymptotically flat \cite{Emparan:2001wk,Tan:2003jz}. However, all these
solutions possess conical singularities reflecting the attraction
between the different black holes. It seems unlikely that the extension
to include off-diagonal metric components (rotation and twists) could
eliminate these singularities and yield balanced solutions.


\subsection{Stability}
\label{subsec:5Dstability}

The linearized perturbations of the black ring metric \eqref{neutral}
have not yielded to analytical study. The apparent absence of a Killing
tensor prevents the separation of variables even for massless scalar
field pertubations. Also, the problem of decoupling the equations to
find a master equation for linearized gravitational perturbations,
already present for the Myers-Perry solutions, is if anything
exacerbated for black rings.

Studies of the classical stability of black rings have therefore been
mostly heuristic. Already the original paper \cite{Emparan:2001wn}
pointed out that very thin black rings locally look like boosted black
strings (this was made precise in \cite{Elvang:2003mj}), which were
expected to suffer from GL-type instabilities. The instability of
boosted black strings was indeed
confirmed in \cite{Hovdebo:2006jy}. Thus, thin black rings are expected
to be unstable to the formation of ripples along their $S^1$ direction.
This issue was examined in further detail in \cite{Elvang:2006dd}, who
found that thin black rings seem to be able to accommodate unstable GL
modes down to values $j\sim O(1)$. Thus, it is conceivable that a large
fraction of black rings in the thin branch, and possibly all of this
branch, suffer from this instability. The ripples rotate with the black
ring and then should emit gravitational radiation. However, the
timescale for this emission is much longer than the timescale of the
fastest GL mode, so the pinch-down created by this instability will
dominate the evolution, at least initially. The final fate of this
instability of black rings depends on the endpoint of the GL
instability, but it is conceivable, and compatible with an increment of
the total area, that the black ring fragments apart into smaller black
holes that then fly away.

Another kind of instability was discussed in \cite{Elvang:2006dd}. By
considering off-shell deformations of the black ring (namely, allowing
for conical singularities), it was possible to compute an effective
potential for radial deformations of the black ring. Fat black rings sit
at maxima of this potential, while thin black rings sit at minima. Thus,
fat black rings are expected to be unstable to variations of their
radius, and presumably collapse to form MP black holes. The analysis in
\cite{Elvang:2006dd} is in fact consistent with a previous, more
abstract analysis of local stability in \cite{Arcioni:2004ww}. This is
based on the `turning-point' method of Poincar\'e, which studies the
equilibrium curves for phases near bifurcation points. For the case of
black rings, one focuses on the cusp where the two branches meet. One
then assumes that these curves correspond to extrema of some potential,
e.g., an entropy, that can be defined on all the plane $(j,a_H)$. The
cusp then corresponds to an inflection point of this potential
where a branch of maxima and a branch of minima meet. By continuity, the
branch with the higher entropy will be the most stable branch, and the
one with lower entropy will be unstable. Thus, for black rings an unstable mode
is added when going from the upper (thin) to the lower (fat) branch.
This is precisely as found in \cite{Elvang:2006dd} from the mechanical
potential for radial deformations.

Thus, a large fraction of all single-spin neutral black rings are expected
to be classically unstable, and it remains an open problem whether a
window of stability exists for thin black rings with $j\sim O(1)$. The
stability, however, can improve greatly with the addition of charges and
dipoles.

Doubly-spinning black rings are expected to suffer from similar
instabilities. Insofar as a fat ring branch exists that meets at a cusp
with a thin ring branch, the fat rings are expected to be unstable. Very
thin rings are also expected to be unstable to GL-perturbations that
form ripples. The angular momentum on the $S^2$ may be redistributed
non-uniformly along the ring, with the larger blobs concentrating more
spin. In addition, although it has been suggested that superradiant ergoregion
instabilities associated to rotation of the $S^2$ might exist
\cite{Dias:2006zv}, a proper account of the asymptotic behavior
of superradiant modes needs to be made before concluding that the
instability is actually present. 

Much of what we can say about the classical stability of black Saturns and multi-rings
follows from what we have said above for each of its components--e.g.,
if their rings are thin enough they are expected to be GL-unstable. We
know essentially nothing about what happens when the gravitational
interactions among the black objects involved is strong. For instance,
we do not know if the GL instability is still present when a thin black
ring lassoes very closely a much larger central MP black hole. 

Massive geodesics on the plane of a black ring (see
\cite{Hoskisson:2007zk}) show that a particle at the center of the $S^1$
is unstable to migrating away towards the black ring. This suggests that
a black Saturn with a small black hole at the center of a larger black
ring should be unstable. A suggestive possibility for a
different instability of black Saturns appears from the analysis of
counter-rotating configurations in \cite{Elvang:2007rd}. For large
enough counter-rotation, the Komar-mass of the central black hole
vanishes and then becomes negative. By itself, this does not imply any
pathology as long as the total ADM mass is positive and the horizon
remains regular, which it does. However, it suggests that the
counterrotation in this regime becomes so extreme that the black hole
might tend to be expelled off the plane of rotation.

Clearly, the classical stability of all, old and new, rotating black hole
solutions of five-dimensional General Relativity remains largely an
open problem where much work remains to be done.

\newpage


\section{Vacuum solutions in more than five dimensions}
\label{section:vachigherd}

With no available techniques to construct asymptotically flat exact
solutions beyond those found by Myers and Perry, the situation in $d\geq
6$ is much less developed than in $d=5$. Recall that the symmetry
requirements of the class of solutions \eqref{aximetrics} are
incompatible with the asymptotic symmetries of globally flat space in
$d>5$: metrics of the type \eqref{aximetrics} necessarily involve
directions with non-flat asymptotics (e.g., infinitely extended black
strings and branes) and/or asymptotic compact Kaluza-Klein circles. In
order to preserve asymptotic flatness one would instead ask for
rotational symmetry around `axes' that are hyperplanes of codimension
larger than two, but the integrability of the resulting equations
\cite{Charmousis:2003wm,Charmousis:2006fx} remains essentially an
unsolved problem.

However, despite the paucity of exact solutions, there are strong
indications that the variety of black holes that populate General
Relativity in $d\geq 6$ is vastly larger than in $d=4,5$. A first
indication came from the conjecture in ref.~\cite{Emparan:2003sy} about
the existence of black holes with spherical horizon topology but with
axially symmetric `ripples' (or `pinches'). The plausible existence of black
rings in any $d\geq 5$ was argued in \cite{Hovdebo:2006jy,Elvang:2006dd}.
More recently,
ref.~\cite{Emparan:2007wm} has constructed approximate solutions for
black rings in any $d\geq 5$ and then exploited the conjecture of
\cite{Emparan:2003sy} to try to draw a phase diagram with connections
and mergers between the different expected phases. In the following we
summarize these results.


\subsection{Approximate solutions from curved thin branes}

In the absence of exact techniques, ref.~\cite{Emparan:2007wm} resorted
to approximate constructions, in particular to the method of matched
asymptotic expansions previously used in the context of black holes
localized in Kaluza-Klein circles in
\cite{Harmark:2003yz,Gorbonos:2004uc,Karasik:2004ds,Gorbonos:2005px}\epubtkFootnote{The 
classical effective field theory of
\cite{Chu:2006ce,Kol:2007rx} is an alternative to matched asymptotic
expansions which presumably should be useful as well in the context
discussed in this section.}.
The basic idea is to find two widely separated scales in the problem,
call them $R_1$ and $R_2$, with $R_2\ll R_1$. Then try to solve the
equations in two limits: first, as a perturbative expansion for small
$R_2$, and then in an expansion in $1/R_1$. The former solves the
equations in the far-region $r\gg R_2$, in which the boundary
conditions, e.g., asymptotic flatness, fix the integration constants.
The second expansion is valid in a near-region $r\ll R_1$. In order to
fix the integration constants in this case, one matches the two
expansions in the overlap region $R_2\ll r\ll R_1$ in which both
approximations are valid. The process can then be iterated to higher
orders in the expansion, see \cite{Gorbonos:2004uc} for an explanation
of the systematics involved.

In order to construct a black ring with horizon topology $S^1\times
S^{d-3}$, we take the scales $R_1,R_2$ to be the radii of the $S^1$ and
$S^{d-3}$ respectively\epubtkFootnote{The $S^{d-3}$ is not round for known
solutions, but one can define an effective scale $R_2$ as the radius of
a round $S^{d-3}$ with the same area.}. To implement the above
procedure, we take $R_2=r_0$, the horizon
radius of the $S^{d-3}$ of a straight boosted black string, and $R_1=R$
the large circle radius of a very thin circular string. Thus, in effect,
to first order in the expansion what one does is: (i) find the solution
within linearized approximation, i.e., for small $r_0/r$, around a
Minkowski background for an infinitely thin circular string with
momentum along the circle; (ii) perturb a straight boosted black string
so as to bend it into an arc of circle of very large radius $1/R$. The
latter step not only requires matching to the previous solution in order to
provide boundary conditions for the homogeneous differential equations:
one also needs to check that the perturbations can be made compatible
with regularity of the horizon. 

It is worth noting that the form of the solution thus found exhibits a
considerable increase in complexity when going from $d=5$, where an
exact solution is available, to $d>5$: simple linear functions of $r$ in
$d=5$ change to hypergeometric functions in $d>5$. We take this as an
indication that exact closed analytical forms for these solutions may
not exist in $d>5$.

We will not dwell here on the details of the perturbative construction
of the solution --see \cite{Emparan:2007wm} for this--, but instead we
shall emphasize that adopting the view that a black object is
approximated by a certain very thin black brane curved into a given
shape can easily yield non-trivial information about new kinds of black
holes. Eventually, of course, the assumption that the horizon remains
regular after curving needs to be checked.

Consider then a stationary black brane, possibly with some momentum
along its worldvolume, with horizon topology $\mathbb{R}^{p+1}\times
S^q$, with $q=d-p-2$. When viewed at distances much larger than the size
$r_0$ of the $S^q$, we can approximate the metric of the black brane
spacetime by the gravitational field created by an `equivalent source'
with distributional energy tensor $T_{\mu}^{\nu}\propto r_0^{q-1}\delta^{(q+1)}(r)$,
with non-zero components only along directions tangent to the worldvolume,
and where $r=0$ corresponds to the location of the brane. Now we want to
put this same source on a curved, compact $p$-dimensional spatial
surface in a given background spacetime (e.g., Minkowski, but possibly
(Anti-)deSitter or others, too). In principle we can obtain the mass $M$
and angular momenta $J_i$ of the new object by integrating $T_{t}^{t}$ and
$T_{t}^{i}$ over the entire spatial section of the brane worldvolume.
Moreover, the total area $\mathcal{A}_H$ is similarly obtained by
replacing the volume of $\mathbb{R}^{p}$ with the volume of the new
surface. Thus, it appears that we can easily obtain the relation
$\mathcal{A}_H(M,J_i)$ in this manner.

There is, however, the problem that having changed the embedding
geometry of the brane, it is not guaranteed that the brane will remain
stationary. Moreover, $\mathcal{A}_H$ will be a function not only of
$(M,J_i)$, but it will also depend explicitly on geometrical parameters
of the surface. However, we would expect than in a situation of
equilibrium some of these geometrical parameters should be fixed dynamically by
the mechanical parameters $(M,J_i)$ of the brane. For instance, take a
boosted string and curve it into a circular ring so the linear velocity
turns into angular rotation. If we fix the mass and the radius then the
ring will {\em not} be in equilibrium for every value of the boost,
i.e., of the angular momentum, so there must exist a fixed relation
$R=f(M,J)$. This is reflected in the fact that in the new situation the
stress-energy tensor is in general not conserved, $\nabla_\mu
T^{\mu\nu}\neq 0$: additional stresses would be required to keep the
brane in place. An efficient way of imposing the brane equations of
motion is in fact to demand conservation of the stress-energy tensor. In the
absence of external forces, the classical equations of motion of the
brane derived in this way are \cite{Carter:2000wv}
\beq\label{cartereq}
{K^\rho}_{\mu\nu}T^{\mu\nu}=0\,, 
\eeq 
where ${K^\rho}_{\mu\nu}$ is the second-fundamental tensor,
characterizing the extrinsic curvature of the embedding surface spanned
by the brane worldvolume. For a string on a circle of radius $R$ in flat
space, parametrized by a coordinate $z\sim z+2\pi R$, this equation
becomes 
\beq
\frac{T_{zz}}{R}=0\,.
\eeq
In $d=5$, this can be seen to correspond to the condition of absence of
conical singularities in the solution \eqref{neutral}, in the limit of a
very thin black ring \cite{Elvang:2003mj}. Ref.~\cite{Emparan:2007wm}
showed that this condition is also required in $d\geq 6$ in order to
avoid curvature singularities on the plane of the ring.

In general, eq.~\eqref{cartereq}
constrains the allowed values of parameters of a
black brane that can be put on a given surface.
Ref.~\cite{Emparan:2007wm} easily derived, for
any $d\geq 5$, that 
the radius $R$ of thin rotating black rings of given $M$ and $J$ is fixed to 
\beq
R=\frac{d-2}{\sqrt{d-3}}\frac{J}{M}
\eeq
so large $R$ corresponds to large spin for fixed mass.
The horizon area of these thin black rings goes like
\begin{equation}\label{stringS}
\mathcal{A}_H (M,J) \propto J^{-\frac{1}{d-4}}\;M^{\frac{d-2}{d-4}}
\,.
\end{equation}
This is to be compared to the value for ultra-spinning MP black holes in
$d \geq 6$ (cf.\ eqs.~\eqref{jnu}, \eqref{ahnu} as $\nu\to 0$),
\begin{equation}
\mathcal{A}_H (M,J) \propto J^{-\frac{2}{d-5}}\;M^{\frac{d-2}{d-5}}
\,.
\end{equation}
This shows that in the ultra-spinning regime the rotating black ring has
larger area than the MP black hole.


\subsection{Phase diagram}
\label{subsec:hidphasediag}

Using the dimensionless area and spin variables \eqref{jaHdef},
equation~\eqref{stringS} allows to compute the asymptotic form of the
curve $a_H(j)$ in the phase diagram at large $j$ for black rings.
However, when $j$ is of order one the approximations in the matched
asymptotic expansion break down, and the gravitational interaction of
the ring with itself becomes important. At present we have no analytical
tools to deal with this regime for generic solutions. In most cases,
numerical analysis may be needed to obtain precise information.

Nevertheless, ref.~\cite{Emparan:2007wm} has advanced heuristic
arguments to propose a completion of the curves that is qualitatively
consistent with all the information available at present. A basic
ingredient is the observation in \cite{Emparan:2003sy}, discussed in
section~\ref{subsec:singlespin}, that
in the ultraspinning regime in $d\geq 6$, MP black holes
approach the geometry of a black membrane $\approx \mathbb{R}^2 \times
S^{d-4}$ spread out along the plane of rotation.

We already discussed how using this analogy, ref.~\cite{Emparan:2003sy}
argued that ultra-spinning MP black holes should exhibit
a Gregory-Laflamme-type of instability. Since
the threshold mode of the GL instability gives rise to a new branch of
static non-uniform black strings and branes
\cite{Gregory:1987nb,Gubser:2001ac,Wiseman:2002zc},
ref.~\cite{Emparan:2003sy} argued that it is natural to
conjecture the existence of new branches of axisymmetric `lumpy' (or
`pinched') black holes, branching off from the MP solutions along the
stationary axisymmetric zero-mode perturbation of the GL-like
instability.

\epubtkImage{}{%
\begin{figure}[h]
  \def\epsfsize#1#2{.4#1}
  \centerline{\epsfbox{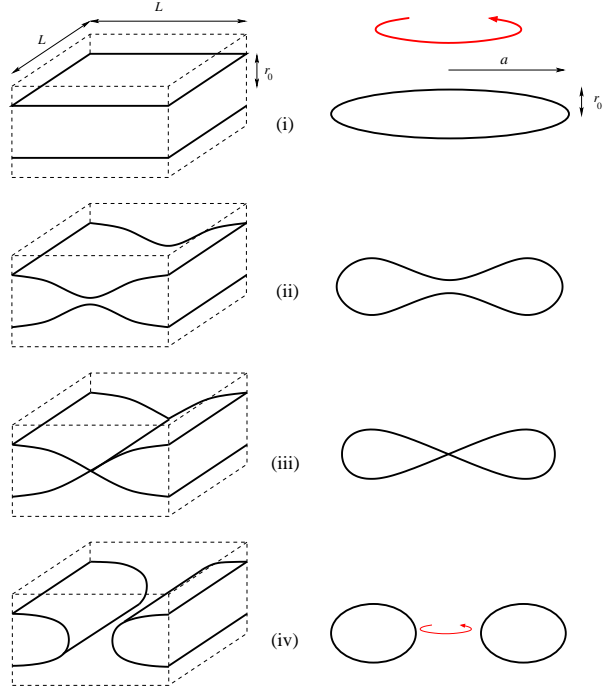}}
\caption{\it  Correspondence
between phases of black membranes wrapped
on a two-torus of side $L$ (left) and fastly-rotating MP black holes with rotation
parameter $a\sim L\geq r_0$ (right: must be rotated along a vertical axis):
(i) Uniform black membrane and MP black hole.
(ii) Non-uniform black membrane and pinched black hole.
(iii) Pinched-off membrane and black hole. (iv) Localized black string
and black ring. (Reproduced from \cite{Emparan:2007wm}).}
  \label{figure:membranes}
\end{figure}}

Ref.~\cite{Emparan:2007wm} developed further this analogy, and drew a
correspondence between the phases of black membranes and the phases of
higher-dimensional black holes, illustrated in figure~\ref{figure:membranes}.
Although the analogy has several limitations, it allows to propose a
phase diagram in $d\geq 6$ of the form depicted in figure~\ref{figure:hidphases},
which should be compared to the much simpler diagram in five dimensions,
figure~\ref{figure:phases5D}.
Observe the presence of an infinite number of black holes with spherical
topology, connected via merger transitions to MP black holes, black
rings, and black Saturns. Of all multi-black hole configurations, the
diagram only includes those phases in which all components of the
horizon have the same surface gravity and angular velocity: presumably,
these are the only ones that can merge to a phase with connected
horizon. Even within this class of solutions, the diagram is not
expected to contain all possible phases with a single angular momentum:
blackfolds with other topologies must likely be included too. The
extension to phases with several angular momenta also remains to be
done.

\epubtkImage{}{%
\begin{figure}[h]
  \def\epsfsize#1#2{.6#1}
  \centerline{\epsfbox{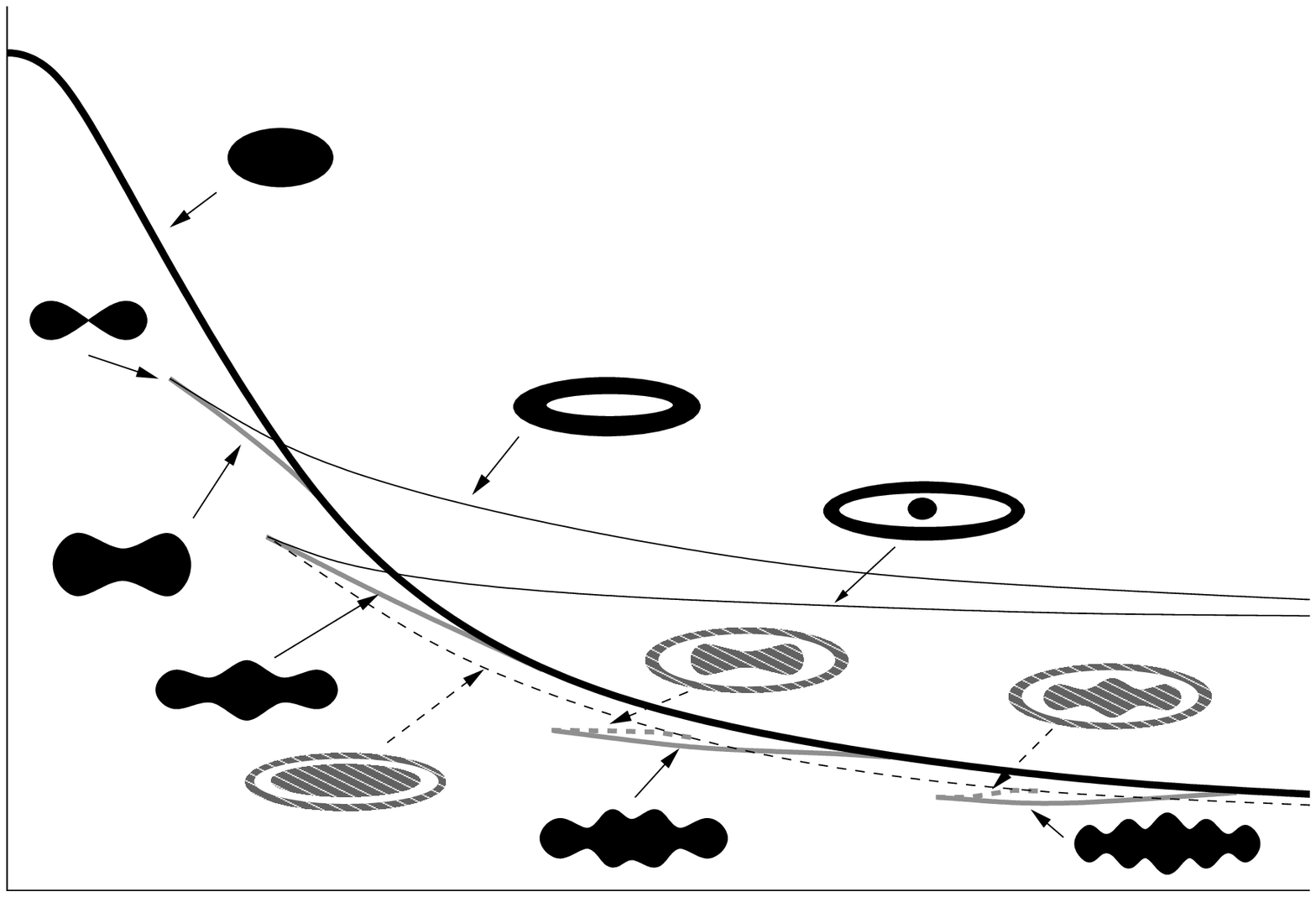}}
  \caption{\it Proposal of ref.~\cite{Emparan:2007wm} for the phase
curves $a_H(j)$ of thermal equilibrium phases in $d\geq 6$. The solid lines and
figures have significant arguments in their favor, while the dashed
lines and figures might not exist and admit conceivable, but more
complicated, alternatives. Some features have been drawn arbitrarily: at
any given bifurcation and in any dimension, smooth connections are
possible instead of swallowtails with
cusps; also, the bifurcation into two black Saturn phases may happen
before, after, or right at the merger with the pinched black hole.
Mergers to di-rings or multi-ring configurations that extend to
asymptotically large $j$ seem unlikely. If thermal equilibrium is not
imposed, the whole semi-infinite strip $0<a_H <a_H(j=0)$, $0\leq
j<\infty$ is covered, and multi-rings are possible.}
  \label{figure:hidphases}
\end{figure}}

Indirect evidence for the existence of black holes with pinched horizons
is provided by the results of \cite{Lahiri:2007ae}, which finds `pinched
plasma ball' solutions of fluid
dynamics that are CFT duals of pinched black holes in
six-dimensional AdS space. The approximations involved in the
construction require that the horizon size of the dual black holes be
larger than the AdS curvature radius, so they do not admit a limit to
flat space space. Nevertheless, their existence provides an example, if
indirect, that pinched horizons make appearance in $d=6$ (and not in $d=5$).


\subsection{Stability}

The situation in $d\geq 6$ is very similar to what we described for
$d=5$ in
section~\ref{subsec:5Dstability}: most of what we know is
deduced by heuristic analogies and approximate methods. The following
prototypic instabilities can be easily identified:
\begin{itemize}

\item GL-type instabilities in ultra-spinning regimes. Both MP black
holes and black rings approach in this regime string or membrane-like
configurations that are expected to be unstable to growing
non-uniformities along the `long' directions on the horizon. Such
instabilities give rise, in the cases where the non-uniformities can
remain stationary, to zero-modes that branch off into new solutions with
a broken symmetry.

\item Turning-point instabilities. Given the phase diagram with
different curves $a_H(j)$ in it, we expect that near a point where
different phases meet, the branch with lower area should have one more
unstable mode than those with higher area. This happens along black ring
curves at the turning points where $j$ reaches a minimum. The
instability is presumably related to radial deformations of the ring. It
also happens at mergers between phases.

\end{itemize}

We finish this section emphasizing that, presumably, new concepts and
tools are required for the characterization of black holes in $d\geq 6$,
let alone for their explicit construction.

The general problem of the dynamical linearized stability of
MP black holes, in particular in the case with a single rotation,
becomes especially acute for the determination of possible black hole
phases in $d\geq 6$. The arguments in favor of an ultra-spinning
instability seem difficult to evade, so a most pressing problem is to
locate the point (i.e. the value of $j$) at which this instability
appears, as a stationary mode, and then perturb the solution along this
mode to determine the direction in which the new branch of solutions
evolves.

\newpage


\section{Solutions with a gauge field}
\label{section:charged}


\subsection{Introduction}

Black holes with non-trivial gauge fields play an important role in
string theory, not least because the types of charge that they carry
helps identify their microscopic constituents (e.g. branes). In this
section we shall review briefly such solutions for $d \ge 5$. We shall
concentrate exclusively on solutions of maximal supergravity theories
arising as consistent toroidal reductions of $d=10,11$ supergravity. 

The bosonic sector of a maximal supergravity theories contains massless
scalars taking values in some coset space $G/H$ where $G$ is noncompact
and $H$ is compact. Since the scalars are massless, their asymptotic
values (for an asymptotically flat solution) are arbitrary, i.e., they
are moduli. $G$ is a global symmetry of the supergravity theory (broken
to a discrete U-duality symmetry in string/M-theory). By acting with $G$
we can choose the moduli to be anything convenient. Given a particular
choice for the moduli, the global symmetry group is broken to $H$. See
e.g.~\cite{polchinski} for a review of this. 

The only type of conserved gauge charge that an asymptotically flat
solution can carry is electric charge with respect to a 2-form field
strength (or magnetic charge with respect to a $d-2$ form field
strength, but this can always be dualized into a 2-form). The group $H$
acts non-trivially on the charges of a solution. Hence the strategy in
looking for black hole solutions is to identify a ``seed" solution with
a small number of parameters, from which a solution with the most
general charge assignments can be obtained by acting with $H$. For
example, dimensional reduction of $d=11$ supergravity on $T^6$ gives the
maximal ${\cal N}=4$, $d=5$ supergravity, which has 27 abelian vectors
and 42 scalars parametrizing the coset $E_{6(6)}/USp(8)$. It turns out
that the a solution with 27 independent charges can be obtained by
acting with $USp(8)$ on a seed solution with just 3 charges
\cite{cvetichull}. 


\subsection{Topologically spherical black holes}

\subsubsection{Non-extremal solutions}

The construction of stationary, charged, topologically spherical, black
hole solutions (``charged Myers-Perry") in maximal supergravity theories
was described in \cite{cvetichull}. It turns out that the seed solutions
are the same as for toroidal compactifications of heterotic string
theory. The latter seed solutions were constructed in
\cite{cveticyoum5d} for $d=5$ and \cite{heteroticseed} for $6 \le d \le
9$. In addition to their mass and angular momenta, they are parametrized
by 3 electric charges in $d=5$ and 2 electric charges for $6\le d \le
9$. For $d=10$ (type IIA theory), the solution describing a rotating
black hole charged with respect to the Ramond-Ramond 2-form field
strength (i.e. D0-brane charge) can be obtained from a general rotating
brane solution given in \cite{elevenauthors}.

In the limit of vanishing angular momenta, all of these solutions reduce to $d>4$ generalizations of the Reissner-Nordstrom solution. This static case is the only case for which linearized stability has been investigated. The spherical symmetry of static solutions permits a scalar/vector/tensor decomposition of perturbations. For Reissner-Nordstrom solutions of $d>4$ Einstein-Maxwell theory, decoupled equations governing each type have been obtained \cite{kodamacharged}. These have been used to prove analytically the stability of tensor and vector perturbations for $d \ge 5$  and scalar perturbations for $d=5$ \cite{kodamacharged}. Numerical studies have revealed that scalar perturbations are also stable for $6 \le d \le 11$ \cite{Konoplya:2007jv}.

\subsubsection{BPS solutions: the BMPV black hole}

Black holes saturating a BPS inequality play an important role in
string theory. The canonical example is the first black hole entropy
calculation \cite{stromingervafa}, in which the BPS condition provided
the justification for relating the Bekenstein-Hawking entropy of a
classical black hole solution to the statistical entropy of a
microscopic brane configuration.

The only known asymptotically flat BPS black hole solutions occur for
$d=4,5$. For $d=5$, one can obtain BPS black hole solutions as a limit
of the non-extremal charged rotating solutions just discussed. Starting
from the 6-parameter seed solution of \cite{cveticyoum5d} one can obtain
a 4-parameter BPS black hole: the BMPV black hole \cite{bmpv,tseytlin}.
The solution has equal angular momenta $J_1=J_2=J$ and the 4 parameters
are $J$ and the 3 electric charges. The mass is fixed by the BPS
relation. Note that one loses {\it two} parameters in the BPS limit:
this is because the BPS limit and the extremal limit are distinct for
rotating black holes - a BPS black hole is necessarily extremal but the
converse is untrue. As for the Myers-Perry solution, the equality of the
angular momenta gives rise to a non-abelian isometry group $\mathbb{R} \times
U(1) \times SU(2)$. Classical properties of the BMPV solution have been
discussed in detail in \cite{gmt,herdeiro}.


\subsection{Black rings with gauge fields}

\subsubsection{Dipole rings}

Black holes can only carry electric charge with respect to a 2-form.
However, higher rank $p$-forms may also be excited by a black hole, even
though there is no net charge associated with them. This occurs
naturally for black rings. Consider a black string in $d$ dimensions. A
string carries electric charge associated with a 3-form field strength
$H$. The electric charge is proportional to the integral of $\star H$
over a $(d-3)$-sphere that links the string. Now consider a black ring
with topology $S^1 \times S^{d-3}$ formed by bending the string into a
loop and giving it angular momentum around the $S^1$. This would be
asymptotically flat and hence the charge associated with $H$ would be
zero. Nevertheless, $H$ would be non-zero. Its strength can be measured
by the flux of $\star H$ through a $S^{d-3}$ linking the ring, which is
no longer a conserved quantity but rather a non-conserved electric {\it
dipole}. Such a ring would have three parameters, but would only have
two conserved charges (mass and angular momentum), hence it would
exhibit {\it continuous} non-uniqueness.

Exact solutions describing such ``dipole rings" have been constructed in
$d=5$ \cite{Emparan:2004wy}. In $d=5$, one can dualize a 3-form $H$ to a
2-form, so these solutions carry {\it magnetic} dipoles with respect to
2-form fields strengths. The dipole rings of \cite{Emparan:2004wy} are
solutions of ${\cal N}=1$ $d=5$ supergravity coupled to two vector
multiplets, which is a theory with $U(1)^3$ gauge group that can be
obtained by consistent truncation of the maximal $d=5$ supergravity
theory. They are characterized by their mass $M$, angular momentum $J_1$
around the $S^1$ of the horizon, and three dipoles $q_i$. The angular
momentum $J_2$ on the $S^2$ of the horizon vanishes. They have the same
$\mathbb{R} \times U(1)^2$ symmetry as vacuum black rings. These solutions are
seed solutions for the construction of solutions of maximal $d=5$
supergravity with 27 independent dipoles obtained by acting with
$USp(8)$ as described above. One would expect the existence of more
general dipole ring solutions with two independent non-zero angular
momenta but these are yet to be discovered. 

\subsubsection{Charged black rings, supersymmetric black rings} 

Black rings can also carry conserved electric charges with respect to a
2-form field strength (the first regular example was found in
\cite{elvang}, which can be regarded as having two charges and one
dipole in a $U(1)^3$ supergravity theory, see also
\cite{Elvang:2003mj}). Hence there is the possibility that they can
saturate a BPS inequality. The first example of a supersymmetric black
ring solution was obtained for minimal $d=5$ supergravity in \cite{eemr}
using a canonical form for supersymmetric solutions of this theory
\cite{gghpr}. This was then generalized to a supersymmetric black ring
solution of the $U(1)^3$ supergravity in \cite{BW,EEMR2,JGJG2}. The
latter solution has 7 independent parameters, which can be taken to be
the 3 charges, 3 dipoles and $J_1$. The mass is fixed by the BPS
relation and $J_2$ is determined by the charges and dipoles. See
\cite{Emparan:2006mm} for more detailed discussion of these solutions. 

The most general stationary black ring solution of the $U(1)^3$
supergravity theory is expected to have 9 parameters since one would
expect the 3 charges, 3 dipoles, 2 angular momenta and the mass to be
independent. This solution has not yet been constructed. Note that the
general non-BPS solution should have 2 more parameters than the general
BPS solution, just as for topologically spherical rotating black holes.
The most general known non-extremal solution \cite{eef} has 7
parameters, and was obtained by applying solution generating
transformations to the dipole ring solutions of \cite{Emparan:2004wy}. This
solution does not have a regular BPS limit, and there is no limit in
which it reduces to a vacuum black ring with two angular momenta.

It has been argued that a 9 parameter black ring solution could {\it
not} be a seed solution for the most general black ring solution of
maximal $d=5$ supergravity \cite{larsen}. By acting with $USp(8)$ one
can construct a solution with 27 independent charges from a seed with 3
independent charges {\it or} one can construct a solution with 27
independent dipoles from a seed with 3 independent dipoles. However, one
cannot do both at once. If one wants to construct a solution with 27
independent charges and 27 independent dipoles from a seed solution with
3 independent charges then this seed must have 15 independent dipoles,
and hence (including the mass and angular momenta) 21 parameters in
total. The seed solution for the most general
supersymmetric black ring in maximal $d=5$ supergravity is expected to
have 19 parameters \cite{larsen}.

\subsection{Solution-generating techniques}

Ref.~\cite{Bouchareb:2007ax} develops solution generating techniques in
minimal $d=5$ supergravity, based on U-duality properties of the latter.
By applying one such transformation to the neutral doubly-spinning black
ring of section~\ref{subsubsec:twospinring} they obtain a new charged
ring solution of five-dimensional supergravity. However, this solution
suffers from the same problem of Dirac-Misner singularities that
\cite{Elvang:2003mj} described when a neutral single-spin ring seed is
used. It appears that the problem could be solved, like in \cite{eef},
by including an additional parameter that is then tuned to cancel the
pathologies. It also seems possible that, like in \cite{Elvang:2003mj},
the neutral doubly-spinning black ring is a good seed for black rings
with two charges, one dipole, and two independent angular momenta in
the $U(1)^3$ supergravity theory.

Besides the solution-generating techniques based on string theory and
supergravity (sequences of boosts and dualities) there have been a
number of analyses of the Einstein-Maxwell(-dilaton) equations leading
to other techniques to generate stationary solutions.
Ref.~\cite{Ida:2003wv} studies general properties of the
Einstein-Maxwell equations in $d$-dimensions with $d-3$ commuting
Killing vector fields. Ref.~\cite{Teo:2003ug} shows how four-dimensional
vacuum stationary, axisymmetric solutions can be used to obtain static,
axisymmetric solutions of five-dimensional dilaton gravity coupled to a
two-form gauge field. Also within the class of five-dimensional
stationary solutions of the Einstein-Maxwell(-dilaton) equations with
two rotational symmetries, solution-generating techniques have been
developed in \cite{Kunduri:2004da} and
\cite{Yazadjiev:2005aw,Yazadjiev:2005pf}. Note that many of these papers
do not take account of the Chern-Simons term present in $d=5$
supergravity. This term is relevant when both electric and magnetic
components of the gauge field are present. 
Thus, it can be ignored for electrically charged static solutions, which
do not give rise to magnetic dipoles. It does not play a role, either,
for the dipole rings of \cite{Emparan:2004wy}, which has allowed a
systematic derivation of these solutions
\cite{Yazadjiev:2006hw,Yazadjiev:2006ew}.

\subsection{Multi-black hole solutions}

The inclusion of electric charge makes it considerably easier to
construct multi-black hole solutions than in the vacuum case discussed
above. In $d=4$, there exist well-known static solutions describing
multiple extremal Reissner-Nordstrom black holes held in equilibrium by
a cancellation of electric repulsion and gravitational attraction
\cite{hartlehawking}. Similar static solutions can be constructed in
$d>4$ \cite{Myers:1986rx}. However, although the $d=4$ solutions have smooth horizons
\cite{hartlehawking}, the $d>4$ solutions have horizons of low
differentiability \cite{welch,candlish}. 

In $d=4,5$, multi-black hole solutions can be supersymmetric.
Supersymmetry makes it easy to write down stationary solutions
corresponding to multiple BMPV black holes \cite{gmt}. However, the
regularity of these solutions has not been investigated. Presumably they
are no smoother than the static solutions just mentioned. Although
electromagnetic and gravitational forces cancel, one might expect
spin-spin interactions to play a role for these solutions, perhaps
leading to even lower smoothness unless the spins are aligned. Note that
in general, superposition of these black holes breaks all symmetries of
a single centre BMPV solution except for time translation invariance. 

Supersymmetric solutions describing stationary superpositions of
multiple concentric black rings have also been constructed
\cite{JGJG,JGJG2}. The rings have a common centre, and can either lie in
the same plane, or in orthogonal planes. The superposition does not
break any symmetries. This may be the reason that these solutions have
smooth horizons. 

Turning to non-extremal solutions, one would certainly expect
generalizations of the solutions of section \ref{subsec:multibhs} with
non-trivial gauge fields. A solution describing a Myers-Perry black hole
with a concentric dipole ring was presented in \cite{yazsaturn}. 


\newpage


\section{General results and open problems}
\label{section:general}


\subsection{Introduction}

In four dimensions, the black hole uniqueness theorem states that there
is at most one stationary, asymptotically flat, vacuum black hole
solution with given mass and angular momentum: the Kerr black hole. The
co-existence of Myers-Perry black holes and black rings shows explicitly
that black hole uniqueness is violated in five dimensions. By now there
is strong evidence that this is even more dramatically true in more than
five dimensions. However, even if higher-dimensional black holes are not
uniquely characterized by their conserved charges, we can still hope to
classify them. A major goal of research in higher-dimensional general
relativity is to solve the {\it classification problem}: determine all
stationary, asymptotically flat black hole solutions of the
higher-dimensional vacuum Einstein equation (or Einstein equation
coupled to appropriate matter). We are still a long way from this goal
but partial progress has been made, as we shall review below.


\subsection{Black hole topology}

Logically, the first step in the proof of the $d=4$ black hole
uniqueness theorem is Hawking's black hole topology theorem
\cite{hawkingellis}, which states that (a spatial cross-section of)
the event horizon must be topologically $S^2$. The strongest version
of this theorem makes use of {\it topological censorship}. Loosely,
this is the statement that, in an asymptotically flat and globally
hyperbolic spacetime obeying the null energy condition, every causal
curve beginning and ending at infinity can be homotopically deformed
to infinity. This can be used to prove the topology theorem for
stationary \cite{chruscielwald}, and even non-stationary
\cite{jacobson,galloway95} black holes.

The existence of black rings demonstrates that topologically
non-spherical horizons are possible for $d>4$. But there are still
restrictions on the topology of the event horizon. It can been shown
\cite{gallowayschoen,galloway06} that a stationary, aymptotically
flat, black hole spacetime obeying the dominant energy condition must
have a horizon that is ``positive Yamabe", i.e. it must admit a metric
of positive Ricci scalar. This restricts the allowed topologies. After
$d=4$, the strongest restriction is for $d=5$, in which case the
topology must be a connected sum of spherical spaces (3-spheres with
identifications) and $S^1 \times S^2$'s.

Another topological restriction arises from cobordism theory. Consider
surrounding the black hole with a large sphere. Let $\Sigma$ denote a
spacelike hypersurface that runs from this sphere down to the event
horizon, which it intersects in some compact surface $H$. Then
$\Sigma$ is a cobordism between $H$ and a sphere. The existence of
such a cobordism imposes topological restrictions on $H$. These have
been discussed in \cite{oz}. The results are much less restrictive
than those just mentioned.

It may be that horizon topology is not very useful for classifying black
holes in $d\geq 6$. As we discussed in the introduction, it is the
combination of extended horizons with the novel possibilities for
rotation that give higher-dimensional black holes much of their
richness. For a black ring with horizon topology $S^1\times S^2$ the two
factors differ in that the $S^1$, being contractible, needs rotation to
be stabilized, whereas the $S^2$ is already a minimal surface. In $d\geq
6$ we can envisage even more complicated situations arising from the
bending into different surfaces of the worldvolume directions of a
variety of black $p$-branes. Not only horizon topology, but also, and
more importantly, extrinsic geometry and dynamical considerations seem
to be relevant to the existence of these black holes.


\subsection{Uniqueness of static black holes}

In $d=4$ dimensions, the Schwarschild solution is the unique static,
asymptotically flat, vacuum black hole solution. The strongest version
of this theorem allows for a possibly disconnected event horizon, and
the proof uses the positive energy theorem \cite{bunting}. This proof
can be extended to $d>4$ dimensions to establish uniqueness of the
$d>4$ dimensional Schwarzschild solution amongst static vacuum
solutions \cite{gibbonsshiromizu1}. The method can also be generalized
to prove a uniqueness theorem for static, asymptotically flat, black
holes solutions of $d>4$ dimensional Einstein-Maxwell-dilaton theory:
such black holes are uniquely characterized by their mass and charge
and are described by generalized Reissner-Nordstrom solutions
\cite{gibbonsshiromizu2}.

These theorems assume that there are no degenerate components of the
horizon. This assumption can be eliminated for $d=4$ vacuum gravity
\cite{chrusciel,crt}. In Einstein-Maxwell theory, one can show that the
only solutions with degenerate horizons are the Majumdar-Papapetrou
multi-Reissner-Nordstrom solutions \cite{ct}. These results have been
generalized to $d>4$ Einstein-Maxwell theory \cite{rogatko}. 

In conclusion, the classification problem for {\it static} black holes
has been solved, at least for the class of theories mentioned. 

It must be noted, though, that the assumption of staticity is stronger
than requiring vanishing total angular momentum. The existence of black
Saturns (sec.~\ref{subsec:multibhs}) shows that there exists an infinite
number of solutions (with disconnected event horizons) characterized by a
given mass and vanishing angular momentum.

\subsection{Stationary black holes}

In $d=4$, the uniqueness theorem stationary black holes relies on
Hawking's result that a stationary black hole must be axisymmetric
\cite{hawkingellis}. This result has been generalized to higher
dimensions \cite{hiw}. More precisely, it can be shown that a
stationary, non-extremal, asymptotically flat, rotating, black hole
solution of $d>4$ dimensional Einstein-Maxwell theory must admit a
spacelike Killing vector field that generates rotations. Here,
``rotating" means that the Killing field that generates time
translations is not null on the event horizon, i.e., the angular
velocity is non-vanishing. However, it can be shown that a
non-rotating black hole must be static for Einstein-Maxwell theory in
$d=4$ \cite{sudarsky} and $d>4$ \cite{rogatkostaticity} so this
assumption can be eliminated. The result of \cite{hiw} also applies in
the presence of a cosmological constant (e.g. asymptotically anti-de
Sitter black holes).

This theorem guarantees the existence of a single rotational symmetry.
However, the {\it known} higher-dimensional black hole solutions (i.e.
Myers-Perry and black rings) admit multiple rotational symmetries: in
$d$ dimensions there are $\lfloor(d-1)/2\rfloor$ rotational Killing
fields. Is there some underlying reason that this must be true, or is
this simply a reflection of the fact that we are only able to find
solutions with a lot of symmetry? 

If there do exist solutions with less symmetry then they must be
non-static (because of the uniqueness theorem for $d>4$ static black
holes). One could look for evidence that such solutions exist by
considering perturbations of known solutions \cite{hsr}. For example,
the existence of non-uniform black strings was first conjectured on the
evidence that uniform strings exhibit a static zero-mode that breaks
translational symmetry. If a Myers-Perry black hole had a stationary
zero-mode that breaks some of its rotational symmetry then that would be
evidence in favour of the existence of a new branch of solutions with
less symmetry \cite{hsr}. Alternatively, bifurcations could occur
without breaking any rotational symmetry. As we have discussed,
refs.~\cite{Emparan:2007wm,Emparan:2003sy} have conjectured that such
bifurcations will happen in $d\geq 6$, see
figure~\ref{figure:hidphases}. Either case could lead to non-uniqueness
within solutions of spherical topology.

The formalism of \cite{kodama:03} allows one to show that the only
regular stationary perturbations of a Schwarzschild black hole lead into
the MP family of black holes \cite{Kodama:2004kz}. Thus, the MP solutions are the only stationary black holes that have a regular static limit.

The issue of how much symmetry a general stationary black hole must
possess is probably the main impediment to progress with the
classification problem. At present, the only uniqueness results for
stationary higher-dimensional black holes {\it assume} the existence
of multiple rotational symmetries. These results are for $d=5$: if one assumes
the existence of two rotational symmetries then it can be shown that
the Myers-Perry solution is the unique stationary, non-extremal,
asymptotically flat, vacuum black hole solution of spherical topology
\cite{idamorisawa}. More generally, it has been shown that stationary,
non-extremal, asymptotically flat, vacuum black hole solutions with
two rotational symmetries are uniquely characterized by their mass,
angular momenta, and {\it rod structure} (see section~\ref{para:rodstructure})
\cite{hollandsyazad}. The remaining step in a full classification of
all $d=5$ stationary vacuum black holes with two rotational symmetries
is to prove that the only rod structures giving rise to regular
black hole solutions are those associated with the known (Myers-Perry
and black ring) solutions.

The situation in $d\geq 6$ is much further away from a complete
classification, even for the class of solutions with the maximal number
of rotational symmetries. For instance, the tools to classify the (yet
to be found) infinite number of families of solutions with `pinched
horizons' proposed in refs.~\cite{Emparan:2007wm,Emparan:2003sy} are
still to be developed. 

It is clear that the notion of black hole uniqueness that holds in four
dimensions, namely that conserved charges serve to fully specify a black
hole, does not admit any simple extension to higher dimensions. This
leaves us with two open questions: (a) what is the simplest and most
convenient set of parameters that fully specify a black hole; (b) how
many black hole solutions with given conserved charges are relevant in a
given physical situation.

Concerning the first question, we note that while the rod structure may
provide the additional data to determine five-dimensional vacuum black
holes, one may still desire a characterization in terms of physical
parameters. In other words, since the dimensionless angular momenta
$j_a$ are insufficient to specify the solutions, an adequate physical
parametrization of the phase space of higher-dimensional black holes is
still missing. Refs.~\cite{Shiromizu:2004jt,Tomizawa:2004fw} have
studied whether higher multipole moments may serve this purpose, but the
results appear to be inconclusive. 

The second question is more vague, as it hinges not only on the answer
to the previous question, but also on the specification of the problem
one is studying. It has been speculated that the conserved charges may
still suffice to select a unique classical stationary configuration, if
supplied with additional conditions, like the specification of horizon
topology or requiring classical stability \cite{Kol:2002dr}. In five
dimensions we already know that horizon topology alone is not enough,
since there are both thin and fat black rings with the same $j$. We have
seen that, very likely, an even larger non-uniqueness occurs in all
$d\geq 6$. Classical dynamical stability to linearized perturbations,
which is not at all an issue in the four-dimensional classification, is
presumably a much more restrictive condition, but even in five
dimensions it is unclear whether it always picks out not more than one
solution for given $(j_1,j_2)$. One must also bear in mind that the
classical instability of a black hole does not {\em per se} rule it out
as a physically relevant solution: the time-scale of the instability
must be compared to the time scale of the situation at hand. Some
classical instabilities (e.g. ergoregion instabilities
\cite{Cardoso:2004nk}) may be very slow. 

It seems possible, although so far we are nowhere near having any
compelling argument, that the requirement of classical linearized
stability, plus possibly horizon topology, suffices to fully specify a
unique vacuum black hole solution with given conserved charges. However,
in the presence of gauge fields this seems less likely, since dipoles
not only introduce larger non-uniqueness but also tend to enhance the
classical stability of the solutions.


\subsection{Supersymmetric black holes}

As mentioned earlier, the study of BPS black holes is 
of special importance in string theory and it is natural to ask whether one
can classify BPS black holes. Asymptotically flat BPS black hole
solutions are known only for $d=4,5$.

In addition to rendering microscopic computations tractable, the
additional ingredient of supersymmetry makes the classification of
black holes easier. A supersymmetric solution admits a globally
defined super-covariantly constant spinor (see
e.g. \cite{gibbonshull}). This is such a strong constraint that it is
often possible to determine the general solution with this
property. This was first done for minimal $d=4$ $N=2$ supergravity,
whose bosonic sector is Einstein-Maxwell theory. It can be shown
\cite{tod} that any supersymmetric solution of this theory must be
either a pp-wave or a member of the well-known Israel-Wilson-Perjes
family of solutions (see e.g. \cite{exactsolutionsbook}). The only
black hole solutions in the IWP family are believed to be the
multi-Reissner-Nordstrom solutions \cite{hartlehawking}, and this can
be proved subject to one assumption\epubtkFootnote{Supersymmetric solutions
admit a globally defined Killing vector field that is timelike or
null. The assumption is that it is non-null everywhere outside the
horizon.} \cite{crt2}. Hence this is a uniqueness theorem for
supersymmetric black hole solutions of minimal $d=4$ $N=2$
supergravity.

This success has been partially extended to minimal $d=5$
supergravity. It can be shown that any supersymmetric black hole must
have near-horizon geometry locally isometric to one of (i) the
near-horizon geometry of the BMPV black hole; (ii) $AdS_3 \times S^2$;
or (iii) flat space \cite{hsr}. Case (iii) would give a black hole
with $T^3$ horizon, which seems unlikely in view of the black hole
topology theorem discussed above (although this does not cover
supersymmetric black holes since they are necessarily extremal). An
explicit form for supersymmetric solutions of this theory is known
\cite{gghpr} and this can be exploited, together with an additional
assumption\epubtkFootnote{The same assumption as for the $d=4$, $N=2$ case
just discussed.} to show that the only black hole of type (i) is the
BMPV black hole itself. The supersymmetric black ring of \cite{eemr}
belongs to class (ii). The remaining step required to complete the
classification would be to prove that this is the only solution in
this class. These results can be extended to minimal $d=5$
supergravity coupled to vector multiplets \cite{gutowski}.

These results show that much more is known about $d=5$ supersymmetric
black holes than about more general stationary $d=5$ black holes. Note
that no assumptions about the topology of the horizon, or the number
of rotational symmetries are required to obtain these results: they
are outputs, not inputs. One might interpret this as weak evidence
that, for general $d=5$ black holes, topologies other than $S^3$ and
$S^1 \times S^2$ cannot occur and that the assumption of two
rotational symmetries is reasonable. However, one should be cautious
in generalizing from BPS to non-BPS black holes since it is known that
many properties of the former (e.g. stability) do not always
generalize to the latter.


\subsection{Algebraic classification}

In $d=4$ dimensions, space-times can be classified according to the
algebraic type of the Weyl tensor. Associated with the Weyl tensor are
four ``principal null vectors" \cite{wald}. In general these are
distinct, but two or more of them coincide in an algebraically special
spacetime. For example, the Kerr-Newman space-time is type D, which
means that it has two pairs of identical principal null vectors.

Given that known $d=4$ black hole solutions are algebraically special,
it is natural to investigate whether the same is true in $d>4$
dimensions. Before doing this, it is necessary to classify possible
algebraic types of the Weyl tensor in higher dimensions. This can be
done using a spinorial approach for the special case of $d=5$
\cite{desmet}. The formalism for all dimensions $d \ge 4$ has been
developed in \cite{algclass} (and reviewed in \cite{coley}). It is based
on ``aligned null directions", which generalize the concept of principal
null vectors in $d=4$. A general $d>4$ spacetime admits no aligned null
directions. The Weyl tensor is said to be algebraically special if one
or more aligned null directions exist.

The algebraic types of some higher-dimensional black hole solutions have
been determined. The Myers-Perry black hole belongs to the
higher-dimensional generalization of the $d=4$ type D class
\cite{desmettyped,ringwands,typed}. The black ring is also known to be
algebraically special, although not as special as the Myers-Perry black
hole \cite{ringwands,typed}. Further analysis of the Weyl tensor and
principal null congruences in higher dimensions can be found in
\cite{Pravda:2004ka}.

In $d=4$ dimensions, interesting new solutions (e.g. the spinning
C-metric) were discovered by solving the Einstein equations to determine
all solutions of type D \cite{kinnersley}. Perhaps the same strategy
would be fruitful in higher dimensions. A particular class of
spacetimes, namely Robinson-Trautman, admitting a hypersurface
orthogonal, non-shearing and expanding geodesic null congruence, has
been studied in \cite{Podolsky:2006du}. However, unlike in four
dimensions, this class does not contain the analogue of the C-metric.

\subsection{Laws of black hole mechanics}

The laws of black hole mechanics are generally valid in any number of
dimensions. The only novelty arises in the first law due to new
possibilities afforded by novel black holes. A non-trivial extension is
to include dipoles charges that are independent of the conserved
charges. An explicit calculation in ref.~\cite{Emparan:2004wy} showed
that black rings with a dipole satisfy a form of the first law with an
additional `work' term $\phi dq$, where $q$ is the dipole charge and
$\phi$ its respective potential. The appearance of the dipole here was
surprising, since the most general derivation of the first law seems to
allow only conserved charges into it. However, ref.~\cite{Copsey:2005se}
showed that a new surface term enters due to the
impossibility of globally defining the dipole potential $\phi$ in such a
way that it is simultaneously regular at the rotation
axis and at the horizon. Then one finds
\beq\label{1stdipole}
dM=\frac{\kappa}{8\pi}\mathcal{A}_H+\Omega_H dJ+ \Phi dQ+
\phi dq
\eeq
where $Q$ and $\Phi$ are the conserved charge and its potential (see
also \cite{Astefanesei:2005ad}). 

The next extension is not truly specific to $d>4$, but it refers
to a situation for which there are no known four-dimensional examples:
stationary multi-black hole solutions with non-degenerate horizons. As
we have seen, there are plenty of these in $d\geq 5$. In
this case, the first law can be easily shown to take the form
\cite{Elvang:2007hg}
\beq
dM=\sum_i \left(\frac{\kappa^{(i)}}{8\pi}
d\mathcal{A}_H^{(i)}
+\Omega_j^{(i)} dJ^{(i)}_j+\Phi^{(i)} dQ^{(i)}\right)\,.
\eeq
Here the index $i$ labels the different disconnected components of the
event horizon, and $j$ their independent angular momenta. The Komar
angular momentum $J^{(i)}_j$ and the charge $Q^{(i)}$ are
computed as integrals on a surface that encloses a single component of
the horizon, generated by the vector $\partial_t+\Omega_j^{(i)}
\partial_{\phi_j}$. The potential $\Phi^{(i)}$ is the difference between the
potential at infinity and the potential on the $i$-th component of the
horizon: in general we cannot choose a globally defined gauge potential that
simultaneously
vanishes on all horizon components. Presumably the result can be extended to
include dipoles but the possible subtleties have not been dealt with yet. A
Smarr relation also exists that relates the total ADM mass to the sums
of the different `heat' and `work' terms on each horizon component
\cite{Elvang:2007hg}.

\subsection{Hawking radiation and black hole thermodynamics}

The extension of Hawking's original calculation to most of the black
holes that we have discussed above presents several difficulties, but we
regard this as mostly a technical issue. In our opinion, there is no
physically reasonable objection to the expectation that Hawking
radiation is essentially unmodified in higher dimensions: a black hole
emits radiation that, up to greybody corrections, has a Planckian
spectrum of temperature $T=\kappa/2\pi$ and chemical potentials
$\Omega$, $\Phi$, etc. 

Some of the technical difficulties relate to the problem of wave
propagation in the black hole background: this can only be dealt with
analytically for Myers-Perry black holes, since only in this case have
the variables been separated (and then only for scalars and vectors in
the general rotating case). There is in fact a considerable body of literature
on the problem of radiation from MP black holes, largely motivated by
their possible detection in scenarios with large extra dimensions. As
mentioned in section~\ref{section:organisation}, this is outside the
scope of this paper and we refer to \cite{Cavaglia:2002si,Kanti:2004nr}
for reviews.

There are also peculiarities with wave propagation that depend on the
parity of the number of dimensions \cite{Ooguri:1985nv,Cardoso:2002pa},
but these are unlikely to modify in any essential way the Planckian
spectrum of radiation. This is in fact confirmed by detailed microscopic
derivations of Hawking radiation in five dimensions based on string
theory \cite{Maldacena:1996ix}. Other approaches to Hawking radiation
that do not require to analyze wave propagation have been applied to
black rings \cite{Miyamoto:2007ue}, confirming the expected results. An
early result was the analysis of vacuum polarization in
higher-dimensional black holes in \cite{Frolov:1989rv}. More recently,
ref.~\cite{Nomura:2005mw} claims that the evolution of a
five-dimensional rotating black hole emitting scalar Hawking radiation
leads, for arbitrary initial values of the two rotation parameters $a$
and $b$, to a fixed asymptotic value of $a/M^2 = b/M^2 = const\neq 0$.

The spectrum of radiation from a multi-black hole configuration will
contain several components with parameters $(T^{(i)}, \Omega^{(i)},
\Phi^{(i)},\dots)$, so it will not be really a thermal distribution unless
all the black holes have equal intensive parameters. This is, of course,
the conventional condition for thermal equilibrium.

The Euclidean formulation of black hole thermodynamics remains largely
the same as in four dimensions. For rotating solutions, it is more
convenient not to continue analytically the angular velocities and
instead work with complex sections that have real actions. In fact,
black rings do not admit non-singular real Euclidean sections
\cite{Elvang:2006dd}. Multi-black hole configurations with disconnected
components of the horizon with different surface gravities, angular
velocities, and electric potentials, clearly do not admit regular Euclidean
sections. Still, the Euclidean periodicity of the horizon
generator yields formally the horizon temperature in the usual fashion.

\subsection{Apparent and isolated horizons, and critical phenomena}

A number of other interesting studies of higher-dimensional black holes
have been made. The properties of higher-dimensional apparent horizons
have been analyzed in ref.~\cite{Senovilla:2002ma}, which provides
simple criteria to determine them. The isoperimetric inequalities and
the hoop conjecture, concerning bounds on the sizes of apparent horizons
through the mass they enclose, involve new features in higher
dimensions. For instance, the four-dimensional hoop conjecture posits
that a necessary and sufficient condition for the formation of a black
hole is that a mass $M$ gets compacted into a region whose circumference
in every direction is ${\mathcal C}\leq 4\pi GM$ \cite{thornehoop}. A
generalization of this conjecture to $d>4$ using a hoop of spatial
dimension 1, in the form ${\mathcal C} \leq \# (GM)^{1/(d-3)}$, is
unfeasible: the existence of black objects whose horizons
have arbitrary extent in some directions (e.g., black strings, black
rings, and ultra-spinning black holes) shows that this condition is not
necessary. It seems possible, however, that plausible necessary and
sufficient conditions exist using the area of hoops of spatial
codimension $d-3$ in the form ${\mathcal C}_{d-3} \leq \# GM$, although
black rings may require hoops with non-spherical topology
\cite{Ida:2002hg,Barrabes:2004rk}. There is also some evidence that the
isoperimetric inequalities, which bound the spatial area of the apparent
horizon by the mass that it encloses \cite{Penrose:1973um}, may be
extended in the form ${\mathcal A} \leq \# (GM)^{\frac{d-2}{d-3}}$
\cite{Ida:2002hg,Barrabes:2004rk}. See
\cite{Yoshino:2002br,Yoo:2005nj,Senovilla:2007dw} for further work on
these subjects.

The study of the possible topologies for black hole event horizons may
be helped by the study of possible apparent horizons in initial
data-sets. Ref.~\cite{Schwartz:2007gj} shows that it is possible to
construct {\it time-symmetric} initial data sets for black holes with
apparent horizon topology of the form of a product of spheres. Time
symmetry, however, implies that these apparent horizons cannot
correspond to rotating black holes and it is likely that their evolution
in time leads to collapse into a spherical horizon.

The formalism of isolated
horizons, and the laws of black hole mechanics for them, have been
extended to higher dimensions in
\cite{Lewandowski:2004sh,Korzynski:2004gr}, and then to five-dimensional Einstein-Maxwell
theory with Chern-Simons term \cite{Liko:2007mu}
and anti-deSitter asymptotics \cite{Ashtekar:2006iw}. 

Finally, the
critical phenomena in the collapse of a massless scalar at the onset of
black hole formation, discovered by Choptuik \cite{Choptuik:1992jv},
have been studied in \cite{Sorkin:2005vz}.

\newpage


\section{Solutions with a cosmological constant}
\label{section:cosconst}


\subsection{Motivation}

The motivation for considering higher-dimensional black holes with a
cosmological constant arises from the AdS/CFT correspondence
\cite{adscftreview}. This is an equivalence between string
theory on space-times asymptotic to $AdS_d \times X$, where $X$ is a
compact manifold, and a conformal field theory (CFT) defined on the
Einstein universe $R \times S^{d-2}$, which is the conformal boundary
of $AdS_d$.  The best understood example is the case of type IIB
string theory on space-times asymptotic to $AdS_5 \times S^5$, which
is dual to ${\cal N}=4$ $SU(N)$ super-Yang-Mills theory on $R \times
S^3$. Type IIB string theory can be replaced by IIB supergravity in
the limit of large $N$ and strong 't Hooft coupling in the Yang-Mills
theory.

Most studies of black holes in the AdS/CFT correspondence involve
dimensional reduction on $X$ to obtain a $d$-dimensional gauged
supergravity theory with a negative cosmological constant. For
example, one can reduce type IIB supergravity on $S^5$ to obtain
$d=5$, ${\cal N}=4$ $SO(6)$ gauged supergravity. One then seeks
asymptotically anti-de Sitter black hole solutions of the gauged
supergravity theory. This is certainly easier than trying to find
solutions in ten or eleven dimensions. However, one should bear in
mind that there may exist asymptotically $AdS_d \times X$ black hole
solutions that cannot be dimensionally reduced to $d$ dimensions. Such
solutions would not be discovered using gauged supergravity.

In this section we shall discuss asymptotically $AdS_d$ black hole
solutions of the $d=4,5,6,7$ gauged supergravity theories arising from
reduction of $d=10$ or $d=11$ supergravity on spheres. The emphasis will
be on classical properties of the solutions rather than their
implications for CFT. In AdS, linearized supergravity perturbations can
be classified as normalizable or non-normalizable according to how they
behave near infinity \cite{adscftreview}. By ``asymptotically AdS", we
mean that we are restricting ourselves to considering solutions that
approach a normalizable deformation of global AdS near infinity. A
non-normalizable perturbation would correspond to a deformation of the
CFT, for instance making it non-conformal. Black hole solutions with
such asymptotics have been constructed but space prevents us from
considering them here.

\subsection{Schwarzschild-AdS}

The simplest example of an asymptotically AdS black hole is the
Schwarschild-AdS solution \cite{kottler,witten2}: 
\be ds^2 = - U(r) dt^2 + U(r)^{-1} dr^2 + r^2 d\Omega_{d-2}^2\,, \qquad U(r) = 
1 - \frac{\mu}{r^{d-3}} + \frac{r^2}{\ell^2} 
\ee 
where $\mu$ is proportional to the mass, 
and $\ell$ is the radius of curvature of the AdS ground
state\epubtkFootnote{The ``topological black holes''
with $U(r) = 
k - \frac{\mu}{r^{d-3}} + \frac{r^2}{\ell^2}$, $k=0,-1$ and toroidal or
hyperbolic horizons \cite{Birmingham:1998nr}, are excluded from our review by
their asymptotics.}. The solution has a regular horizon
for any $\mu>0$.
Definitions of mass and angular momentum for asymptotically AdS
spacetimes have been given in $d=4$ \cite{ashtekarmagnon} and $d \ge 4$
\cite{ashtekardas}. The mass of Schwarzschild-AdS relative to the AdS
ground state is \cite{witten2}
\be
 M = \frac{(d-2)\Omega_{d-2} \mu}{16 \pi G}.
\ee
For $d=4$, the stability of Schwarzschild-AdS against linearized gravitational
perturbations has been proved in \cite{kodama:03a}. For $d>4$, spherical
symmetry enables one to decompose linearized gravitational perturbations
into scalar/vector/tensor types. The equations governing each type can
be reduced to ODEs of Schr\"odinger form, and stability of vector and
tensor perturbations established \cite{kodama:03}. Stability with
respect to scalar gravitational perturbations has not yet been
established.

It is expected that the Schwarzschild-AdS black hole is the unique
static, asymptotically AdS, black hole solution of vacuum gravity with a
negative cosmological constant but this has not been proven.

The thermodynamics of Schwarzschild-AdS were discussed by
Hawking and Page for $d=4$ \cite{hawkingpage} and Witten for $d>4$
\cite{witten2}. Let $r_+$ denote the horizon radius of the
solution. For a small black hole, $r_+ \ll \ell$, the thermodynamic properties
are qualitatively similar to those of an asymptotically flat
Schwarzschild black hole, i.e., the temperature decreases with
increasing $r_+$ so the heat capacity of the hole is negative (as
$r_+$ is a monotonic function of $\mu$).  However, there is an
intermediate value of $r_+ \sim \ell$ at which the temperature reaches
a global minimum $T_{\rm{min}}$ and then becomes an {\it increasing}
function of $r_+$. Hence the heat capacity of large black holes is
positive. This implies that the black hole can reach a stable
equilibrium with its own radiation (which is confined near the hole by
the gravitational potential $\sim r^2/\ell^2$ at large $r$).  Note
that for $T>T_{\rm{min}}$ there are {\it two} black hole solutions
with the same temperature: a large one with positive specific heat and
a small one with negative specific heat.

These properties lead to an interesting phase structure for gravity in
AdS \cite{hawkingpage}. At low temperature, $T<T_{\rm{min}}$, there is
no black hole solution and the preferred phase is thermal radiation in
AdS. At $T \sim T_{\rm{min}}$, black holes exists but
have greater free energy than thermal radiation. However, there is a
critical temperature $T_{\rm{HP}} > T_{\rm{min}}$ beyond which the
large black hole has lower free energy than thermal radiation and the
small black hole. The interpretation is that the canonical ensemble
for gravity in AdS exhibits a (first order) phase transition at
$T=T_{\rm{min}}$.

In the AdS/CFT context, this Hawking-Page phase transition is
interpreted as the gravitational description of a thermal phase
transition of the (strongly coupled) CFT on the Einstein Universe
\cite{witten,witten2}.

When oxidized, to ten or eleven dimensions, the radius $r_+$ of a small
Schwarzschild-AdS black hole will be much less than the radius of
curvature ($\sim \ell$) of the internal space $X$. This suggests that
the black hole will suffer from a classical Gregory-Laflamme type
instability. The probable endpoint of the instability would be a small
black hole localized on $X$, and therefore not admitting a description
in gauged supergravity. Since the radius of curvature of $X$ is typically
$\ell$ and the black hole is much smaller than $\ell$, the geometry near
the hole should be well-approximated by the ten or eleven dimensional
Schwarzschild solution (see e.g. \cite{horowitzhubeny}). However, an
exact solution of this form is not known.


\subsection{Stationary vacuum solutions}

If we consider pure gravity with a negative cosmological constant then
the most general known family of asymptotically AdS black hole
solutions is the generalization of the Kerr-Myers-Perry solutions to
include a cosmological constant. It seems likely that black rings should
exist in asymptotically AdS spacetimes but no exact solutions are
known.\epubtkFootnote{Note that topological censorship can be used to exclude
the existence of topologically non-spherical black holes in $AdS_4$
\cite{gallowayads}.}

The $d=4$ Kerr-AdS solution was
constructed long ago \cite{carter}. It can be parametrized by its mass
$M$ and angular momentum $J$, which have been calculated (using the
definitions of \cite{ashtekarmagnon}) in \cite{gpp}. The region of the
$(M,J)$ plane covered by these black holes is shown in figure
\ref{fig:ads4bh}. Note that, in AdS, angular momentum is a central
charge \cite{gibbonshullwarner}. Hence regular vacuum solutions exhibit
a non-trivial lower bound on their mass: $M \ge |J|/\ell$. The Kerr-AdS
solution never saturates this bound.
\epubtkImage{}{%
\begin{figure}[h]
  \def\epsfsize#1#2{.6#1}
  \centerline{\epsfbox{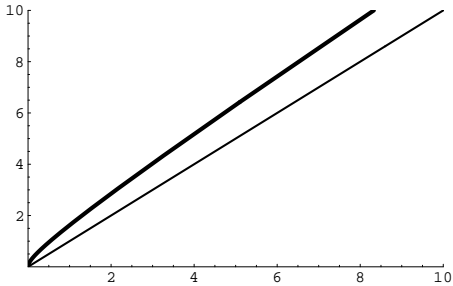}}
\caption{\it $GM/\ell$ against $G|J|/\ell^2$ for $d=4$ Kerr-AdS black
holes. The thick curve corresponds to extremal black holes. Black hole
solutions lie on, or above, this curve (which was determined using
results in \cite{caldarelli99}). The thin line is the BPS bound
$M=|J|/\ell$.}
  \label{fig:ads4bh}
\end{figure}}

The Myers-Perry-AdS solution was
obtained in \cite{hhtr} for $d=5$, and for $d>5$ with rotation in a
single plane. The general $d>5$ solution was obtained in \cite{glpp}. 
They have horizons of spherical topology. There is some
confusion in the literature concerning the conserved charges carried
by these solutions. A careful discussion can be found in \cite{gpp}. The
solutions are uniquely specified by their mass and angular momenta. For
$d=5$, the region of $(M,J_1,J_2)$-space covered by the Myers-Perry-AdS
solution is shown in figure \ref{fig:ads5bh}. 
\epubtkImage{}{%
\begin{figure}[h]
  \def\epsfsize#1#2{.8#1}
  \centerline{\epsfbox{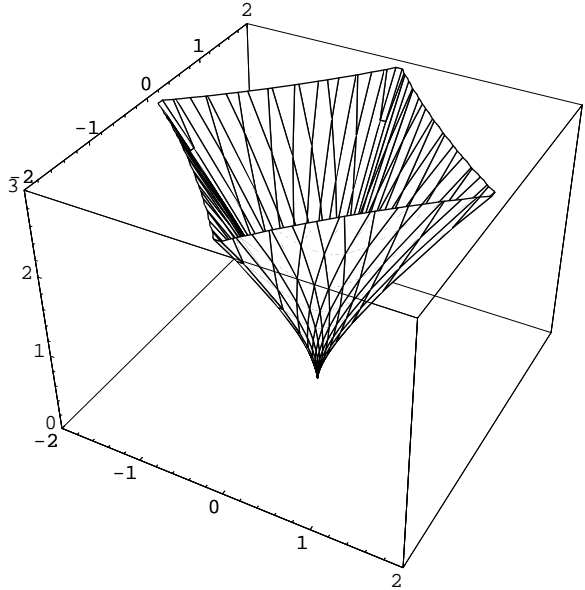}}
\caption{\it $GM/\ell^2$ (vertical) against $G|J_i|/\ell^3$ for $d=5$
Myers-Perry-AdS black holes. Non-extremal black holes fill the region
above the surface. The surface corresponds to extremal black holes,
except when one of the angular momenta vanishes (in which case there is
not a regular horizon, just as in the asymptotically flat case).
This ``extremal surface" lies inside the square based pyramid (with
vertex at the origin) defined by the BPS relation $M=|J_1|/\ell +
|J_2|/\ell$, so none of the black holes are BPS.
}
  \label{fig:ads5bh}
\end{figure}}

Kerr-Myers-Perry-AdS solutions have the same symmetries as their
asymptotically flat cousins, and exhibit similar enhancement of symmetry
in special cases. The integrability of the geodesic equation and
separability of the Klein-Gordon equation also extends to this case
\cite{kubiznakpage,kubiznakfrolovsep}. 

These solutions reduce to the Schwarzschild-AdS solution in the limit of
zero angular momentum. It has been shown that the only regular
stationary perturbations of the Schwarzschild-AdS solution are those
that correspond to taking infinitesimal angular momenta in these
rotating solutions \cite{Kodama:2004kz}. Hence if other stationary
vacuum black hole solutions exist (e.g. black rings) then they are not
continuously connected to the Schwarschild-AdS solution.

These solutions exhibit an important qualitative difference from their
asymptotically flat cousins. Consider the Killing field tangent to the
null geodesic generators of the horizon:
\be
 V = \frac{\partial}{\partial t} + \Omega_i \frac{\partial}{\partial
   \phi_i}.
\ee
In asymptotically flat space-time, this Killing field is spacelike far
from the black hole, which implies that it is impossible for matter to
co-rotate rigidly with the hole (i.e. to move on orbits of
$V$). However, in AdS, if $\Omega_i \ell \le 1$ then $V$  is timelike
everywhere outside the horizon. This implies that rigid co-rotation is
possible: the Killing field $V$ defines a co-rotating reference
frame. Consequently there exists a Hartle-Hawking state describing
thermal equilibrium of the black hole with co-rotating thermal radiation \cite{hhtr}.

The dual CFT interpretation is of CFT matter in thermal equilibrium
rotating around the Einstein universe \cite{hhtr}. There is an
interesting phase structure, generalizing that found for
Schwarzschild-AdS \cite{hhtr,berman,caldarelli99,hawkingreall}. For
sufficiently large black holes, one can study the dual CFT using a fluid
mechanics approximation, which gives quantitative agreement with black
hole thermodynamics \cite{minwallafluids}.

What happens if $\Omega_i \ell > 1$? Such black holes are believed to
be classically unstable. It was observed in \cite{hawkingreall} that
rotating black holes in AdS may suffer from a superradiant
instability, in which energy and angular momentum are extracted from
the black hole by superradiant modes. However, it was proved that this
cannot occur if $\Omega_i \ell \le 1$. But if $\Omega_i \ell > 1$
then an instability may be present. This makes sense from a dual CFT
perspective: configurations with $\Omega_i \ell > 1$ would correspond
to CFT matter rotating faster than light in the Einstein universe \cite{hhtr}. 
The existence of an instability was first demonstrated for small
$d=4$ Kerr-AdS black holes in \cite{cardosodias}. A general analysis
of odd dimensional black holes with equal angular momenta reveals that
the threshold of instability is at $\Omega_i \ell = 1$
\cite{klrperturb}, i.e., precisely where the stability argument of
\cite{hawkingreall} fails. The endpoint of this classical bulk
instability is not known. 

In $d=4$, figure \ref{fig:ads4bh} reveals (using $\Omega = dM/dJ$) that
all extremal Kerr-AdS black holes have $\Omega \ell >1$ and are
therefore expected to be unstable. We have checked that $d=5$ extremal
Myers-Perry-AdS black holes also have $\Omega_i \ell > 1$ and so they
too should be classically unstable. However, the instability should be
very slow when the black hole size is much smaller than the AdS radius
$\ell$, and one expects it to disappear as $\ell\to\infty$: it takes an
increasingly long time for the superradiant modes to bounce back off
the AdS boundary.

Finally, we should mention a subtletly concerning the use of the term
``stationary" in asymptotically AdS spacetimes \cite{klrperturb}.
Consider the $AdS_5$ metric
\be
ds^2 = -\left( 1 + \frac{r^2}{\ell^2} \right) dt^2 + \left( 1 +
\frac{r^2}{\ell^2} \right)^{-1} dr^2 + r^2 \left( d\theta^2 + \sin^2
\theta \, d\phi^2 + \cos^2 \theta \, d\psi^2 \right).
\ee
This admits several types of globally timelike Killing fields. For
example, there is the ``usual" generator of time translations
$k=\partial/\partial t$, which has unbounded norm, but there is also the
``rotating" Killing field $V = \partial/\partial t + \ell^{-1}
\partial/\partial \phi + \ell^{-1} \partial/ \partial \psi$, which has
constant norm. On the conformal boundary, $k$ is timelike and $V$ is
null. Hence, from a boundary perspective, particles following orbits of
$V$ are rotating at the speed of light. These two different types of
timelike Killing vector field allow one to define two distinct notions
of stationarity for asymptotically AdS spacetimes. So far, all known
black hole solutions are stationary with respect to both definitions
because they admit global Killing fields analogous to $\partial/\partial
\phi$, $\partial/\partial \psi$. However, it is conceivable that there
exist AdS black holes (with less symmetry than known solutions) that are
stationary only with respect to the second definition, i.e., they admit
a Killing field that behaves asymptotically like $V$ but not one
behaving asymptotically like $k$. From a boundary CFT perspective, such
black holes would rotate at the speed of light.


\subsection{Gauged supergravity theories}

In order to discuss charged anti-de Sitter black holes we will need to
specify which gauged supergravity theories we are interested in. The
best-understood examples arise from dimensional reduction of $d=10$ or
$d=11$ dimensional supergravity theories on spheres to give theories
with maximal supersymmetry and non-abelian gauge groups. However, when
it comes to looking for explicit black hole solutions, most work has
worked with consistent truncations of these theories, with reduced
supersymmetry, in which the non-abelian gauge group is replaced by its
maximal abelian subgroup. Indeed, there is no known black hole solution
with a non-trivial non-abelian gauge field obeying normalizable boundary
conditions.

There is a consistent dimensional reduction of $d=11$ supergravity on
$S^7$ to give $d=4$, ${\cal N}=8$, $SO(8)$ gauged supergravity
\cite{dewitnicolai}. This non-abelian theory can consistently truncated
to give $d=4$, ${\cal N}=2$, $U(1)^4$ gauged supergravity, whose bosonic
sector is Einstein gravity coupled to four Maxwell fields and three
complex scalars \cite{elevenauthors}. The scalar potential is negative
at its global maximum. The $AdS_4$ ground state of the theory has the
scalars taking constant values at this maximum. One can truncate this
theory further by taking the scalars to sit at the top of their
potential, and setting the Maxwell fields equal to each other. This
gives minimal $d=4$, ${\cal N}=2$ gauged supergravity, whose bosonic
sector is Einstein-Maxwell theory with a cosmological constant. The
embedding of minimal $d=4$, ${\cal N}=2$ gauged supergravity theories
into $d=11$ supergravity can be given explicitly \cite{cejm1}, and is
much simpler than the embedding of the non-abelian ${\cal N}=8$ theory.

The $d=11$ supergravity theory can also be dimensionally reduced on
$S^4$ to give $d=7$, ${\cal N}=2$, $SO(4)$ gauged supergravity
\cite{vann}.

The $d=10$ massive IIA supergravity can be dimensionally reduced on
$S^4$ to give $d=6$ ${\cal N}=2$ $SU(2)$ gauged supergravity
\cite{poperomans}. This theory has half-maximal supersymmetry.

It is believed that the $d=10$ type IIB supergravity theory can be
consistently reduced on $S^5$ to give $d=5$, ${\cal N}=4$, $SO(6)$
gauged supergravity, although this has been established only for a
subsector of the full theory \cite{consistent}. This theory can be
truncated further to give $d=5$, ${\cal N}=1$, $U(1)^3$ gauged
supergravity with $3$ vectors and $2$ scalars. Again, setting the
scalars to constants and taking the vectors equal gives the minimal
$d=5$ gauged supergravity, whose bosonic sector is Einstein-Maxwell
theory with a negative cosmological constant and a Chern-Simons
coupling. The explicit embeddings of these abelian theories into $d=10$
type IIB supergravity are known \cite{cejm1,elevenauthors}.

It is sometimes possible to obtain a given lower dimensional supergravity
theory from several different compactifications of a higher dimensional
theory. For example, minimal $d=5$ gauged supergravity can be
obtained by compactifying type IIB supergravity on any Sasaki-Einstein
space $Y^{p,q}$ \cite{buchelliu}. More generally, if there is a
supersymmetric solution of type IIB supergravity that is a warped
product of $AdS_5$ with some compact manifold $X_5$ then type IIB
supergravity can be dimensionally reduced on $X_5$ to give minimal
$d=5$ gauged supergravity \cite{gauntlettvarela2}. An analagous
statement holds for compactifications of $d=11$ supergravity to give
minimal ${\cal N}=2$, $d=4$ gauged supergravity or minimal $d=5$
gauged supergravity \cite{gauntlettvarela1,gauntlettvarela2}. 


\subsection{Static charged solutions}

The $d=4$ Reissner-Nordstrom-AdS black hole is a solution of minimal
${\cal N}=2$ gauged supergravity. It is parametrized by its mass $M$
and electric and magnetic charges $Q$,$P$. This solution is stable
against linearized perturbations within this (Einstein-Maxwell) theory
\cite{kodamacharged}. Compared with its asymptotically flat counterpart,
perhaps the most surprising feature
of this solution is that it never saturates a BPS bound. If the mass
of the black hole is lowered, it will eventually become extremal, but
the extremal solution is not BPS. If one imposes the BPS condition on
the solution then one obtains a naked singularity rather than a black
hole \cite{romans,london}.

Static, spherically symmetric, charged, black hole solutions of
the ${\cal N}=2$, $d=4$ $U(1)^4$ gauged supergravity theory were obtained in
\cite{duffliu}. The solutions carry only electric charges and are
parameterized by their mass $M$ and electric charges
$Q_i$. Alternatively they can be dualized to give purely magnetic
solutions. Once again, they never saturate a BPS bound. One would
expect the existence of dyonic solutions of this theory but such
solutions have not been constructed.

Static, spherically symmetric, charged, black hole solutions of $d=5$
$U(1)^3$ gauged supergravity were obtained in \cite{behrndt}. They are
parameterized by their mass $M$ and electric charges $Q_i$. If the
charges are set equal to each other then one recovers the $d=5$
Reissner-Nordstrom solution of minimal $d=5$ gauged supergravity. The
solutions never saturate a BPS bound.

A static, spherically symmetric, charged black hole solution of $d=6$
$SU(2)$ gauged supergravity was given in \cite{poperomans}. Only a
single abelian component of the gauge field is excited, and the solution
is parametrized by its charge and mass.

Static, spherically symmetric, charged, black hole solutions of $d=7$
$SO(4)$ gauged supergravity are known \cite{elevenauthors}. They can
be embedded into a truncated version of the full theory in which there
are two abelian vectors and two scalars. They are parameterized by
their mass and electric charges.

Electrically charged, asymptotically AdS, black
hole solutions exhibit a Hawking-Page like phase transition in the
bulk, which entails a corresponding phase transition for the dual CFT
at finite temperature in the presence of chemical potentials for the
R-charge. This has been studied in \cite{cejm1,cveticgubser,cejm2}. 

These black holes exhibit an interesting instability \cite{gubsermitra}.
This is best understood for a black hole so large (compared to the AdS
radius) that the curvature of its horizon can be neglected, i.e., it can be
approximated by a black {\it brane}. The dual CFT interpretation is as a
finite temperature configuration in flat space with finite charge
density. For certain regions of parameter space, it turns out that the
entropy increases if the charge density becomes non-uniform (with the
total charge and energy held fixed). Hence the thermal CFT state
exhibits an instability. Using the AdS/CFT dictionary, this maps to a
{\it classical} instability in the bulk in which the horizon becomes
translationally non-uniform, i.e., a Gregory-Laflamme instability. The
remarkable feature of this argument is that it reveals that a classical
Gregory-Laflamme instability should be present precisely when the black
brane becomes {\it locally} thermodynamically unstable. Here, local
thermodynamic stability means having an entropy which is concave down as
a function of the energy and other conserved charges (if the only
conserved charge is the energy then this is equivalent to positivity of
the heat capacity). The Gubser-Mitra (or ``correlated stability'')
conjecture asserts that this correspondence should apply to {\it any}
black brane, not just asymptotically AdS solutions. See
\cite{Harmark:2007md} for more discussion of this correspondence.

For finite radius black holes, the argument is not so clear-cut because the
dual CFT lives in the Einstein universe rather than flat
spacetime so finite size effects will affect the CFT argument and the
Gubser-Mitra conjecture does not apply. Nevertheless, it
should be a good approximation for sufficiently large black holes and hence
there will be a certain range of parameters for which large charged
black holes are classically unstable.\epubtkFootnote{Note that this does not
disagree with the stability result of \cite{kodamacharged} for $d=4$
Reissner-Nordstrom-AdS since the instability involves charged scalars
and hence cannot be seen within {\it minimal} ${\cal N}=2$ gauged
supergravity.}


\subsection{Stationary charged solutions}

The most general known stationary black hole solution of minimal $d=4$,
${\cal N}=2$ gauged supergravity is the Kerr-Newman-AdS solution, which
is uniquely parametrized by its mass $M$, angular momentum $J$ and
electric and magnetic charges $(Q,P)$. The thermodynamic properties of
this solution, and implications for the dual CFT were investigated in
\cite{caldarelli99}. An important property of this solution is that it
can preserve some supersymmetry. This occurs for a 1-parameter subfamily
specified by the electric charge: $M=M(Q)$, $J(Q)$, $P=0$
\cite{kostelecky,caldarelli98}. Hence supersymmetric black holes can
exist in AdS but they exhibit an important qualitative difference from
the asymptotically flat case: they must rotate. 

Charged rotating black hole solutions of more general $d=4$ gauged
supergravity theories, e.g. ${\cal N}=2$ $U(1)^4$ gauged supergravity,
should also exist. Electrically charged, rotating solutions of the
$U(1)^4$ theory, with the four charges set pairwise equal, were
constructed in \cite{cclp4d}. 

Charged, rotating black hole solutions of $d=7$ $SO(4)$ gauged
supergravity have been constructed by truncating to a $U(1)^2$ theory
\cite{cclp7d,chow}. In this theory, one expects the existence of a
topologically spherical black hole solution parametrized by its mass,
three angular momenta, and two electric charges. This general solution
is not yet known. However, solutions with three equal angular momenta
but unequal charges have been constructed \cite{cclp7d}, as have
solutions with equal charges but unequal angular momenta \cite{chow}.
Both types of solution admit BPS limits. 

Charged, rotating black hole solutions of $d=6$ gauged supergravity have not yet been constructed.

The construction of charged rotating black hole solutions of $d=5$
gauged supergravity has attracted more attention
\cite{gr1,gr2,clpequal,cclpold,cclpmin,klr1,cclpsingle,popetwocharge}.
The most general known black hole solution of the minimal theory is that
of \cite{cclpmin}. This solution is parametrized by the conserved
charges of the theory, i.e., the mass $M$, electric charge $Q$ and two
angular momenta $J_1$, $J_2$. Intuition based on results proved for
asymptotically flat solutions suggests that, for this theory, this is
the most general topologically spherical stationary black hole with two
rotational symmetries. In the BPS limit, these solutions reduce to a
2-parameter family of supersymmetric black holes. In other words, one
loses two parameters in the BPS limit (just as for non-static
asymptotically flat black holes in $d=5$ e.g. the BMPV black hole or BPS
black rings).

Analogous solutions of $d=5$ $U(1)^3$ gauged supergravity are expected
to be parametrized by the 6 conserved quantities $M$, $J_1$, $J_2$,
$Q_1$, $Q_2$, $Q_3$. However, a 6-parameter solution is not yet known.
The most general known solutions are the 4-parameter BPS solution of
\cite{klr1}, and the 5-parameter non-extremal solution of
\cite{popetwocharge}, which has two of the charges $Q_i$ equal. The
former is expected to be the general BPS limit of the yet to be
discovered 6-parameter black hole solution (as one expects to lose two
parameters in the BPS limit). The latter solution should be obtained
from the general 6-parameter solution by setting two of the charges
equal.

Supersymmetric AdS black holes have $\Omega_i \ell = 1$, which implies
that they rotate at the speed of light with respect to the conformal
boundary \cite{gr1}. More precisely, the co-rotating Killing field
becomes null on the conformal boundary. Hence the CFT interpretation of
these black holes involves matter rotating at the speed of light on the
Einstein universe. The main motivation for studying supersymmetric AdS
black holes is the expectation that it should be possible to perform a
microscopic CFT calculation of their entropy. The idea is to count
states in weakly coupled CFT and then extrapolate to strong coupling. In
doing this, one has to count only states in short BPS multiplets that do
not combine into long multiplets as the coupling is increased. One way
of trying to do this is to work with an index that receives vanishing
contributions from states in multiplets that can combine into long
multiplets. Unfortunately, such indices do not give agreement with the
black hole entropy \cite{malmin}. This is not a contradiction: indices
count bosonic and fermionic states with opposite signs, and there must
be a cancellation between the contributions from bosonic and fermionic
black hole microstates.

We mentioned above that asymptotically AdS black ring solutions have not
yet been constructed. The simplest starting point is probably to look
for supersymmetric AdS black rings. However, it has been shown that such
solutions do not exist in minimal $d=5$ gauged supergravity \cite{klr2}.
The proof involves classifying supersymmetric near-horizon geometries
(with two rotational symmetries), and showing that $S^1 \times S^2$
topology horizons always suffer from a conical singularity, except in
the limit in which the cosmological constant vanishes. Analogous results
for the $U(1)^3$ theory have also been obtained \cite{kl}. 

\newpage


\section{Acknowledgments} \label{section:acknowledgments} 

RE is supported in part by DURSI 2005 SGR 00082, CICYT FPA
2004-04582-C02-02 and FPA 2007-66665C02-02, and the European Community
FP6 program MRTN-CT-2004-005104. HSR is a Royal Society University
Research Fellow.

\newpage


\end{document}